\documentclass[twocolumn]{aastex631}

\shorttitle{HD 35843}
\shortauthors{Hesse et al.}
\graphicspath{{./}{figures/}}

\begin{document}

\title{HD 35843: A Sun-like star hosting a long period sub-Neptune and inner super-Earth}

\author[0000-0002-2135-9018]{Katharine Hesse}
\affiliation{Kavli Institute for Astrophysics and Space Research 
Massachusetts Institute of Technology 
70 Vassar St 
Cambridge, MA 02139}

\author[0000-0002-4510-2268]{Ismael Mireles}
\affiliation{Department of Physics and Astronomy 
The University of New Mexico 
210 Yale Blvd NE 
Albuquerque, NM 87106}

\author{François Bouchy}
\affiliation{Observatoire astronomique de l\'{}Universit\'{e} de Gen\`{e}ve, 51 Chemin Pegasi, 1290 Versoix, Switzerland}

\author[0000-0003-2313-467X]{Diana Dragomir}
\affiliation{Department of Physics and Astronomy 
The University of New Mexico 
210 Yale Blvd NE 
Albuquerque, NM 87106}

\author[0000-0003-2417-7006]{Sol\`{e}ne Ulmer-Moll}
\affiliation{Observatoire astronomique de l\'{}Universit\'{e} de Gen\`{e}ve, 51 Chemin Pegasi, 1290 Versoix, Switzerland}
\affiliation{Physikalisches Institut, University of Bern, Gesellschaftsstrasse 6, 3012 Bern, Switzerland}

\author{Nora L. Eisner}
\affiliation{Center for Computational Astrophysics, Flatiron Institute,
New York, NY 10010, USA}

\author[0000-0002-3481-9052]{Keivan G.\ Stassun}
\affiliation{Department of Physics and Astronomy, Vanderbilt University, Nashville, TN 37235, USA}

\author[0000-0002-8964-8377]{Samuel N. Quinn}
\affiliation{Center for Astrophysics \textbar \ Harvard \& Smithsonian, 60 Garden Street, Cambridge, MA 02138, USA}

\author[0000-0002-4047-4724]{Hugh P. Osborn}
\affiliation{Physikalisches Institut, University of Bern, Gesellsschaftstrasse 6, 3012 Bern, Switzerland}

\author[0000-0001-9047-2965]{Sergio G.\ Sousa}
\affiliation{Instituto de Astrof\'{i}sica e Ci\^{e}ncias do Espa\c{c}o, Universidade do Porto, CAUP,
Rua das Estrelas, 4150-762 Porto, Portugal}

\author[0000-0001-8621-6731]{Cristilyn N.\ Watkins}
\affiliation{Center for Astrophysics \textbar \ Harvard \& Smithsonian, 60 Garden Street, Cambridge, MA 02138, USA}

\author[0000-0001-6588-9574]{Karen A.\ Collins}
\affiliation{Center for Astrophysics \textbar \ Harvard \& Smithsonian, 60 Garden Street, Cambridge, MA 02138, USA}

\author[0000-0001-7904-4441]{Edward M. Bryant}
\affiliation{Mullard Space Science Laboratory, University College London, Holmbury St Mary, Dorking, Surrey RH5 6NT, UK}

\author{Jonathan~M.~Irwin}
\affiliation{Institute of Astronomy, University of Cambridge, Madingley
  Road, Cambridge, CB3 0HA, United Kingdom}

\author[0000-0002-3439-1439]{Coel Hellier}
\affiliation{Astrophysics Group, Keele University, Staffordshire ST5 5BG, UK}

\author[0000-0002-5099-8185]{Marshall C. Johnson}
\affiliation{Department of Astronomy, The Ohio State University, 4055 McPherson Laboratory, 140 West 18$^{\mathrm{th}}$ Ave., Columbus, OH 43210 USA}

\author{Carl Ziegler}
\affiliation{Department of Physics, Engineering and Astronomy, Stephen F. Austin State University, 1936 North St, Nacogdoches, TX 75962, USA}

\author[0000-0002-2532-2853]{Steve~B.~Howell}
\affiliation{NASA Ames Research Center, Moffett Field, CA 94035, USA}

\author[0000-0001-7416-7522]{David R. Anderson}
\affiliation{Instituto de Astronom\'ia, Universidad Cat\'olica del Norte, Angamos 0610, 1270709, Antofagasta, Chile}

\author[0000-0001-6023-1335]{Daniel Bayliss}
\affiliation{Department of Physics, University of Warwick, Coventry CV4 7AL, UK}
\affiliation{Centre for Exoplanets and Habitability, University of Warwick, Coventry CV4 7AL, UK}

\author[0000-0001-6637-5401]{Allyson~Bieryla} 
\affiliation{Center for Astrophysics ${\rm \mid}$ Harvard {\rm \&} Smithsonian, 60 Garden Street, Cambridge, MA 02138, USA}

\author{C\'{e}sar Brice\~{n}o}
\affiliation{Cerro Tololo Inter-American Observatory, Casilla 603, La Serena, Chile}

\author{R. Paul Butler}
\affiliation{Department of Terrestrial Magnetism, Carnegie Institution for Science, 5241 Broad Branch Road, NW, Washington, DC 20015, USA}

\author{David~Charbonneau}
\affiliation{Center for Astrophysics \textbar \ Harvard \& Smithsonian, 60 Garden Street, Cambridge, MA 02138, USA}

\author[0000-0001-5383-9393]{Ryan Cloutier}
\affiliation{Department of Physics \& Astronomy, McMaster University, 1280 Main St West, Hamilton, ON, L8S 4L8, Canada}

\author{Jeffrey Crane}
\affiliation{Observatories of the Carnegie Institution for Science, 813 Santa Barbara Street, Pasadena, CA 91101, USA}

\author{Jason Dittmann}
\affiliation{Department of Astronomy, University of Florida, 211 Bryant Space Science Center, Gainesville, FL, 32611, USA}

\author[0000-0003-3773-5142]{Jason D.\ Eastman}
\affiliation{Center for Astrophysics \textbar \ Harvard \& Smithsonian, 60 Garden Street, Cambridge, MA 02138, USA}

\author{Sebastián A. Freigeiro}
\affiliation{Citizen Scientist, Zooniverse c/o University of Oxford, Keble Road, Oxford OX1 3RH, UK}

\author{Benjamin J. Fulton}
\affiliation{Cahill Center for Astronomy \& Astrophysics, California Institute of Technology, Pasadena, CA 91125, USA}
\affiliation{IPAC-NASA Exoplanet Science Institute, Pasadena, CA 91125, USA}

\author[0000-0002-4259-0155]{Samuel Gill}
\affiliation{Department of Physics, University of Warwick, Coventry CV4 7AL, UK}
\affiliation{Centre for Exoplanets and Habitability, University of Warwick, Coventry CV4 7AL, UK}

\author[0000-0002-3164-9086]{Maximilian G{\"u}nther}
\affiliation{European Space Agency (ESA), European Space Research and Technology Centre (ESTEC), Keplerlaan 1, 2201 AZ Noordwijk, The Netherlands}

\author[0009-0006-0871-1618]{Haedam Im}
\affiliation{Kavli Institute for Astrophysics and Space Research 
Massachusetts Institute of Technology 
70 Vassar St 
Cambridge, MA 02139}

\author[0000-0002-4715-9460]{Jon~M.~Jenkins}
\affiliation{NASA Ames Research Center, Moffett Field, CA 94035, USA}

\author[0000-0001-9269-8060]{Michelle Kunimoto}
\affiliation{Department of Physics and Astronomy, University of British Columbia, 6224 Agricultural Road, Vancouver, BC V6T 1Z1, Canada}

\author{Baptiste Lavie}
\affiliation{Observatoire astronomique de l\'{}Universit\'{e} de Gen\`{e}ve, 51 Chemin Pegasi, 1290 Versoix, Switzerland}

\author[0000-0001-9699-1459]{Monika Lendl}
\affiliation{Observatoire astronomique de l\'{}Universit\'{e} de Gen\`{e}ve, 51 Chemin Pegasi, 1290 Versoix, Switzerland}

\author[0000-0003-2527-1598]{Michael B. Lund}
\affiliation{NASA Exoplanet Science Institute, Caltech/IPAC, Mail Code 100-22, 1200 E. California Blvd., Pasadena, CA 91125, USA}

\author[0000-0003-3654-1602]{Andrew W. Mann}
\affiliation{Department of Physics and Astronomy, The University of North Carolina at Chapel Hill, Chapel Hill, NC 27599-3255, USA}

\author{Belinda Nicholson}
\affil{University of Southern Queensland, Centre for Astrophysics, West Street, Toowoomba, QLD 4350 Australia}

\author{David Osip}
\affiliation{Las Campanas Observatory, Chile}

\author[0000-0001-8120-7457]{Martin Paegert}
\affiliation{Center for Astrophysics \textbar \ Harvard \& Smithsonian, 60 Garden Street, Cambridge, MA 02138, USA}

\author[0000-0003-4422-2919]{Nuno C.\ Santos}
\affiliation{Instituto de Astrof\'{i}sica e Ci\^{e}ncias do Espa\c{c}o, Universidade do Porto, CAUP,
Rua das Estrelas, 4150-762 Porto, Portugal}
\affiliation{ Departamento de F\'{i}sica e Astronomia, Faculdade de Ci\^{e}ncias, Universidade do Porto, Rua do Campo Alegre, 4169-007 Porto, Portugal}

\author[0000-0001-8227-1020]{Richard P. Schwarz}
\affiliation{Center for Astrophysics \textbar \ Harvard \& Smithsonian, 60 Garden Street, Cambridge, MA 02138, USA}

\author[0000-0002-6892-6948]{Sara Seager}
\affiliation{Department of Earth, Atmospheric, and Planetary Science, Massachusetts Institute of Technology, Cambridge, MA 02139, USA}

\author{Stephen Shectman}
\affiliation{Observatories of the Carnegie Institution for Science, 813 Santa Barbara Street, Pasadena, CA 91101, USA}

\author{Johanna Teske}
\affiliation{Department of Terrestrial Magnetism, Carnegie Institution for Science, 5241 Broad Branch Road, NW, Washington, DC 20015, USA}

\author[0000-0002-6778-7552]{Joseph D. Twicken}
\affiliation{SETI Institute, Mountain View, CA 94043 USA/NASA Ames Research Center, Moffett Field, CA 94035 USA}

\author{St\'ephane Udry}
\affiliation{Observatoire astronomique de l\'{}Universit\'{e} de Gen\`{e}ve, 51 Chemin Pegasi, 1290 Versoix, Switzerland}

\author[0000-0001-5542-8870]{Vincent Van Eylen}
\affiliation{Mullard Space Science Laboratory, University College London, Holmbury St. Mary, Dorking, Surrey RH5 6NT, UK}

\author[0000-0002-1896-2377]{Jos\'{e} Vin\'{e}s L\'{o}pez}
\affiliation{Instituto de Astronom\'ia, Universidad Cat\'olica del Norte, Angamos 0610, 1270709, Antofagasta, Chile}

\author[0000-0002-6937-9034]{Sharon X.~Wang}
\affiliation{Department of Astronomy, Tsinghua University, Beijing 100084, People's Republic of China}

\author[0000-0003-1452-2240]{Peter J.\ Wheatley}
\affiliation{Department of Physics, University of Warwick, Coventry CV4 7AL, UK}
\affiliation{Centre for Exoplanets and Habitability, University of Warwick, Coventry CV4 7AL, UK}

\author[0000-0002-4265-047X]{Joshua N.\ Winn}
\affiliation{Department of Astrophysical Sciences, Princeton University, 4 Ivy Lane, Princeton, NJ 08544, USA}

\author{Edward E. Zuidema}
\affiliation{Citizen Scientist, Zooniverse c/o University of Oxford, Keble Road, Oxford OX1 3RH, UK}




\begin{abstract}

We report the discovery and confirmation of two planets orbiting the metal-poor Sun-like star, HD 35843 (TOI 4189). HD 35843 c is a temperate sub-Neptune transiting planet with an orbital period of 46.96 days that was first identified by Planet Hunters TESS. We combine data from TESS and follow-up observations to rule out false-positive scenarios and validate the planet. We then use ESPRESSO radial velocities to confirm the planetary nature and characterize the planet's mass and orbit. Further analysis of these RVs reveals the presence of an additional planet, HD 35843 b, with a period of 9.90 days and a minimum mass of $5.84\pm0.84$ $M_{\oplus}$. For HD 35843 c, a joint photometric and spectroscopic analysis yields a radius of $2.54 \pm 0.08 R_{\oplus}$, a mass of $11.32 \pm 1.60 M_{\oplus}$, and an orbital eccentricity of $e = 0.15\pm0.07$. With a bulk density of $3.80 \pm 0.70$ g/cm$^3$, the planet might be rocky with a substantial H$_2$ atmosphere or it might be a ``water world". With an equilibrium temperature of $\sim$480 K, HD 35843 c is among the coolest $\sim 5\%$  of planets discovered by TESS. Combined with the host star's relative brightness (V= 9.4), HD 35843 c is a promising target for atmospheric characterization that will probe this sparse population of temperate sub-Neptunes.

\end{abstract}



\section{Introduction \label{sec:intro}}

In the last two decades, the number of confirmed exoplanets has grown considerably, rising from less than 500 in the early 2000s to more than 5000 today and significantly modifying our understanding of planet composition and system architecture. Originally, it was anticipated that planets would be similar to those found in the solar system, but the planets discovered so far show that this is anything but the case. Few exoplanets have been found that are true analogs to those in our solar system. In particular, solar system planets have much longer periods than the majority of the exoplanets detected thus far, due to observational biases in current exoplanet detection methods \citep{Winn_2015}. In terms of composition, exoplanets have been discovered that are similar to the planets of our solar system but also those with composition unparalleled in our solar system counterparts like super-Earths and sub-Neptunes. Prior to the \textit{Kepler} mission, exoplanet detections smaller than Neptune were relatively rare \citep{Borucki_2010}. However, \textit{Kepler} has contributed 91\% of sub-Neptune size planets to date\footnote{\url{https://exoplanetarchive.ipac.caltech.edu/} accessed 20 May 2024}, with the majority of targets falling into two radius regimes: Earth-size planets (R$_p$ $<$ 1.5 R$_\oplus$) and sub-Neptune size planets with R$_p$ between 2 and 4 R$_\oplus$. Since its launch in 2018, the Transiting Exoplanet Survey Satellite (\textit{TESS}) has discovered more than 7000 planet candidates and confirmed more than 400 planets, with more than 30\% of those confirmations fitting into the sub-Neptune radius regime. Though great strides have been made in exoplanet research, there are still many questions, particularly about exoplanets unlike those within the solar system as well as those at longer orbital periods.

Of the more than 4000 transiting exoplanets confirmed so far, only $\sim$12\% have a period greater than 40 days. There are even fewer confirmed transiting sub-Neptunes with a period longer than 40 days and a well-constrained mass, with only 24 at the time of this writing. Relatively little is known about this population of long-period, temperate sub-Neptune planets. This is especially true in terms of composition. Planets in this category might have compositions similar to Neptune, or similar to the Earth. Their composition may even be something new and unlike anything in our solar system, such as a ``water world". This theoretical type of planet consists of a significant fraction of water exceeding that found on Earth, making these planets less dense than terrestrial planets like Earth. The focus in differentiating the significant number of exoplanet candidates larger than the Earth but smaller than Neptune has been on the radius boundaries of this region, but recent work suggests it may be planet density, not radius alone, that determines the composition of planets in this radius range \citep{Luque_2022}. Particularly, planets with radii between 1.5 R$_\oplus$ and 2.5 R$_\oplus$ can have bulk densities consistent with a wide variety of compositions \citep{Benneke_2024}.  These less dense sub-Neptunes are particularly interesting candidates for atmospheric follow-up as they may hold the key to understanding the formation of planets in this radius range.

In addition to composition, system architecture may hold clues about a planet system's formation history. Only a handful of confirmed multi-planet systems so far have been found to consist of a non-transiting planet interior to an outer transiting planet and this could indicate a non-negligible difference in the orbital inclinations of the two planets. In addition to orbital eccentricities, any differences in inclination between planets may help provide information about possible planet formation scenarios that could explain these types of planetary systems. This could include planets potentially having been perturbed, particularly when misalignments and eccentricities are significant. 

Here, we present the discovery and confirmation of the HD\,35843\ system, which consists of a metal-poor Sun-like star hosting two sub-Neptunes: HD\,35843\ b and c, with orbital periods of 9.90 and 46.96 days, respectively. HD\,35843\ b has a minimum mass of $5.84\pm0.84$ $M_{\oplus}$ and has not been observed to transit. HD\,35843\ c has a radius of $2.54 \pm 0.08$ $R_{\oplus}$ and a mass measurement of $11.32 \pm 1.60 M_{\oplus}$ and a slightly eccentric orbit ($e = 0.15 \pm 0.07$). In Section \ref{sec:observations}, we describe the TESS and follow-up observations used to detect and confirm the two planets. We present a detailed analysis of the data used to characterize the host star, detect the non-transiting HD\,35843\ b, and determine the physical and orbital parameters of the two planets in Section \ref{sec:analysis}. We discuss the broader context of this system, including the possible composition of the outer planet, and prospects for atmospheric characterization in Section \ref{sec:discussion}.

\section{Observations} \label{sec:observations}

\subsection{TESS Photometry\label{subsec:tess}}

\begin{figure*}
    \centering
    \includegraphics[width=0.32\linewidth]{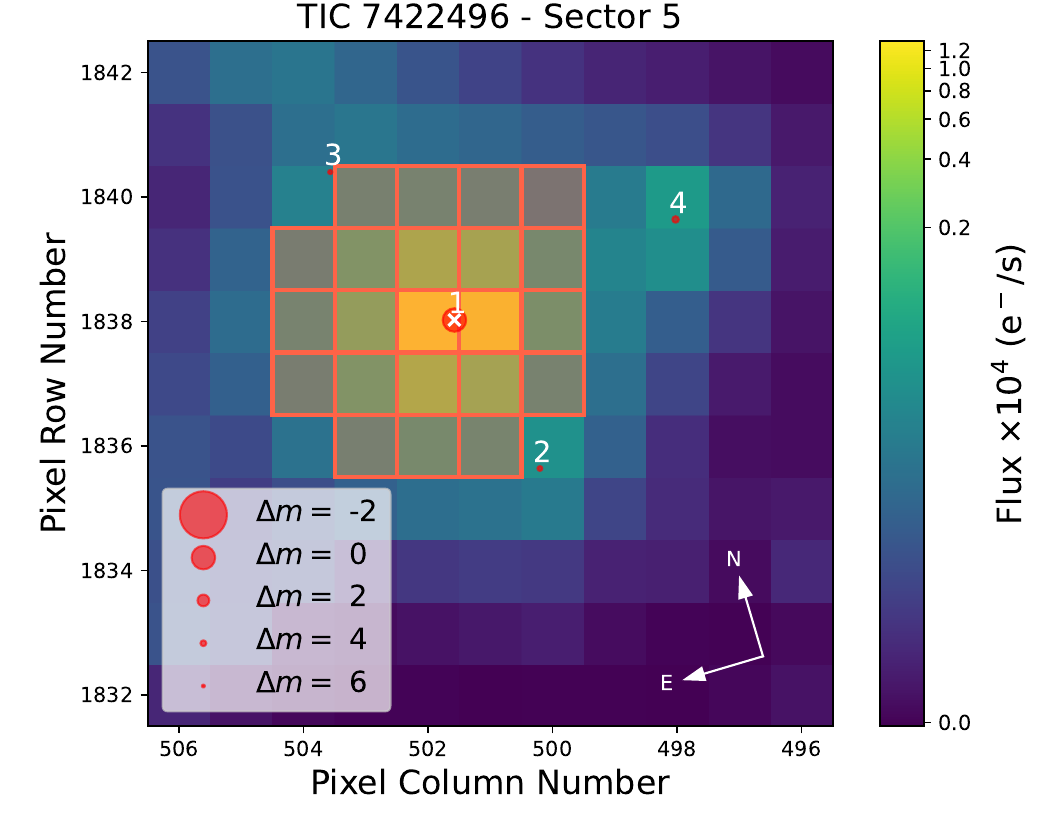}
    \includegraphics[width=0.32\linewidth]{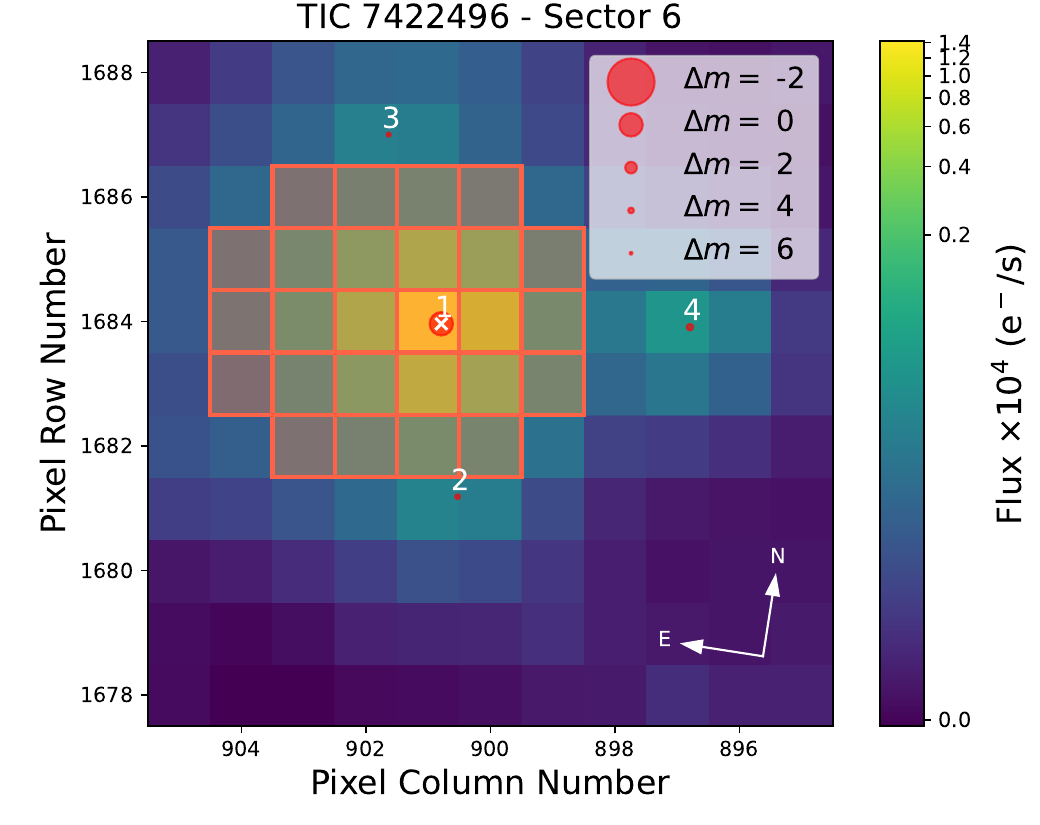}
    \includegraphics[width=0.32\linewidth]{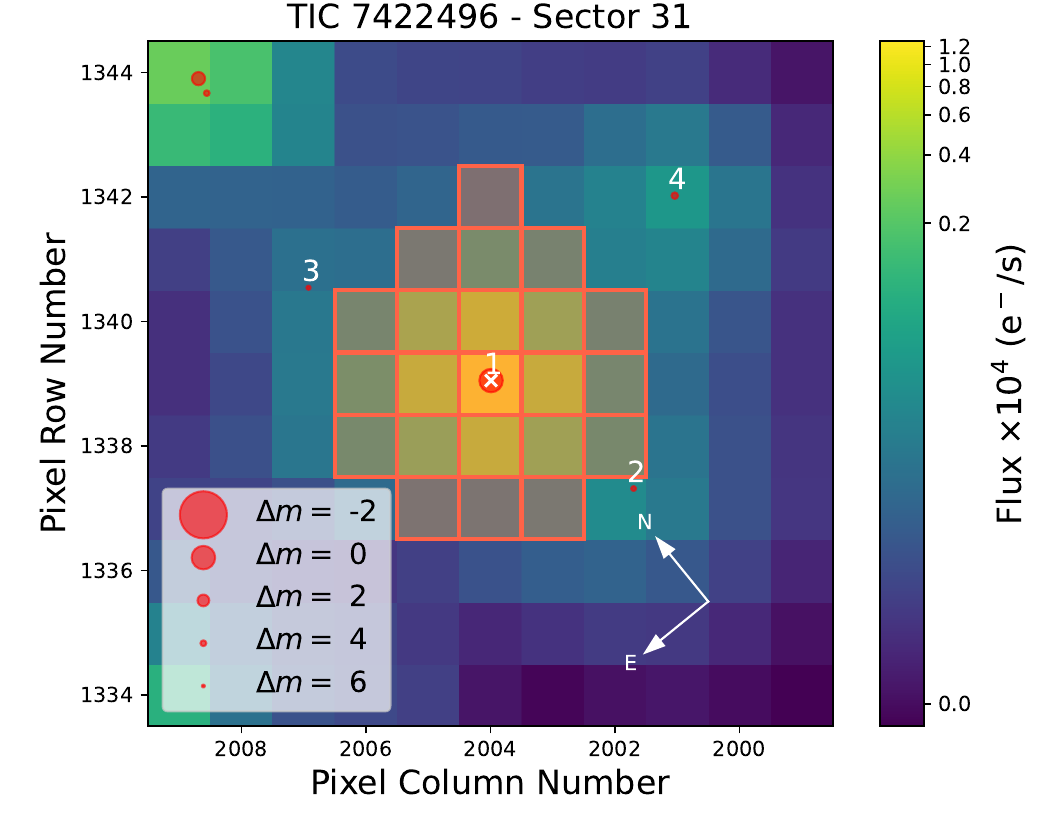}
    \includegraphics[width=0.32\linewidth]{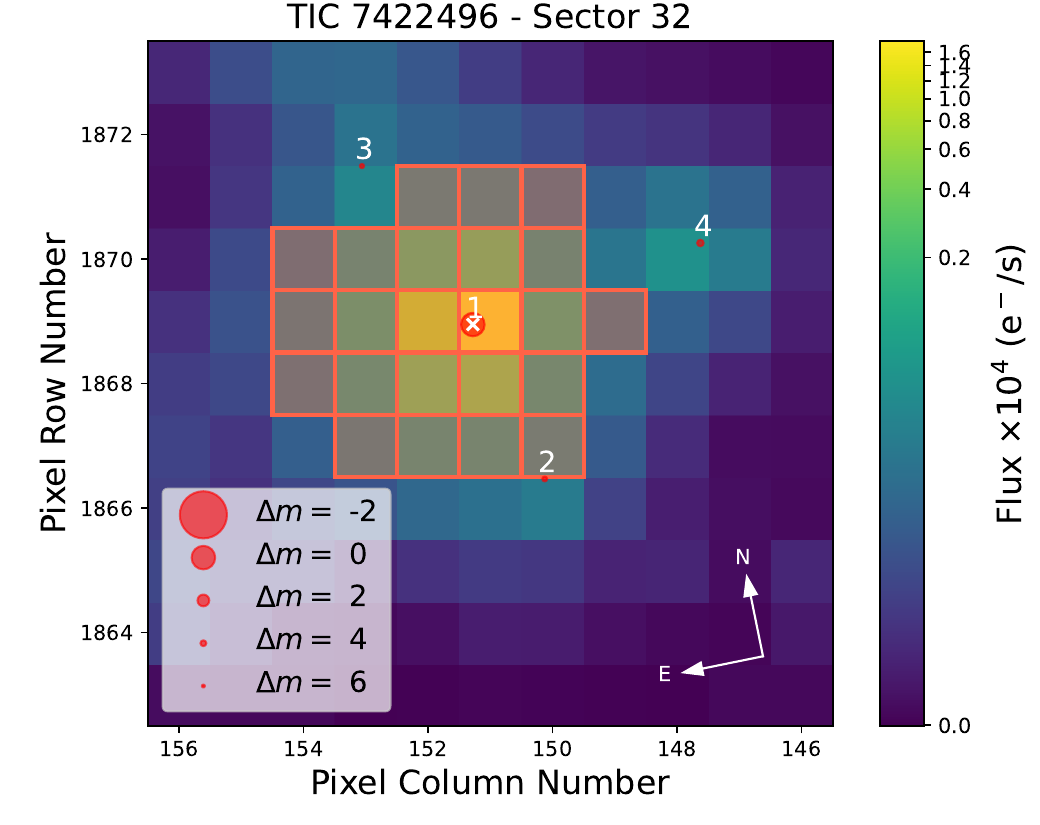}
    \includegraphics[width=0.32\linewidth]{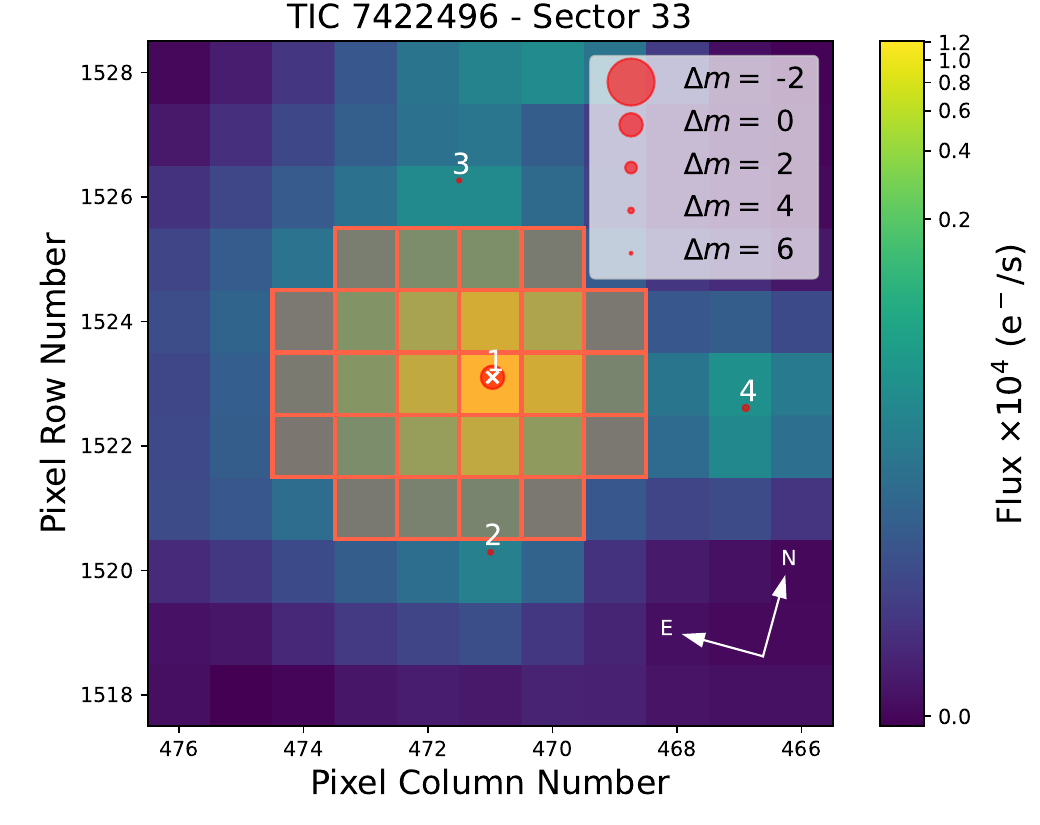}
    \caption{TESS target pixel files of HD 35843 (TIC 7422496) for all 5 sectors of observations made using \texttt{tpfplotter} \cite{2020A&A...635A.128A}. The SPOC photometric aperture is highlighted in each panel. }
    \label{fig:tpf_ap}
\end{figure*}

HD 35843 (TOI 4189, TIC 7422496; V = 9.36) was observed by \textit{TESS} Camera 3 for a total of 5 sectors. It was observed in 2-minute cadence in Sectors 5 and 6  (UT 2018 November 15 to UT 2019 January 7) and Sectors 31-33 (UT 2020 October 22 to UT 2021 January 13). \textit{TESS} observations were interrupted between each of the 13.7 day long orbits of the satellite when data were downloaded to Earth. The data were processed by the Science Processing Operations Center (SPOC; \citet{2016SPIE.9913E..3EJ}) pipeline which produced two light curves per sector called Simple Aperture Photometry (SAP; \citet{Twicken_2010,Morris_2020}) and Presearch Data Conditioning Simple Aperture Photometry (PDCSAP; \citet{2012PASP..124.1000S, 2012PASP..124..985S, 2014PASP..126..100S}). Figure \ref{fig:tpf_ap} shows the photometric aperture used in each sector along with nearby stars marked from the Gaia DR2 catalog, showing that the target is relatively isolated.

\begin{figure*}
    \centering
    \includegraphics[width=\textwidth]{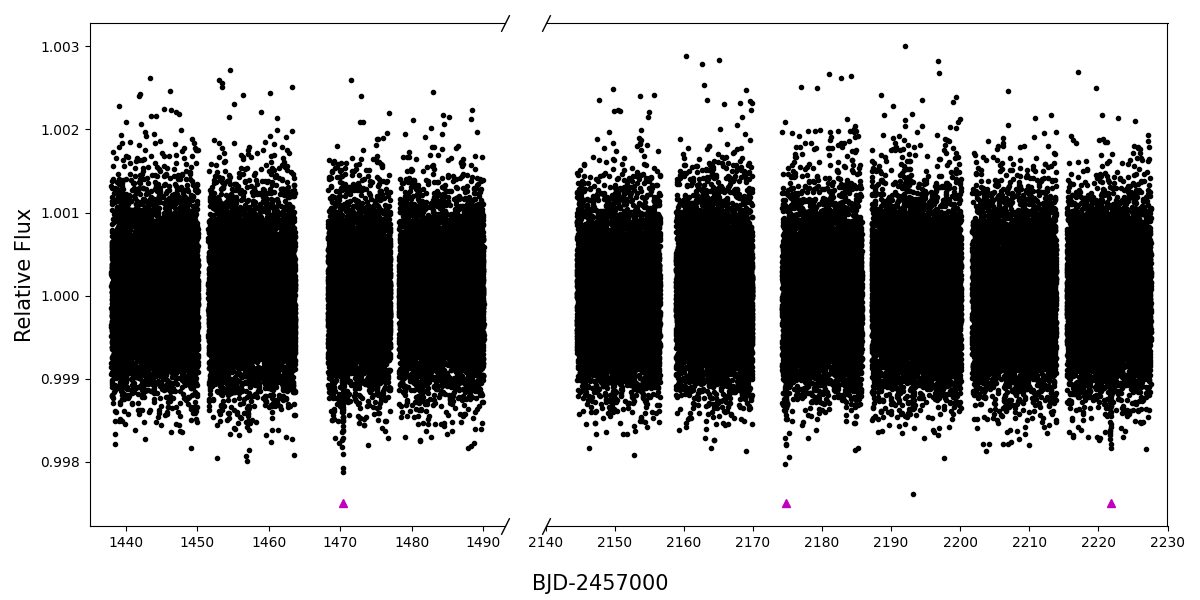}
    \includegraphics[width=\textwidth]{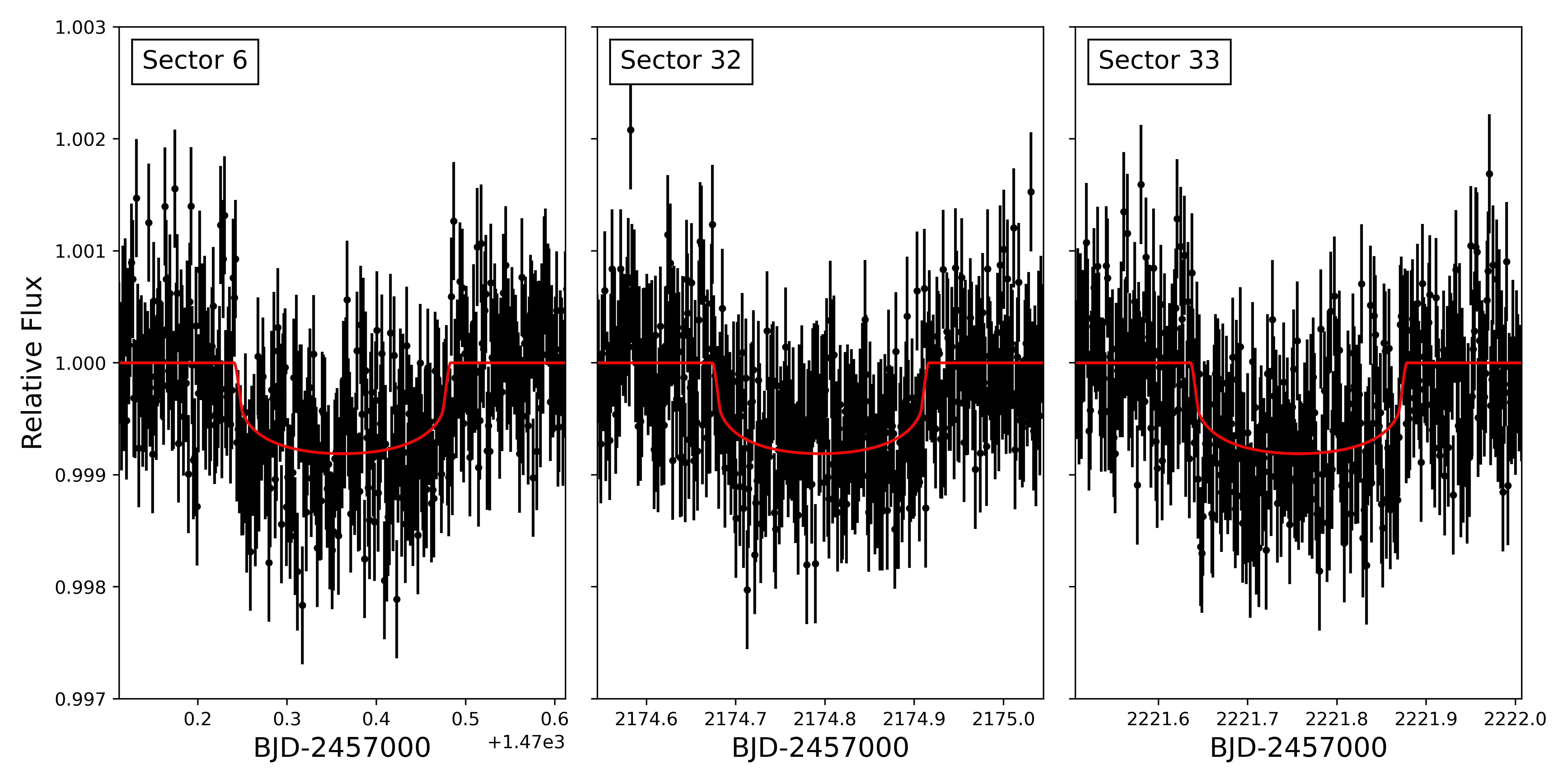}
    \caption{\textit{Top:} Full TESS PDCSAP lightcurve of HD 35843. Transits of HD 35843 c are marked with dashed magenta lines. \textit{Bottom:} Zoom-in of the 3 individual transits from Sectors 6, 32, and 33 along with the best-fit transit light curve.}
    \label{fig:tess_lc}
\end{figure*}

HD 35843 c was first identified as a planet candidate nearly 18 months prior to becoming a \textit{TESS} Object of Interest (TOI), by the Planet Hunters TESS citizen science project \citep{Eisner_2021}. In brief, Planet Hunters TESS harnesses the power of over 40,000 registered citizen scientists who visually inspect all of the TESS two-minute cadence SPOC light curves in search of transit events. As of February 2024, Planet Hunters TESS has contributed 183 cTOIs (Community TESS Objects of Interest), 109 of which have been promoted to TOI status, and 19 of which have become confirmed planets. HD 35843 c was initially identified by citizen scientists who visually inspected the Sector 6 TESS data. Out of the 15 citizen scientists who visually inspected this light curve, 14 identified the transit event. The candidate was vetted for instrumental and astrophysical false positives using the LATTE vetting suite \citep[for details see ][]{latte2020eisner} before being uploaded to ExoFOP as a cTOI on 2020-01-24. Following review by the TESS vetting team of the SPOC Sector 1-36 multi-sector data validation report \citep{Twicken_2018, Li_2019} (made with TESS data from Sectors 1 through 13 and Sectors 26 through 36), the target was promoted to TOI status on 2021-07-08 with master priority 1 (where 1 is highest priority and 5 lowest priority). Figure \ref {fig:tess_lc} shows the full detrended light curve for HD 35843 and zoom-ins of the 3 individual transits along with best-fit model.

\textit{TESS} has not re-observed HD 35843 (TIC 7422496) since Sector 33 but is scheduled to observe it in Sector 87 in December 2024 and Sector 96 in December 2025 according to the \textit{TESS-point Web Tool}\footnote{\url{https://heasarc.gsfc.nasa.gov/wsgi-scripts/TESS/TESS-point_Web_Tool/TESS-point_Web_Tool/wtv_v2.0.py/}}. A transit of HD 35843\,c is expected to occur during the second half of the Sector 87 observations.

\subsection{Ground-based Photometry\label{subsec:ground}}

The \textit{TESS} pixel scale is $\sim 21\arcsec$ pixel$^{-1}$ and photometric apertures typically extend out to roughly 1 arcminute, generally causing multiple stars to blend in the \textit{TESS} photometric aperture. To attempt to determine the true source of the \textit{TESS} detection, we acquired ground-based time-series follow-up photometry of the field around TOI-4189 as part of the \textit{TESS} Follow-up Observing Program \citep[TFOP;][]{collins:2019}\footnote{https://tess.mit.edu/followup}. We used the {\tt TESS Transit Finder}, which is a customized version of the {\tt Tapir} software package \citep{Jensen:2013}, to schedule our transit observations.

\subsubsection{Las Cumbres Observatory (LCOGT)\label{subsubsec:lco}}

\begin{figure*}
    \centering
    \includegraphics[width=0.32\textwidth]{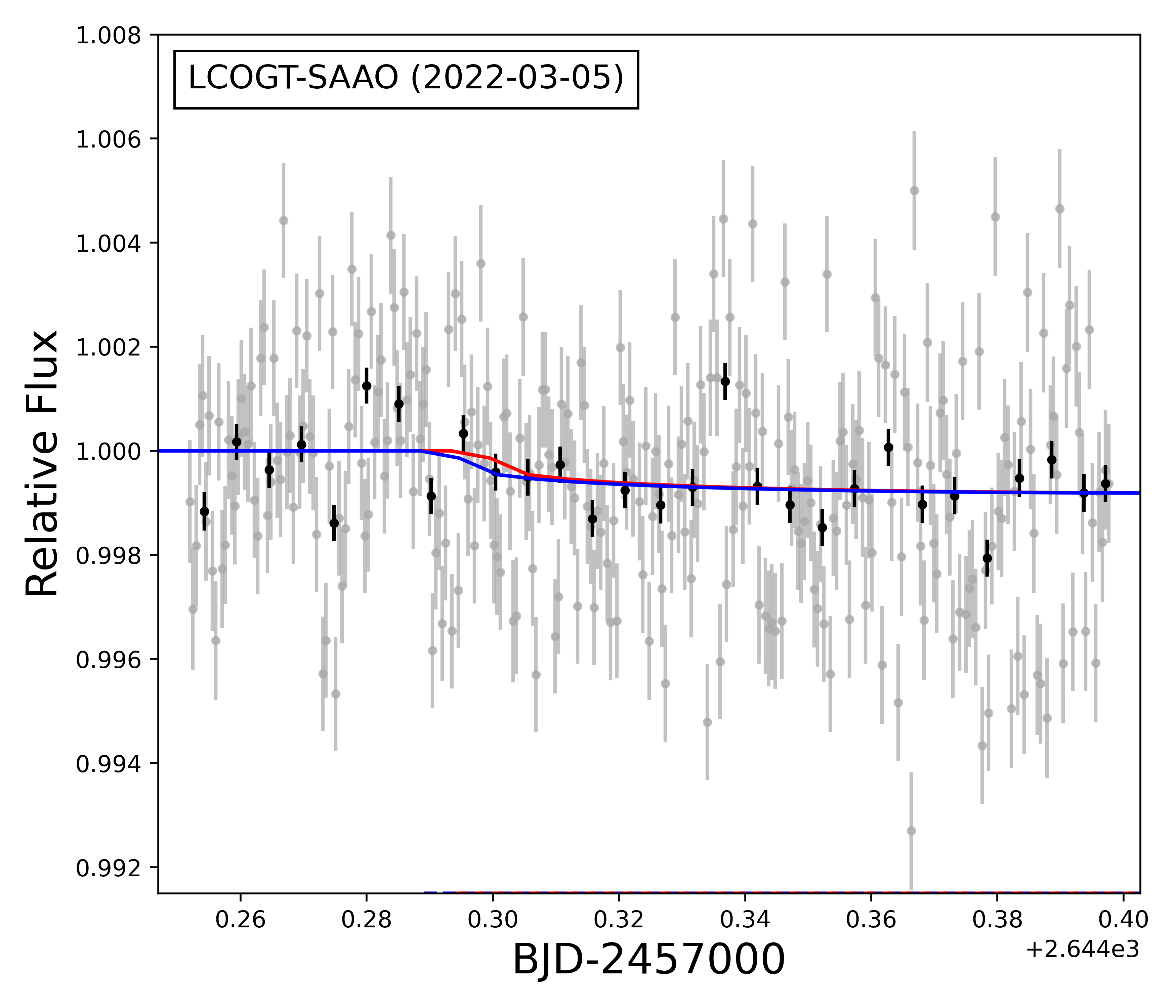}
    \includegraphics[width=0.32\textwidth]{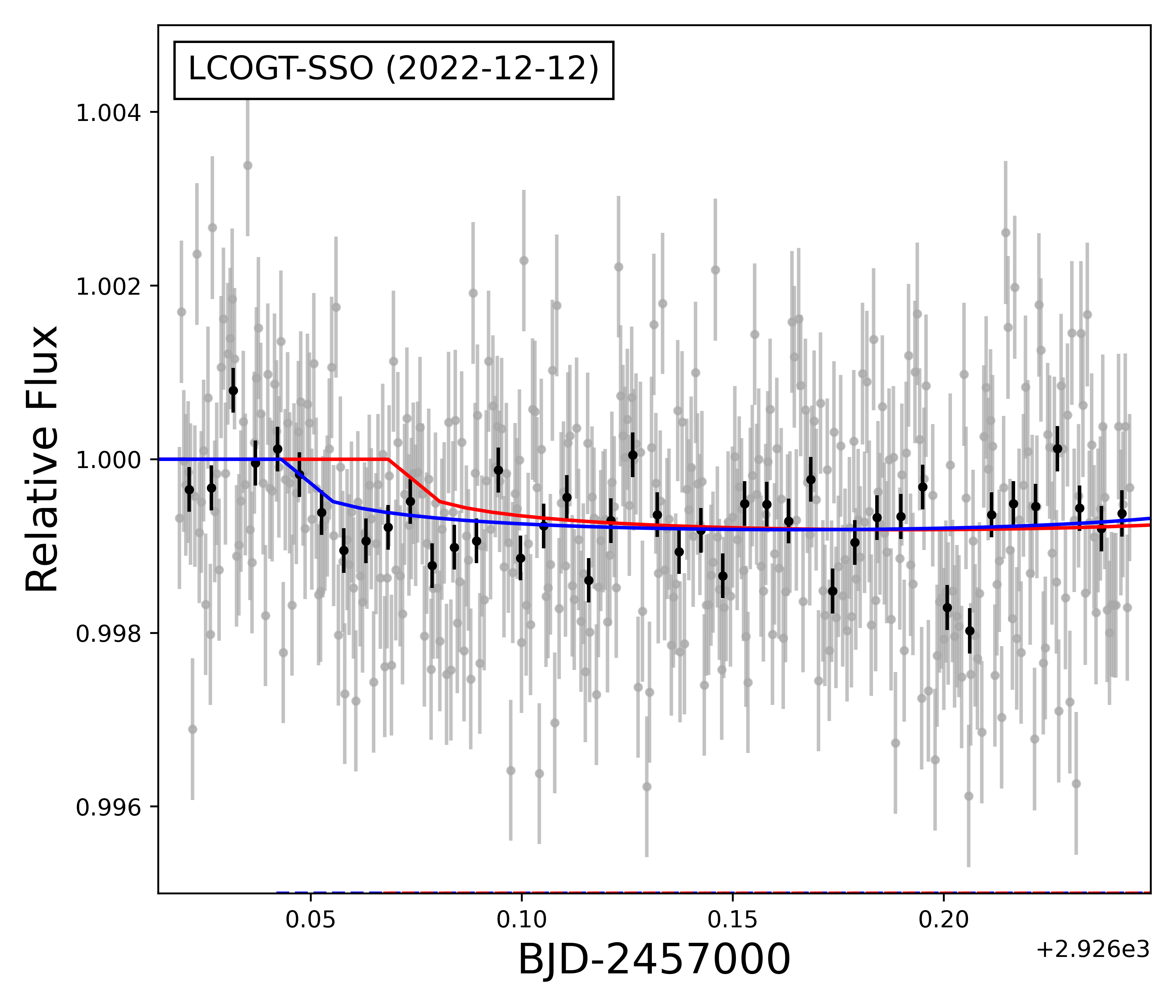}
    \includegraphics[width=0.32\textwidth]{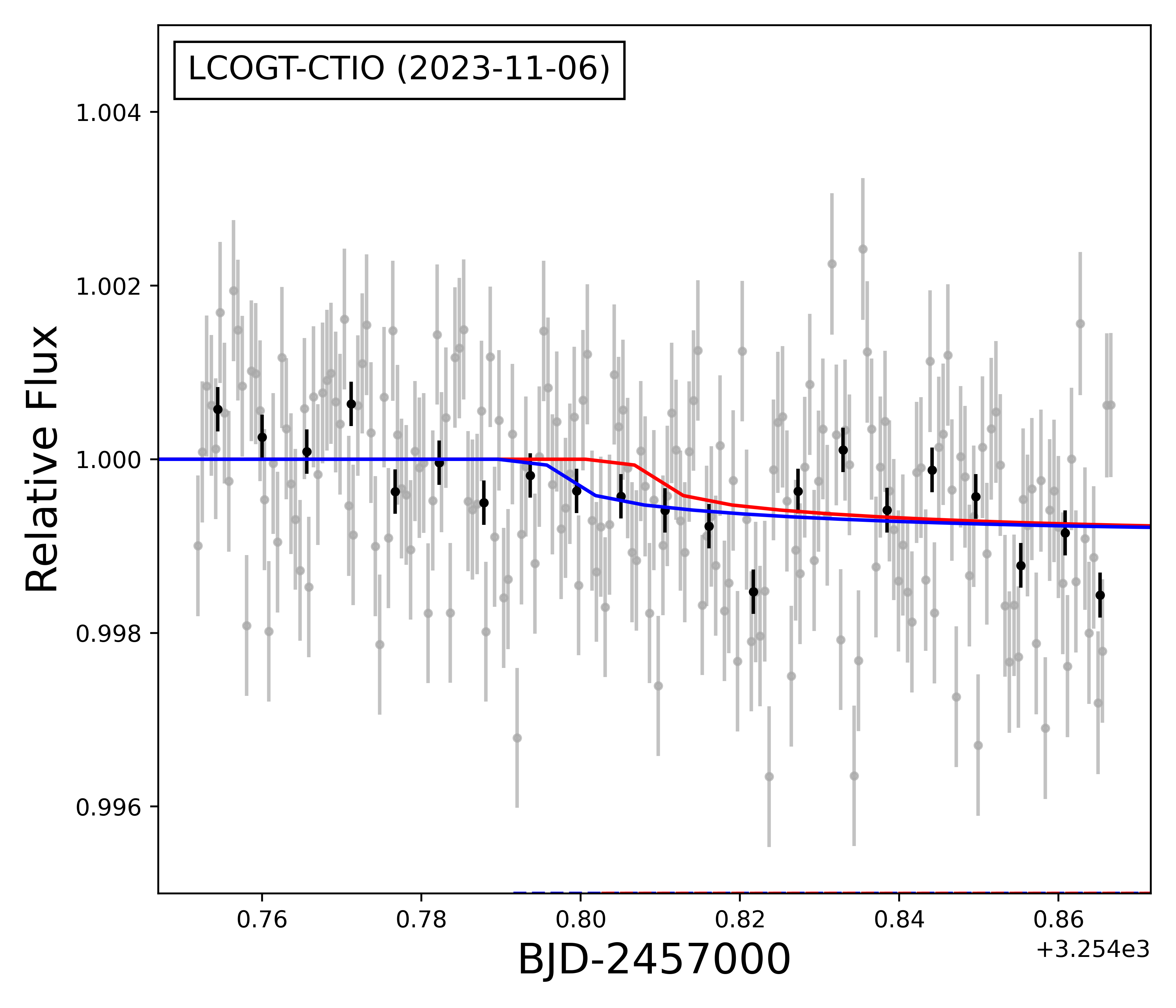}
    \includegraphics[width=0.32\textwidth]{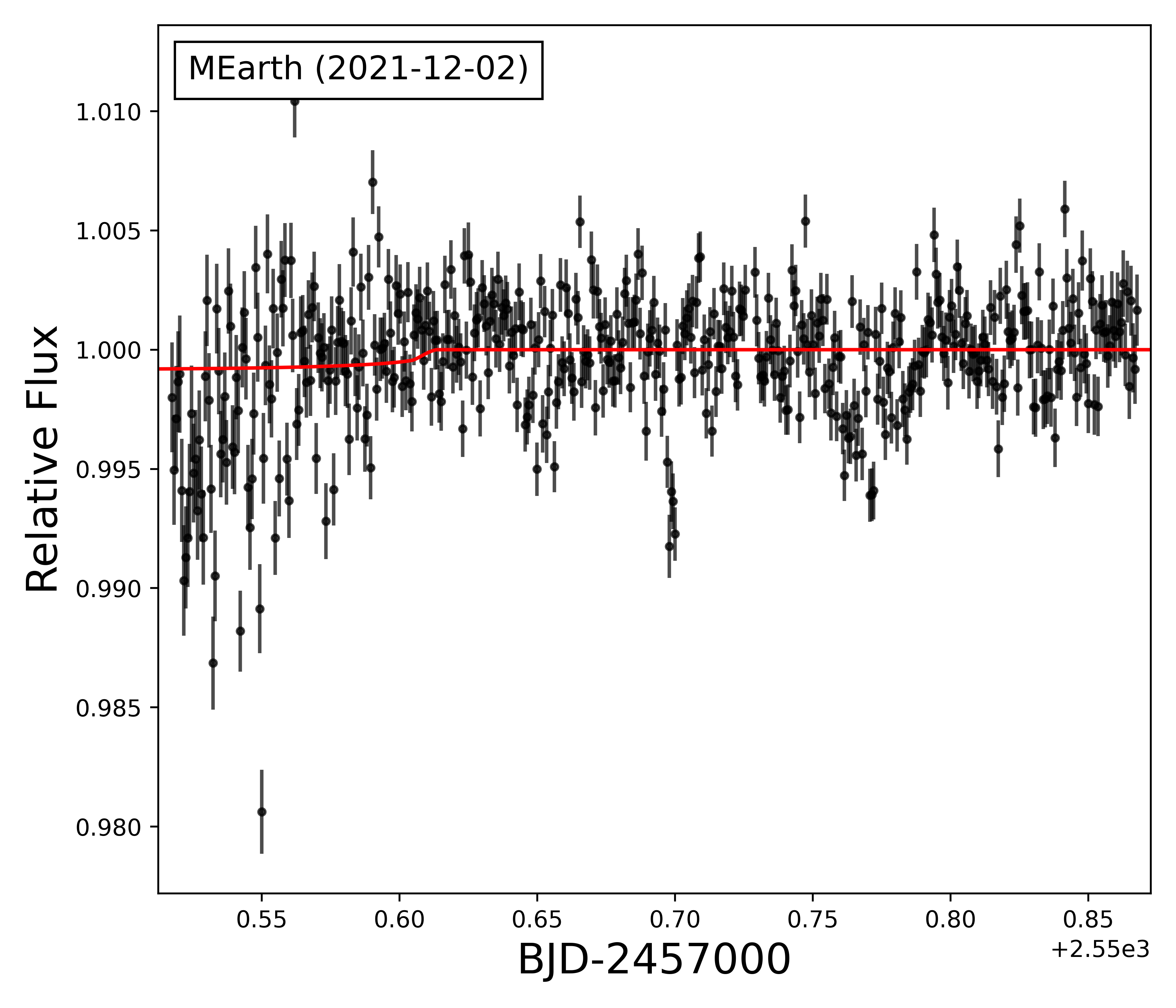}
    \includegraphics[width=0.32\textwidth]{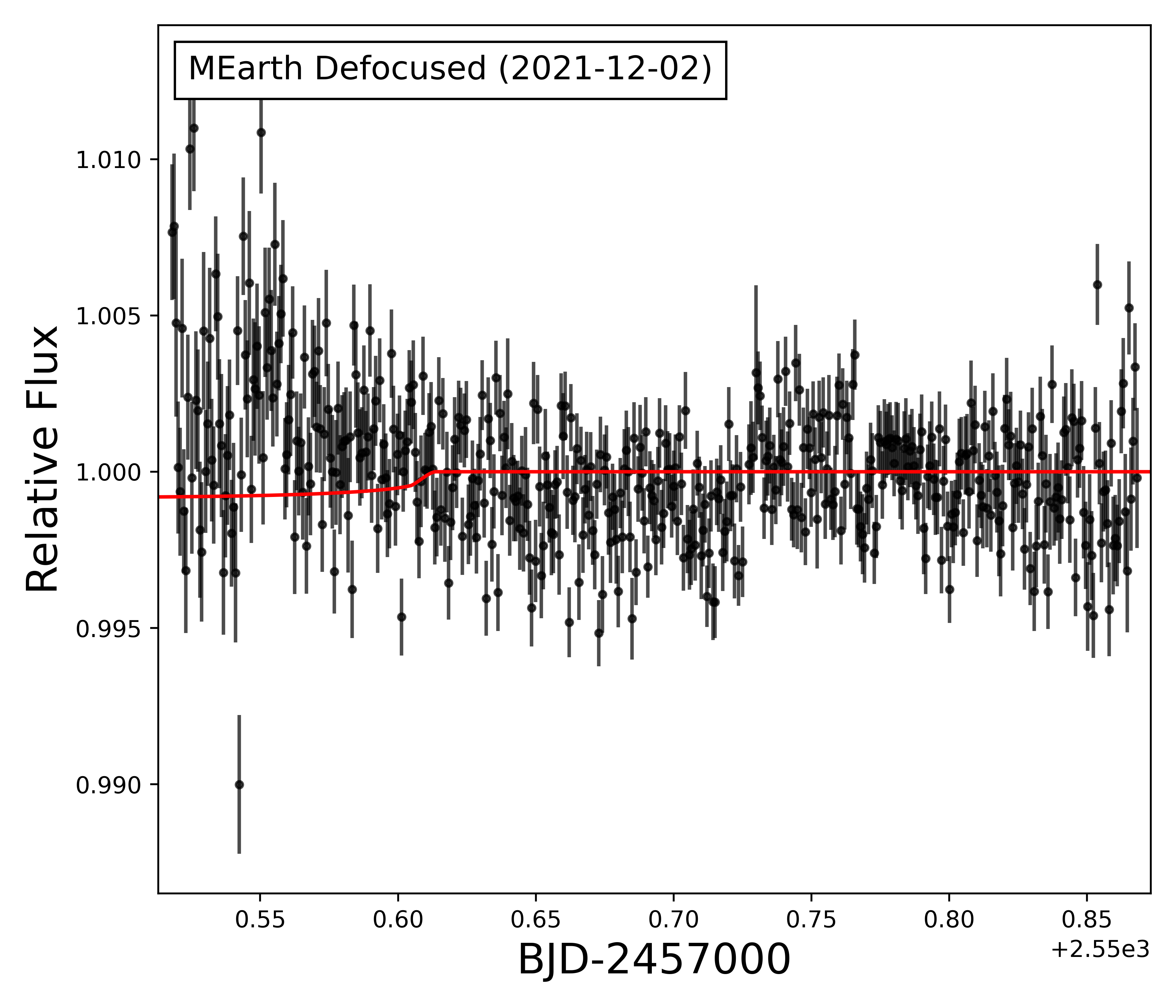}
    \includegraphics[width=0.32\textwidth]{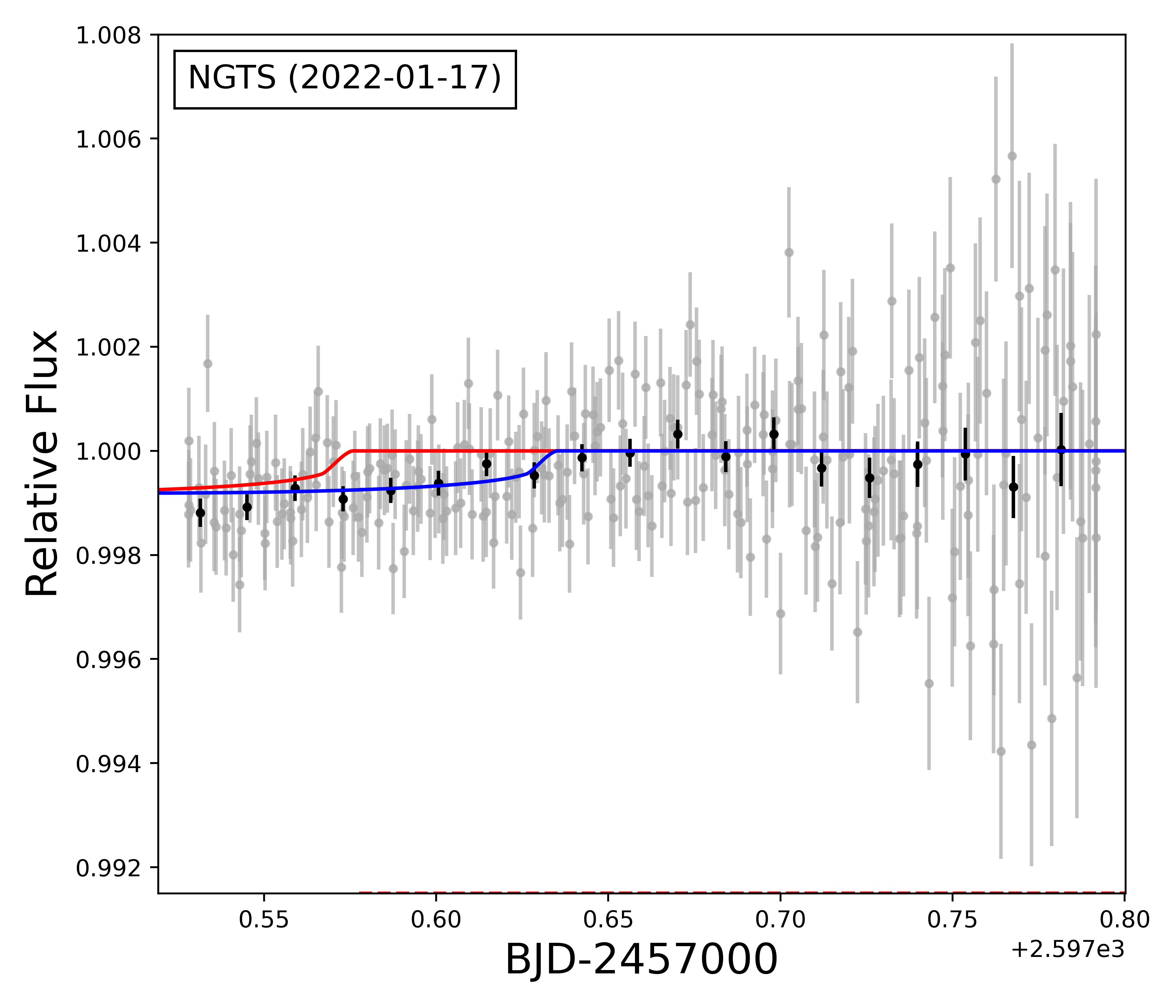}
    \caption{Ground-based photometry of HD 35843 b from LCOGT, MEarth, and NGTS along with the best-fit transit light curve from the joint TESS+ESPRESSO fit in red. Blue curves account for the potential TTVs discussed in Section \ref{subsec:ttvs}.}
    \label{fig:ground_photometry}
\end{figure*}

Three partial transit windows of TOI-4189.01 were observed on UTC 2022 March 05, 2022 December 12, and 2023 November 06 in Pan-STARRS $z$-short band from the Las Cumbres Observatory Global Telescope \citep[LCOGT;][]{Brown:2013} 1\,m network nodes at South Africa Astronomical Observatory near Sutherland, South Africa (SAAO), Siding Spring Observatory near Coonabarabran, Australia (SSO), and Cerro Tololo Inter-American Observatory in Chile (CTIO), respectively. The 1\,m telescopes are equipped with a $4096\times4096$ SINISTRO camera having an image scale of $0\farcs389$ per pixel, resulting in a $26\arcmin\times26\arcmin$ field of view and the images were calibrated by the standard LCOGT {\tt BANZAI} pipeline \citep{McCully:2018}, and differential photometric data were extracted using {\tt AstroImageJ} \citep{Collins:2017}. We used circular photometric apertures with radius $5\farcs5$--$7\farcs8$ that excluded all of the flux from the nearest known neighbor in the Gaia DR3 catalog (Gaia DR3 4799563663371481472), which is $13\farcs6$ southwest of TOI-4189. The light curve data are available on the {\tt EXOFOP-TESS} website\footnote{\href{https://exofop.ipac.caltech.edu/tess/target.php?id=7422496}https://exofop.ipac.caltech.edu/tess/target.php?id=7422496} and are shown in Figure \ref{fig:ground_photometry}. 

\subsubsection{MEarth-South\label{subsubsec:mearth}}

A predicted egress of TOI~4189 was observed using the MEarth-South telescope array \citep{2015csss...18..767I} at Cerro Tololo Inter-American Observatory (CTIO), Chile on UT 2021 December 2, with the aim of ruling out false positives due to eclipse events occurring in nearby contaminating stars that were unresolved from the target in the TESS photometric aperture. 4 telescopes were operated defocused to 12 pixels half flux diameter (HFD), or approximately 10.1 arcsec (where the detector pixel scale is approximately 0.84 arcsec/pix) to obtain a light curve for the target star and any bright, isolated surrounding field stars. 1 telescope was operated in focus using the same exposure time, saturating the target star but obtaining better angular resolution and sensitivity to faint nearby contaminants. The exposure times were 38s for all telescopes, observing continuously for the entire night. A meridian flip occurred at approximately UT 05:30, well after the predicted egress, and was taken into account in the analysis by allowing for separate magnitude zero-points on either side of the meridian to remove residual flat fielding errors.

Data were reduced using the standard MEarth reduction pipeline
(e.g. \citealt{2007MNRAS.375.1449I,2012AJ....144..145B}) using
aperture photometry with an extraction aperture of $r = 17$ pixels (14.3 arcsec) for the defocused time series, and $r = 6$ pixels (5 arcsec) for the in-focus time series. Combining these two time series, we did not detect any deep event on field stars within 2.5 arcminutes of the target star, which is sufficient, given the relatively low uncertainty in the predicted time of egress, to rule these out as the source of the event, consistent with the TESS detected event having occurred on target.

We do not consider these data to provide positive confirmation of an event on target due to the difficulty of detecting shallow transits in time series with partial coverage of the transit window, particularly during times when the target star is rising, where any observed slopes or trends might be due to the effects of differential atmospheric extinction rather than astrophysical events. Further follow-up observations targeting better-placed transit windows were therefore scheduled.

\subsubsection{Next Generation Transit Survey (NGTS)\label{ngts}}
The Next Generation Transit Survey \citep[NGTS;][]{wheatley18ngts} is a ground-based photometric facility located at ESO's Paranal Observatory in Chile. NGTS consists of twelve robotic telescopes each with an aperture diameter of 20\,cm and a field-of-view of 2.6$\times$2.6 degrees. Using multiple NGTS telescopes to simultaneously observe the same star yields a significant improvement in the photometric precision achieved allowing for high-precision photometry of exoplanet host stars \citep{bryant2020ngtsmulticam}.

TOI-4189 was observed using seven NGTS telescopes on 2022 January 17, with all telescopes observing at a cadence of 13\,s (10\,s exposures and 3\,s read-out time) and with the custom NGTS filter (520 -- 890\,nm). The NGTS observations were reduced and the light curves were extracted using a custom aperture photometry pipeline using a photometric aperture with a radius of 7 pixels (35\arcsec). To create relative time series photometry we select unblended comparison stars similar in magnitude, colour, and CCD position to TOI-4189 using \textit{Gaia} DR2 \citep{gaiaDR2_2018}.  The NGTS lightcurves for TOI-4189 are presented in Figure~\ref{fig:ground_photometry} and the data are available in full via the ExoFOP TESS database (\url{https://exofop.ipac.caltech.edu/tess}). 

\subsubsection{WASP Photometry}
\begin{figure}
    \centering
    \includegraphics[width=\linewidth]{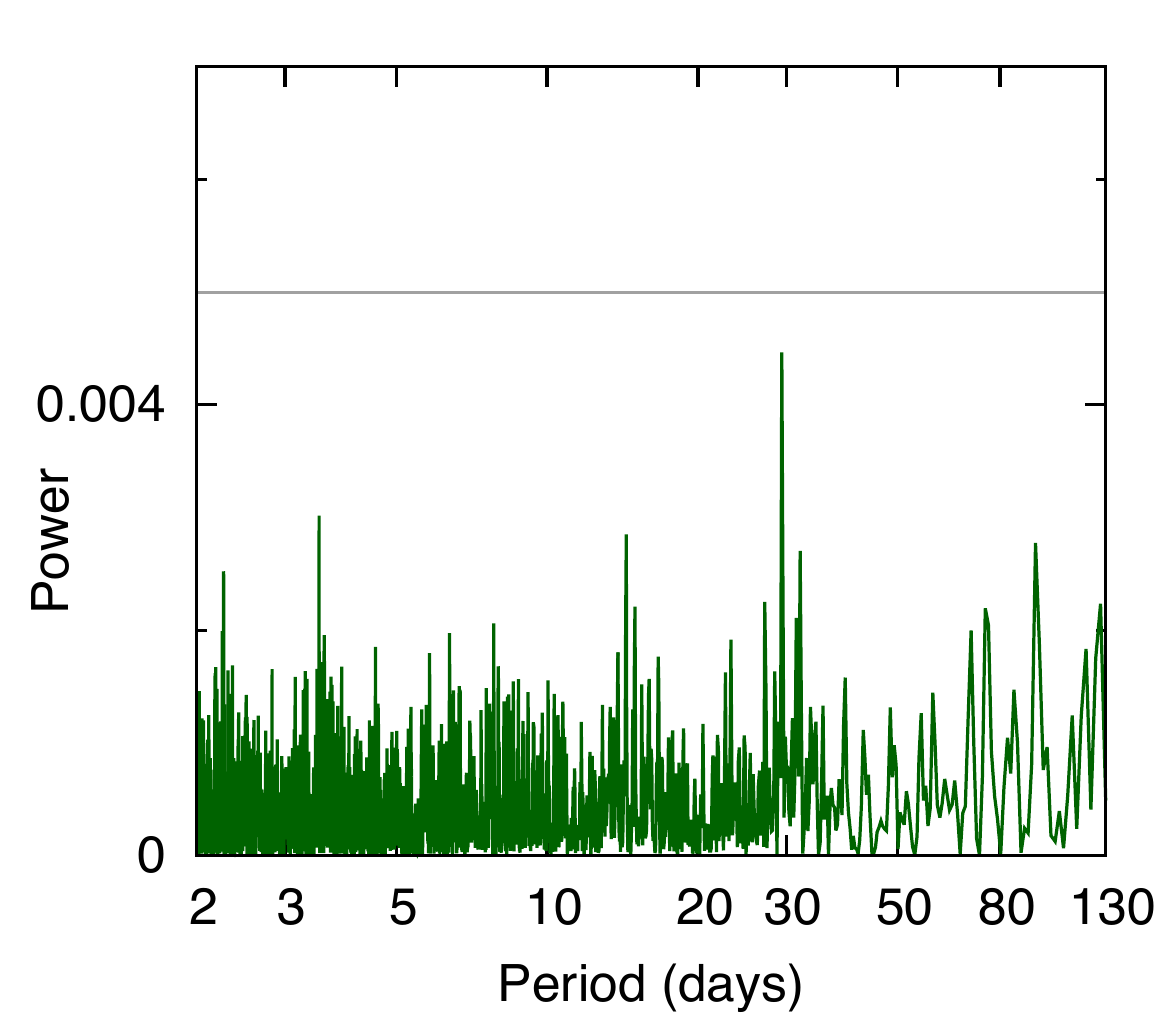}
    \caption{The periodogram of the accumulated WASP-South data, showing the absence of any rotational modulation. The horizontal line is the estimated 1\%-likelihood false-alarm level (the highest peak near 30 d is caused by the residual effects of moonlight, propagating through the reduction pipeline at a low level).}
    \label{fig:wasp_phot}
\end{figure}

TOI-4189 was observed by the WASP-South transit search between 2006 and 2011, when WASP-South was equipped with 200-mm lenses, and then from 2012 to 2014 when equipped with 85-mm lenses. In each year the observing season lasted 140 days, observing on each clear night.  In total, 53\,000 photometric data points were obtained (for more details of WASP see \citealt{2006PASP..118.1407P}). TOI-4189 is more than 5 magnitudes brighter than other stars in the photometric extraction aperture. We searched the lightcurves for a rotational modulation using the methods outlined in \citet{2011PASP..123..547M}. We find no significant periodicity over the range 1 d to 100 d, with a 95\%-confidence upper limit on the amplitude of 0.7 mmag, as displayed in the periodogram for the WASP-South data shown in Figure \ref{fig:wasp_phot}. Thus TOI-4189 appears to be magnetically inactive.

\subsection{Spectroscopic Follow-up\label{subsec:spectroscopy}}

\subsubsection{MINERVA-Australis}

4 observations were obtained between 18th December 2019 and 12th January 2020 using MINERVA Australis Telescope facility, located at Mount Kent Observatory in Queensland, Australia \citep{addison2019}. MINERVA Australis is made up of four 0.7-m CDK700 telescopes, which individually feed light via optic fiber into a KiwiSpec high-resolution (R = 80 000) stabilized spectrograph \citep{barnes2012} that covers wavelengths from 480 to 620 nm. Based on these observations, the target was erroneously reported as an SB2 (double-lined spectroscopic binary) in \citep{Eisner_2021}, due to consecutive observations being taken of a different field star with similar spectral properties.

\subsubsection{CHIRON}

We obtained three spectra of TOI-4189 between UT 2021-Mar-12 and 2023-Feb-04 using the CHIRON high-resolution echelle spectrograph on the 1.5\,m SMARTS telescope at the CTIO \citep{tokovinin:2013}. We used a fiber-fed image slicer to feed the instrument, yielding a resolving power of $R\sim80,000$\ over the wavelength range between ($4100$\,\AA\ and $8700$\,\AA). The reduced spectra were provided by the CHIRON instrument team, optimally extracted according to the procedure described in \citet{paredes:2021}, in which the instrument is also shown to produce radial velocity measurements at the level of $5\,{\rm m s}^{-1}$\ for bright, slowly rotating, K dwarfs. 

We derive radial velocities by fitting the spectral line profiles, measured via least-squares deconvolution (LSD) of the observed spectra against synthetic templates \citep{donati:1997,Zhou:2018}, yielding $23\,{\rm m s}^{-1}$\ mean uncertainties. We find no evidence for composite line profiles or line profile shape variations, and we find the star to have a stable radial velocity to the precision of our observations, with a standard deviation of only $9\,{\rm m s}^{-1}$. We measure no significant rotational broadening, and we conclude that the star is well suited for mass measurement using more precise facilities.

\subsubsection{LCOGT-NRES}

We obtained reconnaissance spectroscopy of HD 35843 using the Network of Robotic Echelle Spectrographs \citep[NRES;][]{Siverd:2018} on LCOGT. We obtained two sets of observations, each one with $3\times1200$ second exposures. The first observation, on 2021 Sept. 02 UT, resulted in lower than expected SNR, and so a second set of spectra was taken on 2021 Sept. 09 UT. Both observations were obtained with the NRES unit at the South African Astronomical Observatory, South Africa. The data were reduced using the \texttt{BANZAI-NRES} pipeline \citep{McCully:2022}, and we estimated stellar parameters using the SpecMatch-Synth code\footnote{\url{https://github.com/petigura/specmatch-syn/tree/master}}. These indicated a slowly-rotating G star amenable to follow-up, and no significant RV shift between the two epochs to a precision of $\sim100$ m s$^{-1}$.

\subsubsection{PFS}

We observed TOI-4189 with the Planet Finder
Spectrograph \citep[PFS;][]{Crane:2006,Crane:2008,Crane:2010}, which is mounted on the 6.5 m Magellan II (Clay) Telescope at Las Campanas Observatory in Chile. PFS is a slit-fed echelle spectrograph with a wavelength coverage of $3910$—$7340$\ \AA. We used a 0.3\arcsec\ slit and $1 \times 2$\ binning, which yields a resolving power of $R \approx 110,000$. Wavelength calibration is achieved via an iodine gas cell, which also allows characterization of the instrumental profile. We obtained spectra on 18 epochs, observed through iodine, between UT 2021-Sep-13 and 2022-Nov-14, with typical exposure times between 15--20 minutes and SNR near 550~nm around 50--70. We also obtained a high-SNR, iodine-free template observation. The radial velocities were extracted using a custom IDL pipeline following the prescriptions of \citet{Marcy:1992} and \citet{Butler:1996}, and achieved a mean internal precision of $1.05\ {\rm m\,s}^{-1}$. The radial velocities are presented in Table \ref{tab:recrvs}.

\subsubsection{CORALIE}
We observed HD 35843 with the high-resolution CORALIE spectrograph that is installed at the Swiss 1.2-m Leonhard Euler Telescope at ESO’s La Silla Observatory \citep{Queloz_2001}. CORALIE has a resolving power of R $\sim$ 60,000 and is fed by a 2 arcsec fiber \citep{Segransan_2010}. A total of 4 RVs were obtained from 2022 August 30 to 2022 October 10 using exposure times of 1200 s which translated in spectra with a signal-to-noise ratio per resolution element (S/N) around 50 at 550nm. We derived the RV of each epoch by cross-correlating the spectrum with a binary G2 mask \citep{Baranne_1996,Pepe_2002} with uncertainties of about 6 m/s. These observations allow us to exclude any kind of binaries, massive planetary companions and to exclude fast rotating stars.  

\subsubsection{ESPRESSO}

We acquired a total of 70 high-resolution spectroscopic observations of HD 35843 using ESPRESSO \citep{Pepe_2021} on the 8.2 m Very Large Telescope (VLT) 
located in Paranal, Chile. The observations were carried out between 2022 October 01 and 2023 March 27 as part of the observing programs 110.243Y.001 (PI: Hesse) and 110.2481.001 (PI: Bouchy), both dedicated to the characterization of warm and temperate sub-Neptune transiting exoplanets. The exposure time was fixed to 900 s and 600 s for programs 110.243Y and 110.2481, respectively. Both programs used the singleHR mode with 1 × 1 binning, allowing it to reach a median resolving power of 140,000 and a wavelength range of 380-788 nm. 
The RVs and activity indicators were extracted using version 3.0.0. of the ESPRESSO pipeline, and we computed the RVs by cross-correlating the Echelle spectra with a G2 numerical mask.
The average uncertainty of the RV data is 0.3 and 0.4 m/s for programs 110.243Y and 110.2481, respectively and the RMS is 2.6 m/s for both data sets. We report the ESPRESSO RV measurements and their uncertainties, along with the full width at half maximum (FWHM), bisector, contrast, S-index, H$\alpha$, Na, Ca, and log R'$_{HK}$ activity indexes in Tables \ref{tab:espresso_vals_kh} and \ref{tab:espresso_vals_fb}. 

\begin{figure*}
    \centering
    \includegraphics[width=\linewidth]{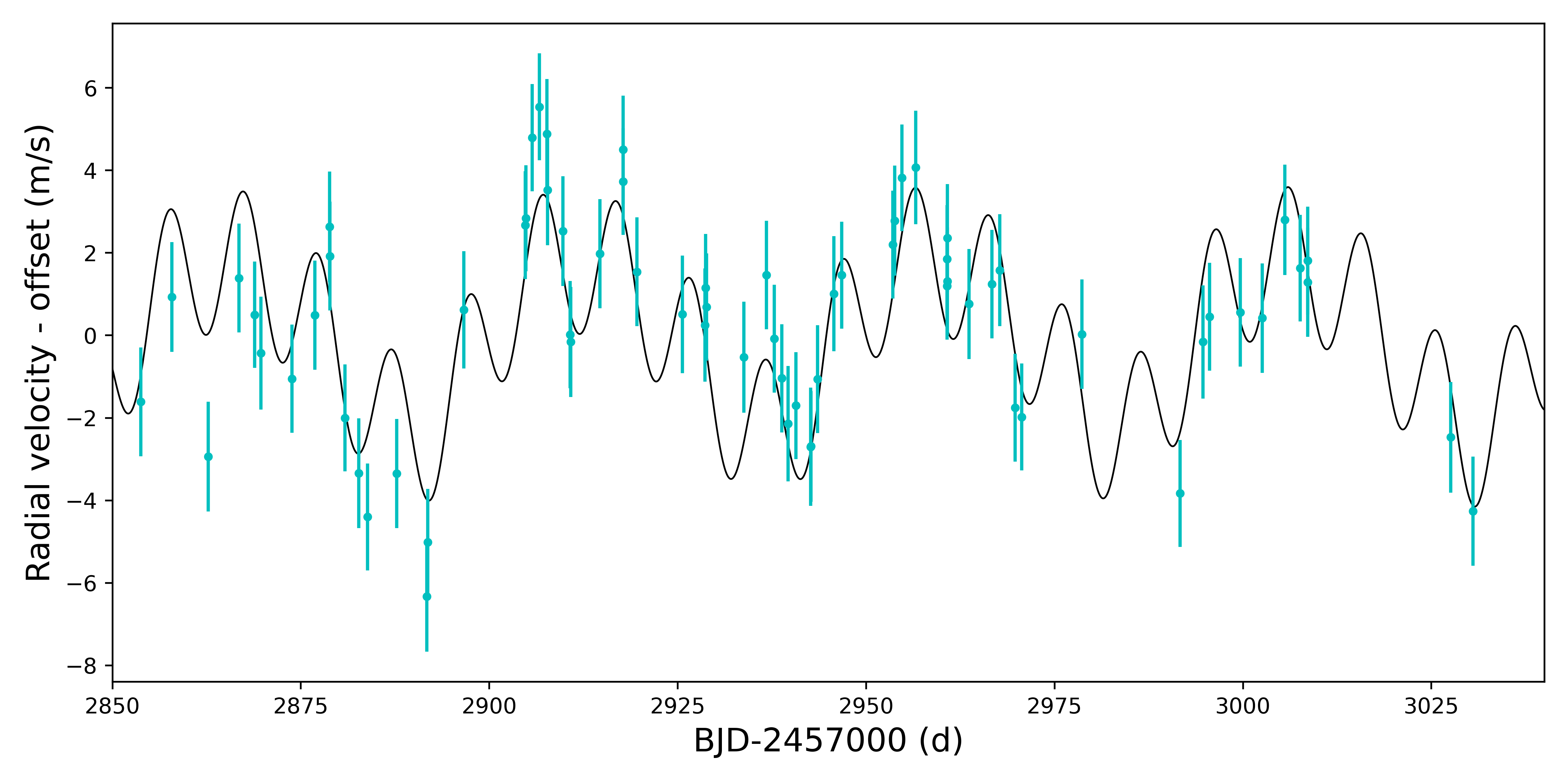}
    \caption{ESPRESSO RVs for HD 35843 and best-fit RV curve for the 2-planet system.}
    \label{fig:full_rvs}
\end{figure*}

The time series of both sets of ESPRESSO RVs for HD 35843 are shown in Figure \ref{fig:full_rvs}, over-plotted with the best-fit RV curve for the system. The solution behind this best-fit curve is described in Section \ref{sec:analysis}.

\subsection{High Angular Resolution Imaging\label{highres}}

\subsubsection{SOAR}

High-angular resolution imaging is needed to search for nearby sources that can contaminate the TESS photometry, resulting in an underestimated planetary radius, or be the source of astrophysical false positives, such as background eclipsing binaries. We searched for stellar companions to HD 35843 with speckle imaging on the 4.1-m Southern Astrophysical Research (SOAR) telescope \citep{2018PASP..130c5002T} on 1 October 2021 UT, observing in Cousins I-band, a similar visible bandpass as TESS. This observation was sensitive to a 5.3-magnitude fainter star at an angular distance of 1 arcsec from the target. More details of the observations within the SOAR TESS survey are available in \citet{2020AJ....159...19Z}. The 5$\sigma$ detection sensitivity and speckle auto-correlation functions from the observation are shown in Figure \ref{fig:gemini_soar_contrast}. No nearby stars were detected within 3\arcsec of HD 35843 in the SOAR observations.

\subsubsection{Gemini}

Spatially close stellar companions to exoplanet host stars (bound or line of sight) can create a false-positive transit signal and/or Third-light flux contamination, leading to an underestimated planetary radius if not accounted for in the transit model \citep{2015ApJ...805...16C}. Binary host stars also are known to cause non-detections of small planets residing with the same exoplanetary system \citep{Lester_2021} due to transit dilution. Additionally, the discovery of close, bound companion stars, which exist in nearly one-half of FGK type stars \citep{2018AJ....156...31M} provide crucial information toward our understanding of exoplanetary formation, dynamics and evolution \citep{2021FrASS...8...10H}. Thus, to search for close-in bound companions unresolved in TESS photometry, we obtained high-resolution speckle imaging observations of TOI-4189.

\begin{figure*}
    \centering
    \includegraphics[width=0.49\linewidth]{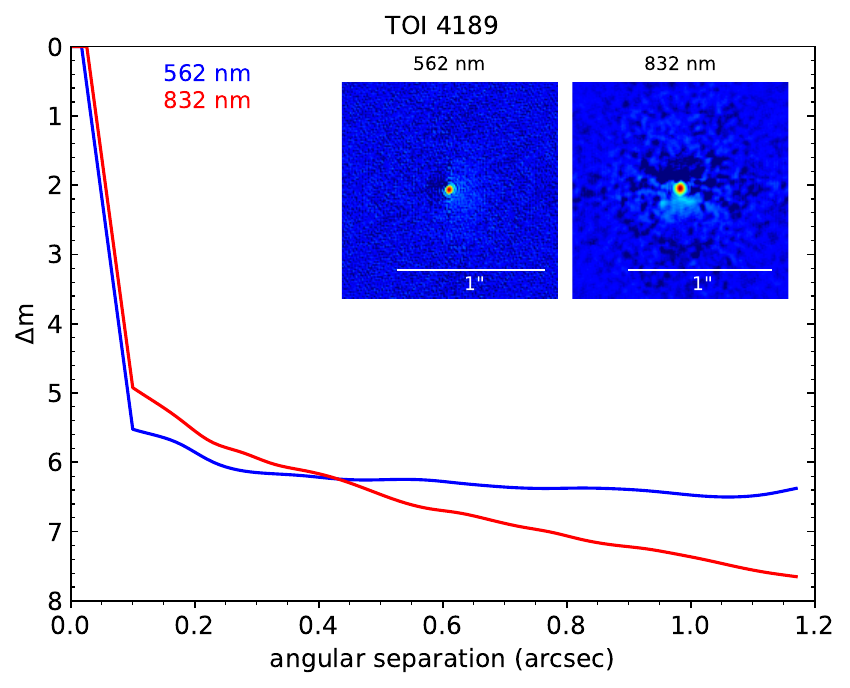}
    \includegraphics[width=0.49\linewidth]{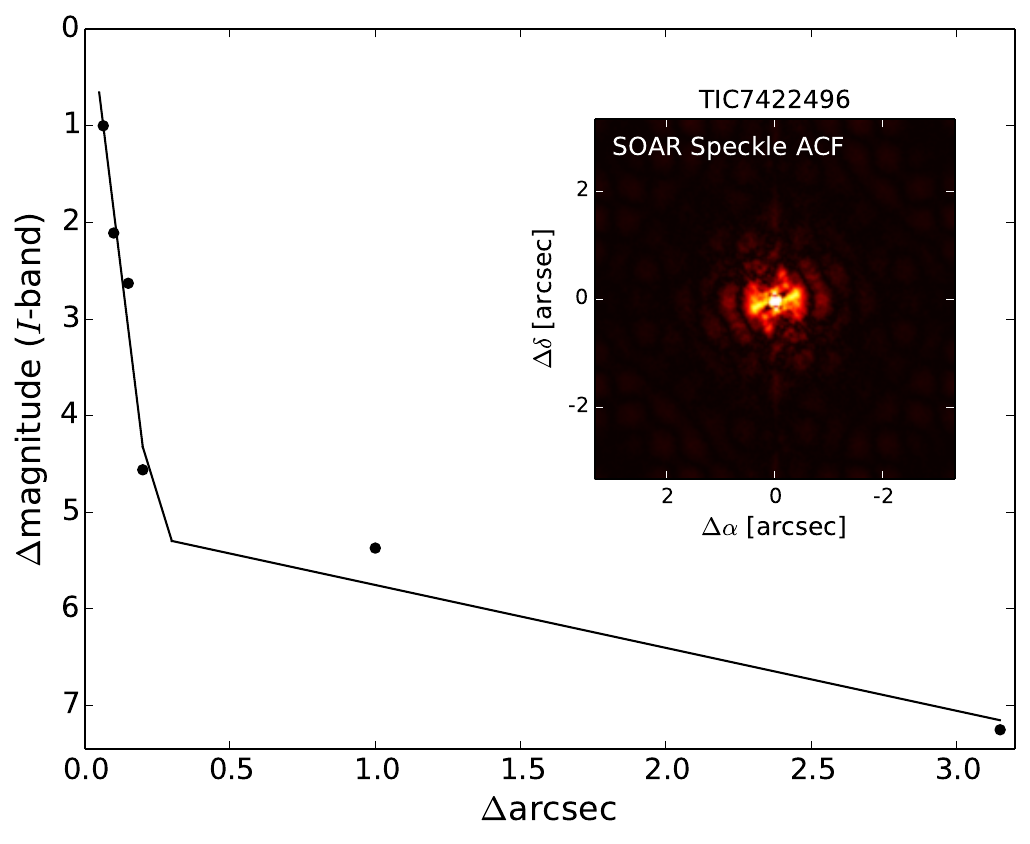}
    \caption{\textit{Left:} Gemini speckle imaging contrast curves of HD 35843 (TOI 4189), with insets showing the images for each filter. \textit{Right:} High resolution speckle imaging and contrast curve from SOAR.}
    \label{fig:gemini_soar_contrast}
\end{figure*}

TOI-4189 was observed on 2021 October 21 UT using the Zorro speckle instrument on the Gemini South 8-m telescope\footnote {\url{https://www.gemini.edu/sciops/instruments/alopeke-zorro/}}.  Zorro provides simultaneous speckle imaging in two bands (562 nm and 832 nm) with output data products including a reconstructed image with robust contrast limits on companion detections \citep{2021FrASS...8..138S, 2022FrASS...9.1163H}. Three sets of 1000 $\times$ 0.06 sec exposures were collected and subjected to Fourier analysis in our standard reduction pipeline \citep[see][]{2011AJ....142...19H}. Figure \ref{fig:gemini_soar_contrast} shows our final contrast curves and the reconstructed speckle images. We find that TOI-4189 is a single star revealing no close companions brighter than 5-7.5 magnitudes below that of the target star within the angular and 5$\sigma$ magnitude contrasts achieved from the diffraction limit (20 mas) out to 1.2”. At the distance of TOI-4189 (d=70 pc) these angular limits correspond to spatial limits of  1.4 to 84 au.

\section{Data Analysis} \label{sec:analysis}

\subsection{Host Star Parameters\label{hoststar}}
\subsubsection{Spectroscopic Parameters}

The stellar spectroscopic parameters ($T_{\mathrm{eff}}$, $\log g$, microturbulence, [Fe/H]) were estimated using the ARES+MOOG methodology which is described in detail in \citet[][]{Sousa-21, Sousa-14, Santos-13}. Here we used the latest version of ARES \footnote{The last version, ARES v2, can be downloaded at \url{https://github.com/sousasag/ARES}} \citep{Sousa-07, Sousa-15} to consistently measure the equivalent widths (EW) on the list of iron lines presented in \citet[][]{Sousa-08}. In this process we used the combined ESPRESSO data in a higher SNR spectrum. The analysis involves a minimization process to find the ionization and excitation equilibrium to converge for the best set of spectroscopic parameters. This process makes use of a grid of Kurucz model atmospheres \citep{Kurucz-93} and the radiative transfer code MOOG \citep{Sneden-73}. We also derived a more accurate trigonometric surface gravity using recent Gaia data following the same procedure as described in \citet[][]{Sousa-21}. The obtained value (4.47 $\pm$ 0.03 dex) is consistent with the spectroscopic surface gravity (4.44 $\pm$ 0.10 dex).

We analyzed several stellar activity indicators from the ESPRESSO data using a Generalized Lomb-Scargle (GLS) periodogram\footnote{\url{https://github.com/mzechmeister/GLS}}, as implemented by \citet{Zechmeister&Kurster_2009}, in order to determine the stellar rotation period. There are no significant peaks in the periodograms of H-$\alpha$, sodium, or calcium.
We find peaks of varying statistical significance in the periodograms of the bisector span, full-width half maximum (FHWM), and contrast at 25.49$\pm$0.42, 27.09$\pm$0.89, and 25.54$\pm$1.21 days with False Alarm Probabilities (FAP) of 1.1$\times$10$^{-8}$, 0.13, and 0.11, respectively. We also find a peak in the residual RVs at 26.95$\pm$0.77 days with a FAP of 0.02. 

Given the significantly lower FAP for the period derived from the bisector span, we adopt that value for the rotation period, which is in good agreement with the inferred rotation period of 21$\pm$9 days based on the $v\, \text{sin} i$ value and the assumption $\sin i = 1$, described further in Section \ref{subsubsec:sed}.

\def\arraystretch{1.15}
\begin{deluxetable}{lcc}
\tablewidth{0pc}
\tabletypesize{\scriptsize}
\tablecaption{
    Target Information
    \label{tab:stellar_info}
}
\tablehead{
    \multicolumn{1}{c}{Parameter} &
    \multicolumn{1}{c}{Value} &
    \multicolumn{1}{c}{Source}
}
\startdata
TIC & 7422496 &  TIC V8$^a$\\
R.A. & 05:25:23.73 & Gaia DR2$^b$ \\
Dec. & 05:25:23.73 & Gaia DR2$^b$ \\
$\mu_{ra}$ (mas yr$^{-1}$)  & -43.937 $\pm$ 0.014 & Gaia DR3$^b$ \\
$\mu_{dec}$ (mas yr$^{-1}$) & -110.240 $\pm$ 0.016 & Gaia DR3$^b$ \\
Parallax (mas) & 14.4269 $\pm$ 0.0133 & Gaia DR3$^b$ \\
Epoch & 2016.0 &  Gaia EDR3$^b$\\
$B$ (mag)    & 10.098 $\pm$ 0.187 & AAVSO DR9$^c$ \\
$V$ (mag)    & 9.36 $\pm$ 0.03 & AAVSO DR9$^c$ \\
$Gaia$ (mag) & 9.19551 $\pm$ 0.00032 & Gaia DR3$^b$ \\
$B_P$ & 9.527 $\pm$ 0.003 & Gaia DR3$^b$ \\
$R_P$ & 8.706 $\pm$ 0.003 & Gaia DR3$^b$ \\
\textit{TESS} (mag) & 8.7601 $\pm$ 0.006 & TIC V8$^a$\\
$J$ (mag)    & 8.151 $\pm$ 0.024 & 2MASS$^d$ \\
$H$ (mag)    & 7.859 $\pm$ 0.026 & 2MASS$^d$ \\
$K_S$ (mag)  & 7.782 $\pm$ 0.017 & 2MASS$^d$ \\
& & \\
\multicolumn{3}{l}{\textit{Derived parameters}} \\
$M_{\star}$ ($M_{\odot}$)  & 0.94 $\pm$ 0.06 & This work \\
$R_{\star}$ ($R_{\odot}$)   & 0.897 $\pm$ 0.019 & This work \\ 
$\rho_{\star}$ (g cm$^{-3}$) & 1.84 $\pm$ 0.16 & This work \\
$L_{\star}$ ($L_{\odot}$)   & 0.75 $\pm$ 0.05 & This work \\ 
$T_{eff}$ (K)      & 5666 $\pm$ 61 & This work \\ 
$[$Fe/H$]$         & -0.27 $\pm$ 0.04 & This work \\
log $g$          & 4.47 $\pm$ 0.03 & This work \\
$v_t$ (km s$^{-1}$) & 0.83 $\pm$ 0.03 & This work \\
Age (Gyr)       & 2.5 $\pm$ 0.5 & This work \\ 
log $R'_{HK}$   & -5.01 $\pm$ 0.02 & This work \\ 
$v\, \text{sin} i$ (km s$^{-1}$) & 2.09 $\pm$ 0.91 & This work \\
P$_{rot}$ (d) & 25.49 $\pm$ 0.42 & This work \\
\enddata
\tablenotetext{a}{\cite{2018AJ....156..102S}.}
\tablenotetext{b}{\cite{2023AA...674A...1G}.}
\tablenotetext{c}{\cite{2016yCat.2336....0H}.}
\tablenotetext{d}{\cite{2003yCat.2246....0C}.}
\label{target_info}
\end{deluxetable}

\subsubsection{Spectral Energy Distribution Analysis} \label{subsubsec:sed}

\begin{figure}
\centering
\includegraphics[width=0.75\linewidth,trim=70 80 80 100,clip,angle=90]{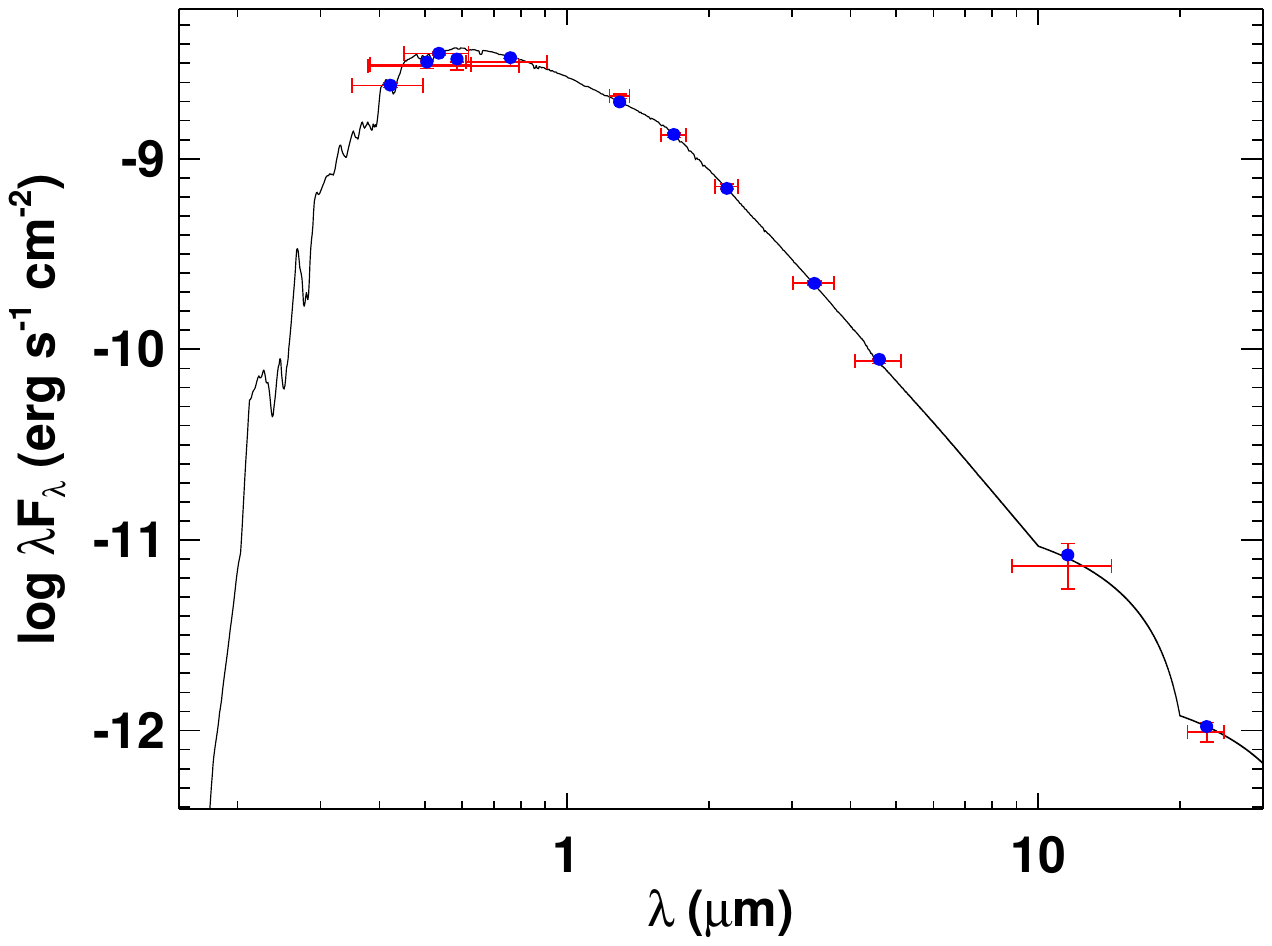}
\caption{Spectral energy distribution of TOI-4189. Red symbols represent the observed photometric measurements, where the horizontal bars represent the effective width of the passband. Blue symbols are the model fluxes from the best-fit Kurucz atmosphere model (black).  \label{fig:sed}}
\end{figure}

As an independent determination of the basic stellar parameters, we performed an analysis of the broadband spectral energy distribution (SED) of the star together with the {\it Gaia\/} DR3 parallax \citep[with no systematic offset applied; see, e.g.,][]{StassunTorres:2021}, in order to determine an empirical measurement of the stellar radius, following the procedures described in \citet{Stassun:2016,Stassun:2017,Stassun:2018}. We pulled the $JHK_S$ magnitudes from {\it 2MASS}, the W1--W4 magnitudes from {\it WISE}, and the $G G_{\rm BP} G_{\rm RP}$ magnitudes from {\it Gaia}. Together, the available photometry spans the full stellar SED over the wavelength range 0.4--20~$\mu$m (see Figure~\ref{fig:sed}).  

We performed a fit using Kurucz stellar atmosphere models, with the principal parameters being the effective temperature ($T_{\rm eff}$), surface gravity ($\log g$) and metallicity ([Fe/H]), adopted from the spectroscopic analysis. The extinction $A_V$ was fixed at zero due to the close proximity of the system. The resulting fit (Figure~\ref{fig:sed}) has a reduced $\chi^2$ of 1.4. Integrating the model SED gives the bolometric flux at Earth, $F_{\rm bol} = 4.977 \pm 0.058 \times 10^{-9}$ erg~s$^{-1}$~cm$^{-2}$. Taking the $F_{\rm bol}$ and $T_{\rm eff}$ together with the {\it Gaia\/} parallax, gives the stellar radius, $R_\star = 0.897 \pm 0.019$~R$_\odot$. In addition, we can estimate the stellar mass from the empirical relations of \citet{Torres:2010}, giving $M_\star = 0.94 \pm 0.06$~M$_\odot$. 

We can also estimate the stellar rotation period from the combination of the spectroscopically determined $v\sin i$ and the estimated value of $R_\star$, giving $P_{\rm rot}/\sin i = 21.7 \pm 9.4$~d. In addition, we can estimate the system age from this rotation period and the empirical age-rotation relations of \citet{Mamajek:2008}, giving $\tau_\star = 2.5 \pm 0.5$~Gyr. This value as well as previously mentioned derived stellar parameters are listed in Table \ref{tab:stellar_info}.

\subsection{False Positive Scenarios \& Statistical Validation\label{subsec:validation}}

We use a combination of the TESS photometry and follow-up ground-based photometry, spectroscopy, and high-resolution imaging to rule out false positive scenarios for the 46.96 d candidate. SPOC analysis of the difference image centroid offsets (\citep{Twicken_2018}) places the signal to within 14.0 $\pm$ 7.6 arcsec of the target star. We rule out neighboring stars inside this region as the source of the signal using ground-based photometry. We also rule out an eclipsing binary on-target using the reconnaissance spectra. While \citet{Eisner_2021} designated this candidate a double-lined spectroscopic binary based on their MINERVA observations, we find no evidence of a double-lined spectrum in the CHIRON, LCO-NRES, CORALIE, PFS, or ESPRESSO spectra.

We use $triceratops$ \citep{Giacalone&Dressing_2020} to calculate the false positive probability (FPP) and nearby false-positive probability (NFPP) for the candidate, including the contrast curve from our high-resolution imaging in order to provide additional constraints. False-positive scenarios include an eclipsing binary on target, on a background star, or an unseen companion as well as a transiting planet on a background star or unseen companion. Nearby false-positive scenarios include a transiting planet or an eclipsing binary on a nearby star. We calculate an NFPP of (2.7 $\pm$ 0.7) $\times\, 10^{-8}$. We calculate an FPP of 0.05 $\pm$ 0.01 \%, well below the nominal limit of 1.5\% required to be statistically validated as a planet \citep{Lester_2021}. 

\subsection{Search for Additional Planets\label{subsec:moreplanets}}

We search for signals of additional planets in both the TESS photometry and ESPRESSO RVs. We search for transits in the former using the Transit Least Squares (TLS) algorithm from \citet{2019A&A...623A..39H}, after masking the transits of the 47 d planet. We are able to rule out a second transiting planet out to an orbital period of 50 d.

\begin{figure}
    \centering
    \includegraphics[width=\linewidth]{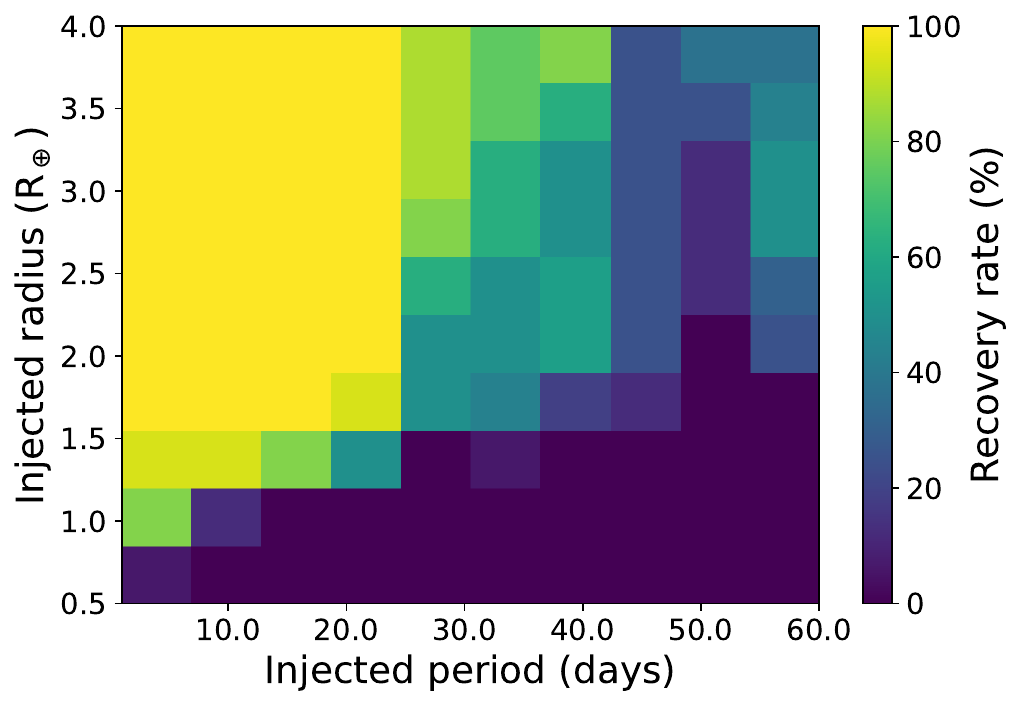}
    \caption{Transit injection and recovery for the TESS light curve of HD 35843 made using \texttt{matrix} \citep{2022zndo...6570831D} showing that an additional transiting planet at a period shorter than 20 days or a radius larger than 1.5 $R_{\oplus}$ should have been detected if present.}
    \label{fig:tr_inj_rec}
\end{figure}

As a check, we also perform transit injection and recovery using the masked light curve to determine if we might have missed a transiting planet. We use \texttt{tkmatrix} \citep{2022zndo...6570831D}, which uses the aforementioned TLS algorithm, to perform the injection and recovery test, and find that planets within 20 d and larger than 1.5 $R_{\oplus}$ should have been detected if they were present in the system (Figure \ref{fig:tr_inj_rec}). However, we note that longer periods and smaller planets might have been missed. 

\begin{figure}
    \centering
    \includegraphics[width=\linewidth]{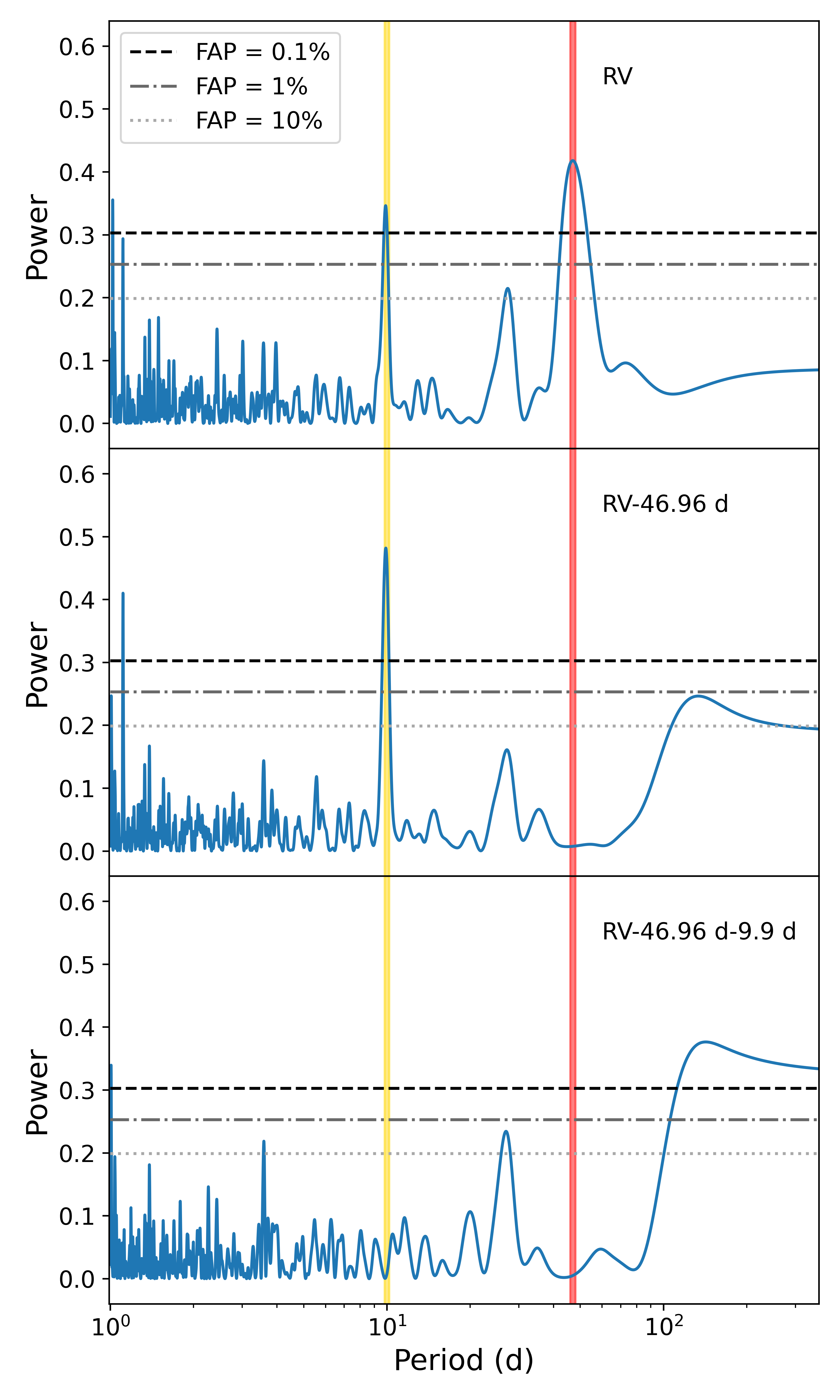}
    \caption{Generalized Lomb-Scargle periodogram for the ESPRESSO RVs before (top) and after subtracting out the 46.96 day planet (middle) and both planets (bottom). The peak near 10 days becomes more significant after this subtraction. The peaks near 1 day and the past 100 days are artifacts of the window function of the observation times. The magenta line marks the 46.96 d planet and the yellow line marks the peak near 10 days.}
    \label{fig:gls_per_bc}
\end{figure}

\begin{figure}
    \centering
    \includegraphics[width=\linewidth]{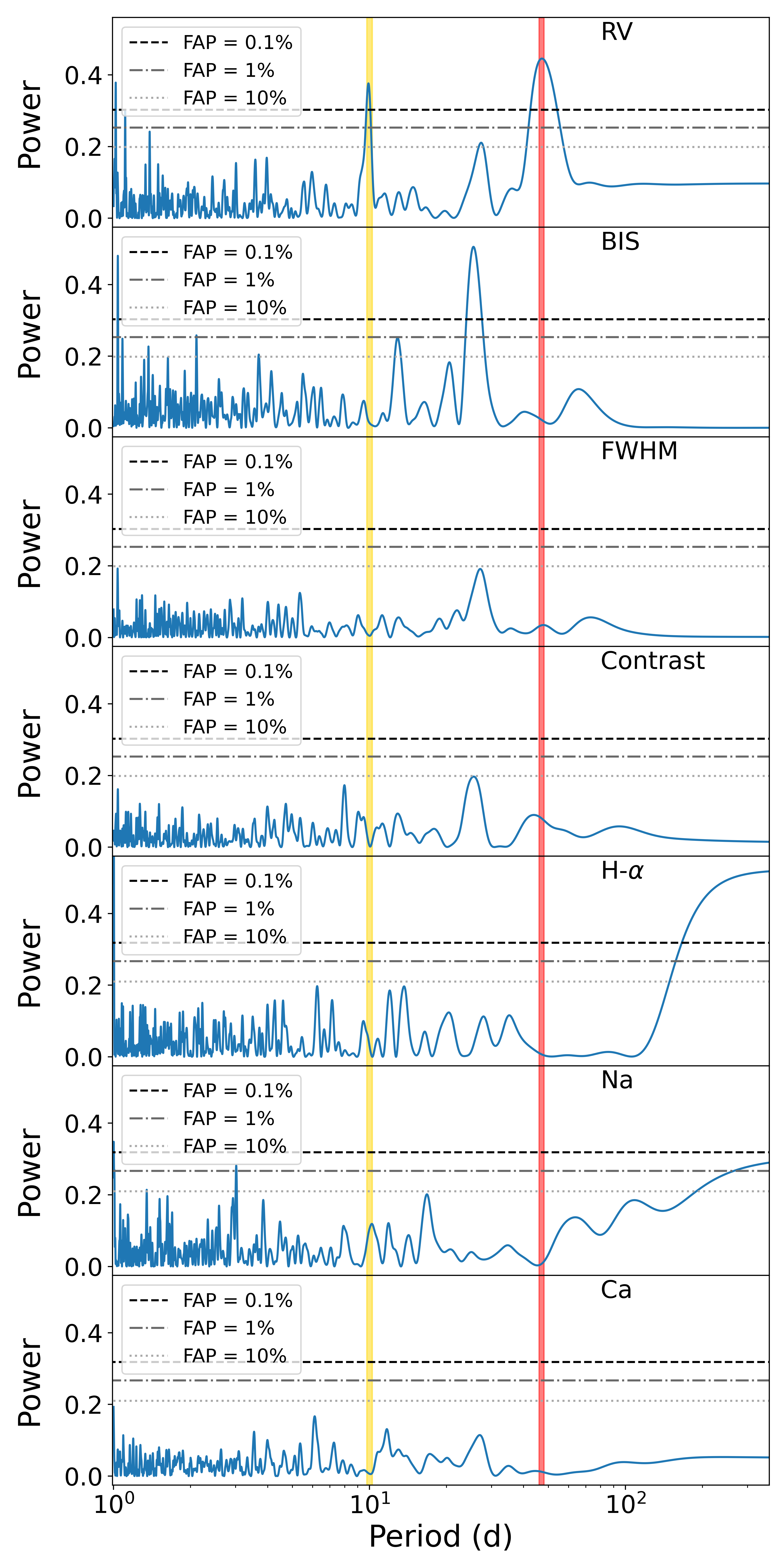}
    \caption{Generalized Lomb-Scargle periodogram for the ESPRESSO RVs and various activity indicators. There is a significant peak near 47 days in the RVs (marked in red) and less significant peak near 10 days (marked in yellow). The bisector span, FWHM, and contrast all show peaks of varying significance near 25 days.}
    \label{fig:gls_per}
\end{figure}

\begin{figure*}
    \centering
    \includegraphics[width=\linewidth]{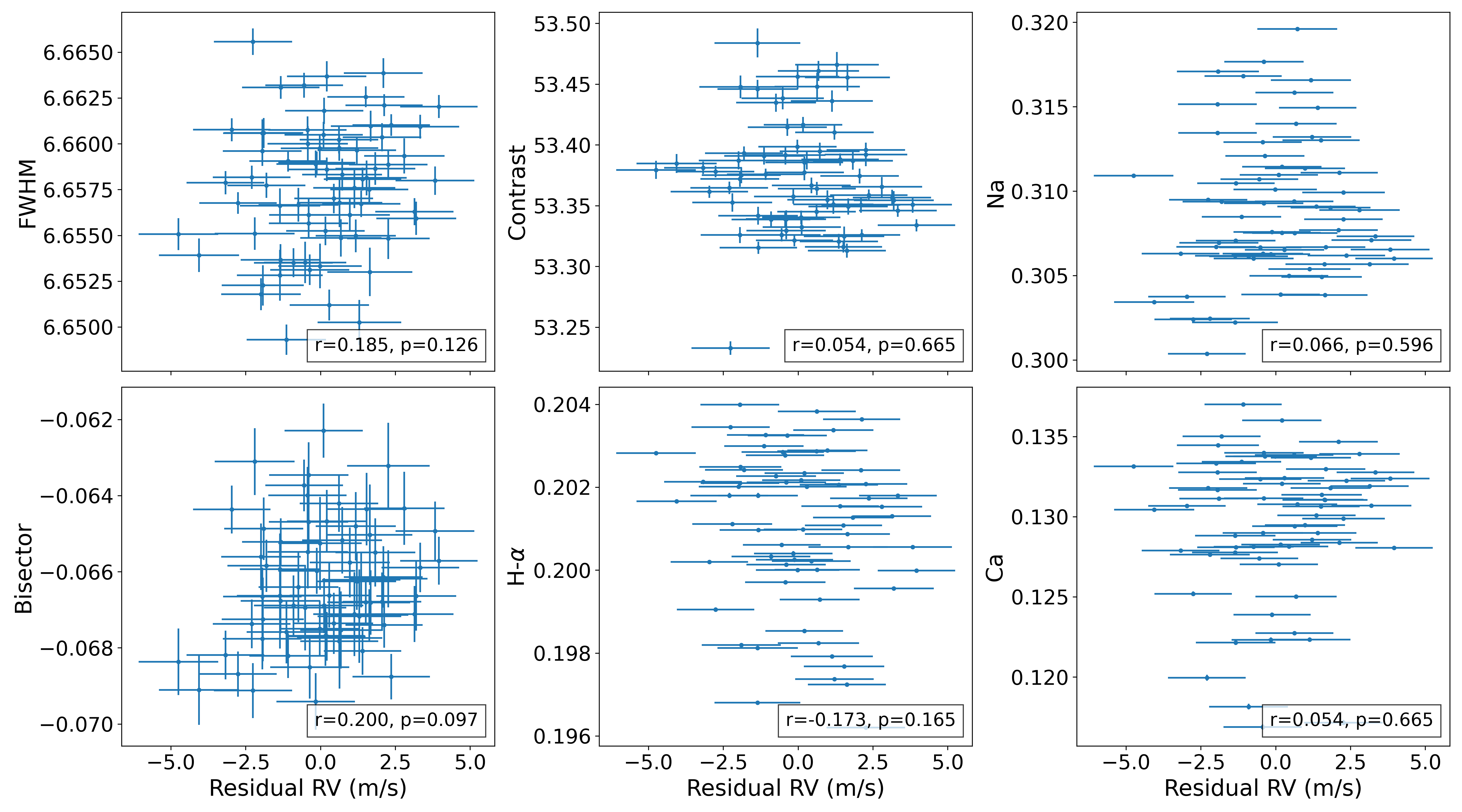}
    \caption{ESPRESSO residual RVs (46.96 d planet subtracted) vs various activity indicators showing no significant correlations. The plots show the Pearson correlation coefficient and corresponding p-value.}
    \label{fig:res_corr}
\end{figure*}

We search for planets in the RV data using a GLS periodogram. We run it before and after subtracting the best-fit model of the 47 d planet, and detect a peak at $\sim9.9$ d with a FAP below 0.1\%, as shown in Figure \ref{fig:gls_per_bc}. In order to rule out stellar variability as the source of this peak, we apply a GLS periodogram to activity indicators, including the bisector span, contrast, FWHM, H-$\alpha$, sodium, and calcium, as shown in Figure \ref{fig:gls_per}. We find no statistically significant peak from stellar activity indicators at 9.9 d. Additionally, we examine the correlations between the previously mentioned activity indicators and the residual RVs after subtracting out the 47 d planet. If stellar activity is the source of an RV signal, a strong correlation is expected between the observed RVs and activity indicators. The residual RVs exhibit no such correlation with the activity indicators (see Figure \ref{fig:res_corr}), thus strongly suggesting that the detected RV signal at 9.9 days is not due to stellar activity. Given our analysis of the TESS photometry, if this planet is real, it is either small ($<1.5 R_{\oplus}$) or non-transiting. 

\subsubsection{RV Analysis\label{subsec:rvanalysis}}

We use \texttt{radvel} \citep{fulton2018} to compare four scenarios to explain the ESPRESSO RVs: single planet (47 d) with a flat RV baseline, single planet with a linear RV trend, two planets with a flat RV baseline, and two planets with a linear RV trend.

For all four scenarios, we use Markov chain Monte Carlo (MCMC) sampling to best fit the following parameters:

\begin{itemize}
    \itemsep0pt 
    \item orbital period, $P$,
    \item time of inferior conjunction, $T_{0}$,
    \item radial velocity semi-amplitude, $K$, 
    \item RV zero point, $\gamma$, 
    \item RV jitter term, $\sigma$, 
    \item eccentricity, $e$, and argument of periastron, $\omega$. 

\end{itemize}

In the scenarios with a linear RV trend, we also include the linear term, $\dot{\gamma}$. For the scenarios with two planets, we fix the eccentricity of the inner planet to 0. 

\begin{deluxetable}{lllll}
\tablewidth{0pc}
\tabletypesize{\scriptsize}
\tablecaption{
   Bayesian information criterion and Akaike information criterion values from \texttt{radvel} scenario comparison. The preferred model of 2 planets with a flat RV trend is marked with bold text and delta BIC and delta AIC are calculated in comparison to this model.
    \label{tab:radvel_scenarios}
}
\tablehead{
    \multicolumn{1}{c}{Scenario} & 
    \multicolumn{1}{c}{BIC} &
    \multicolumn{1}{c}{$\Delta \mathrm{BIC}$} &
    \multicolumn{1}{c}{AIC} &
    \multicolumn{1}{c}{$\Delta \mathrm{AIC}$}
}
\startdata
1 planet + flat  & 303.15 & 30.35 & 292.84 &  35.67    \\
1 planet + linear  & 307.21 & 34.41 & 295.05 &   37.88   \\
\textbf{2 planets + flat} & \textbf{272.80} & \textbf{0} & \textbf{257.17} &   \textbf{0}   \\
2 planets + linear & 275.35 & 2.55 & 258.11 &  0.94    \\
\enddata
\label{radvel_scenarios}
\end{deluxetable}

Based on the Bayesian information criterion (BIC) and Akaike information criterion values from the four scenarios, the preferred scenario with the lowest BIC is the 2-planet model with a flat baseline (Table \ref{tab:radvel_scenarios}). There is significant preference for the 2-planet models over the 1-planet models. The 2-planet model with a linear baseline finds a nearly indistinguishable preference between a fit with a linear RV term and without, indicating the significance of a linear trend in the RV points is negligible.   

\begin{figure*}
    \centering
    \includegraphics[width=0.32\textwidth]{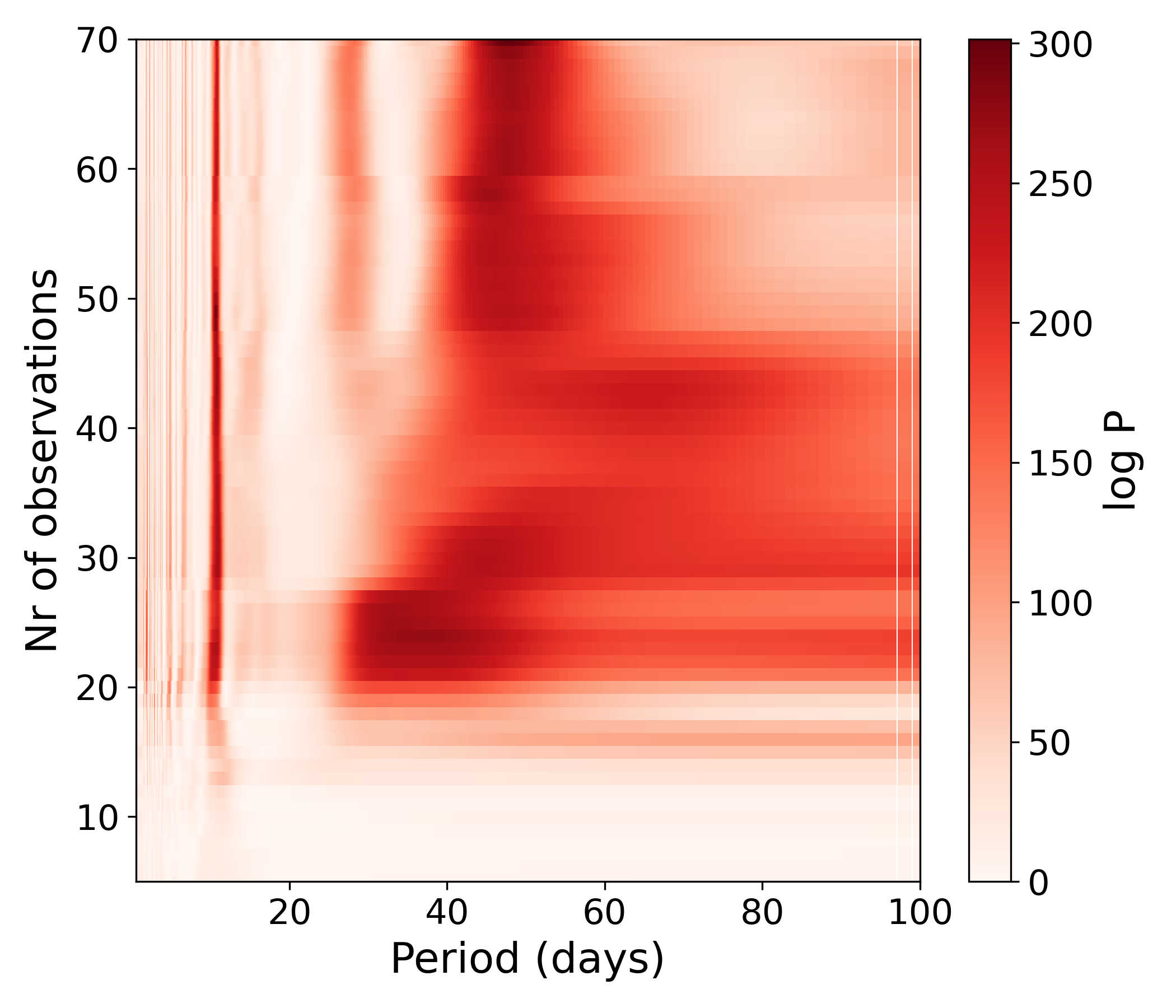}
    \includegraphics[width=0.32\textwidth]{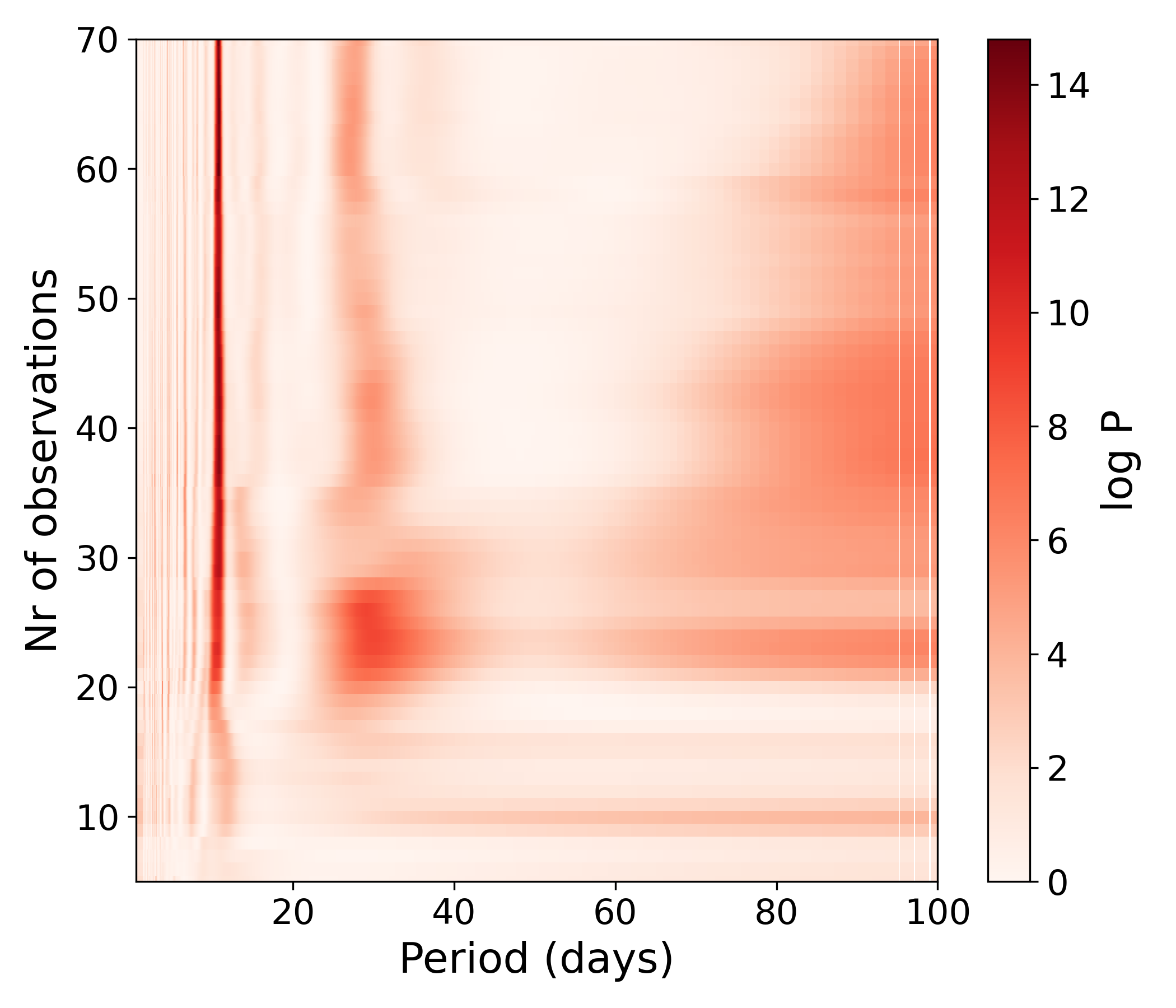}
    \includegraphics[width=0.32\textwidth]{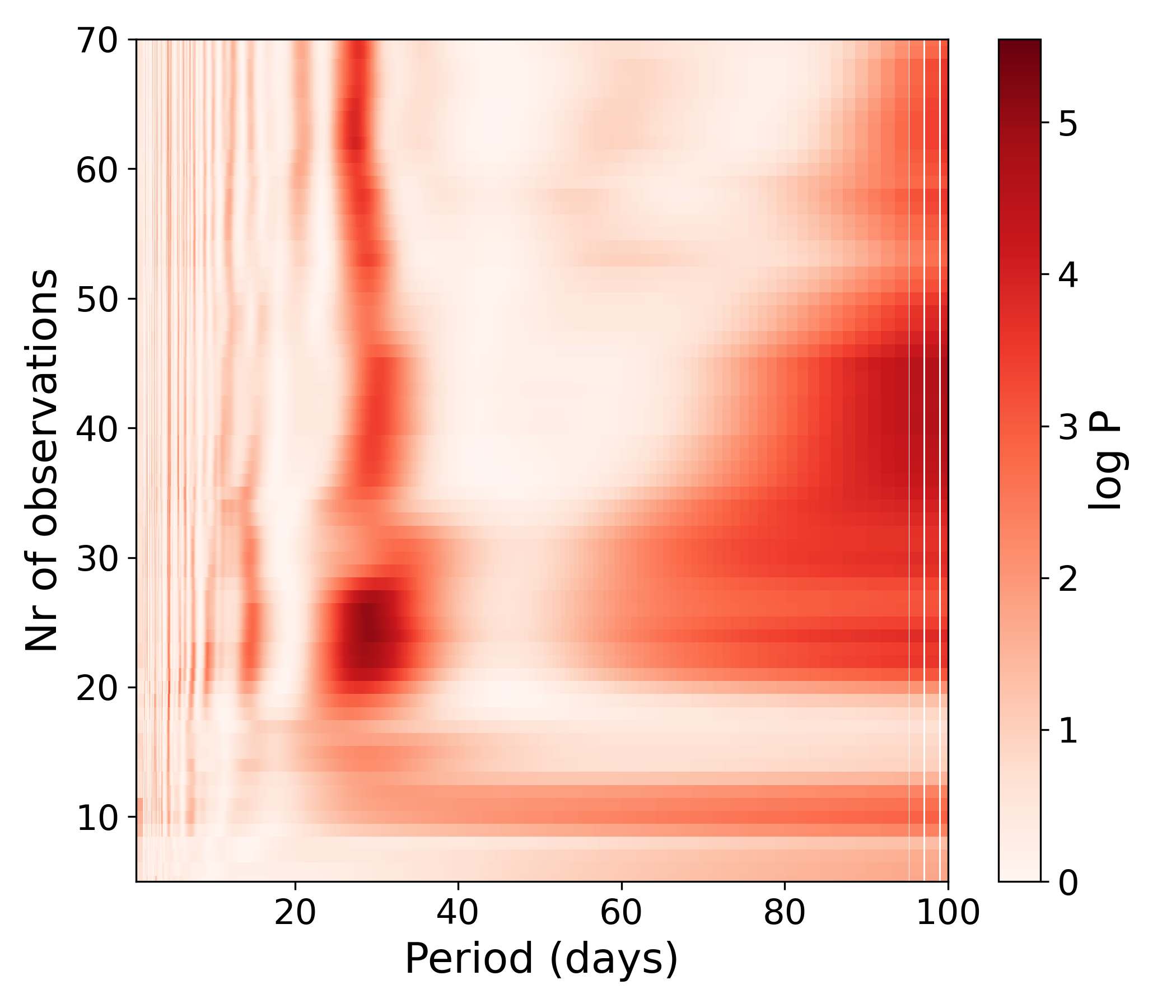}
    \includegraphics[width=0.32\textwidth]{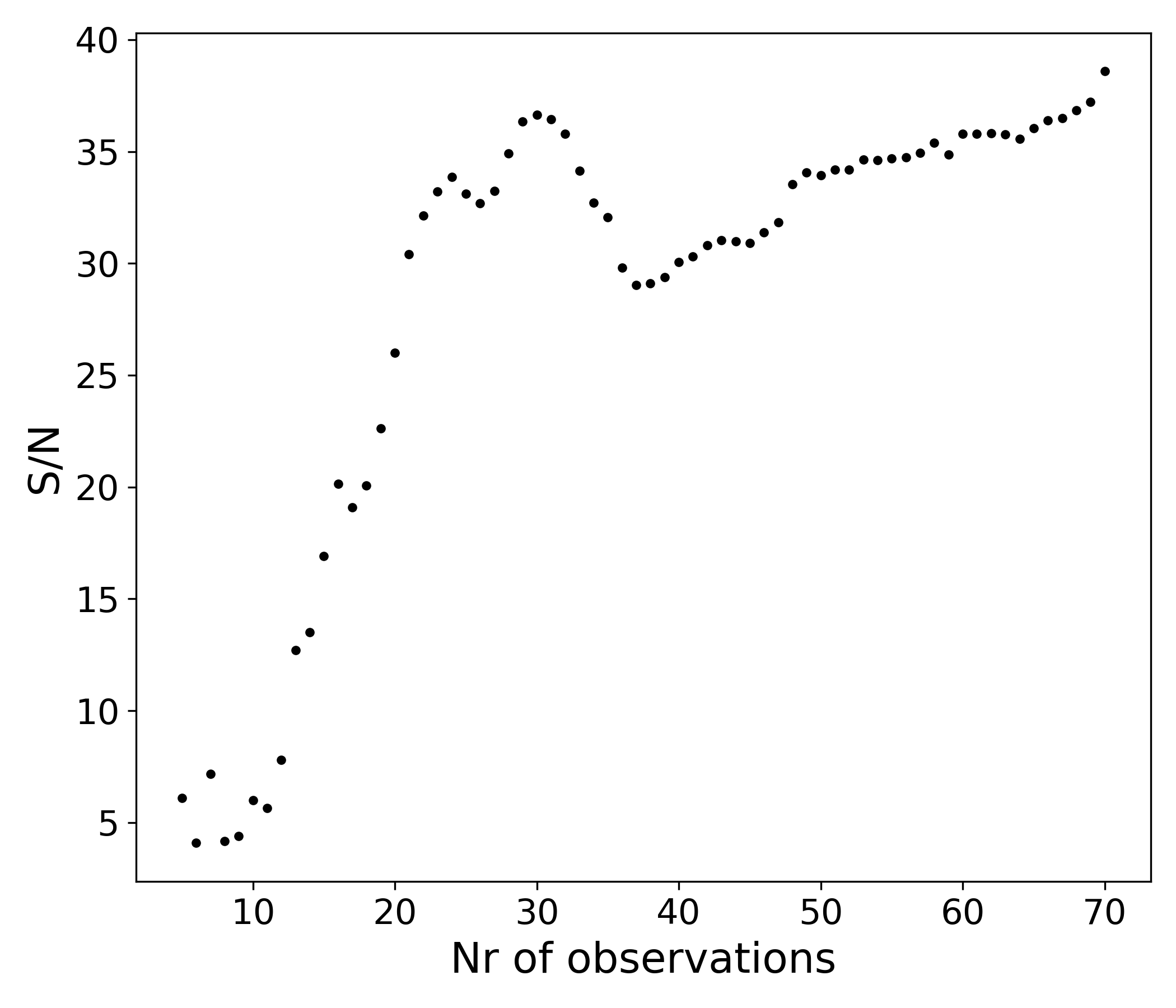}
    \includegraphics[width=0.32\textwidth]{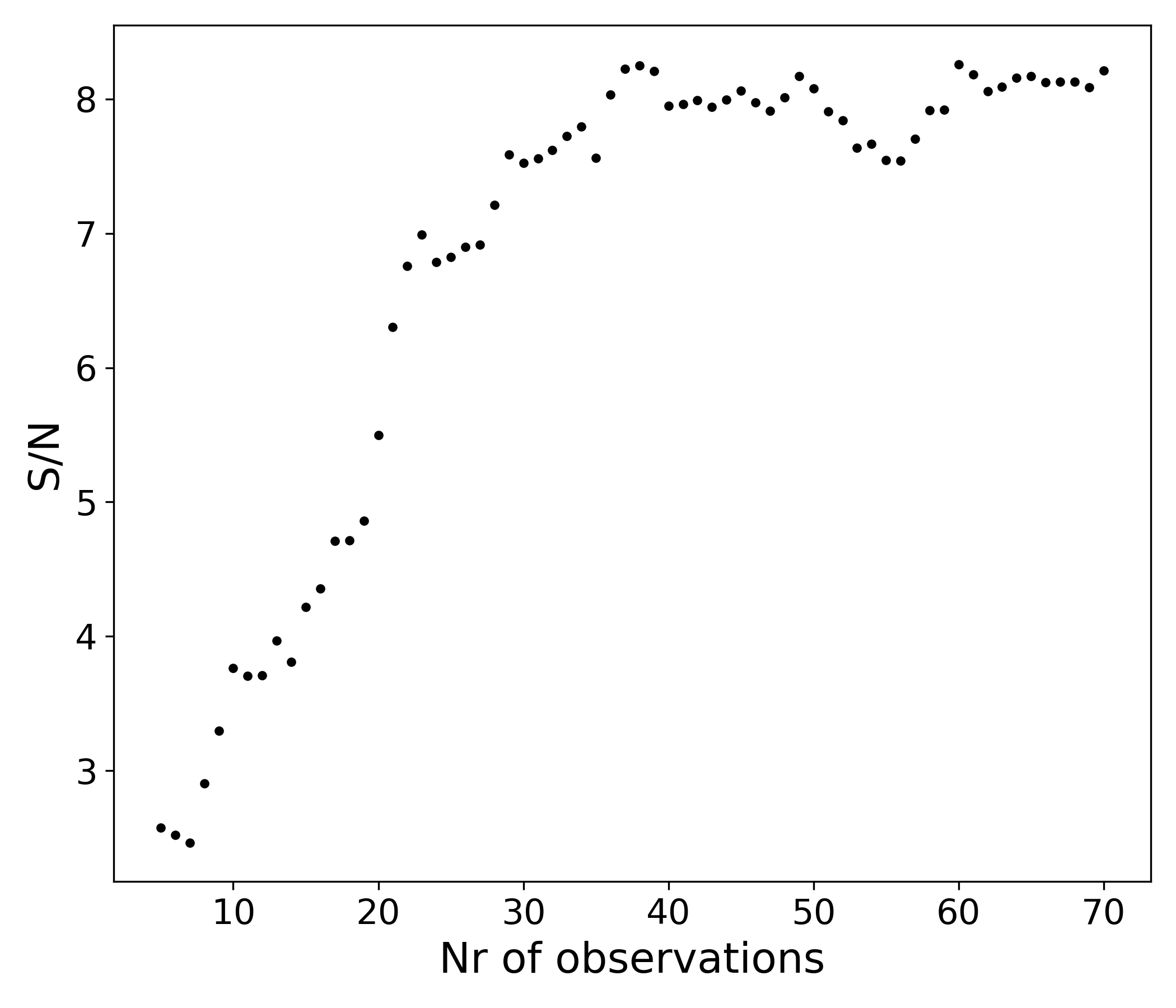}
    \includegraphics[width=0.32\textwidth]{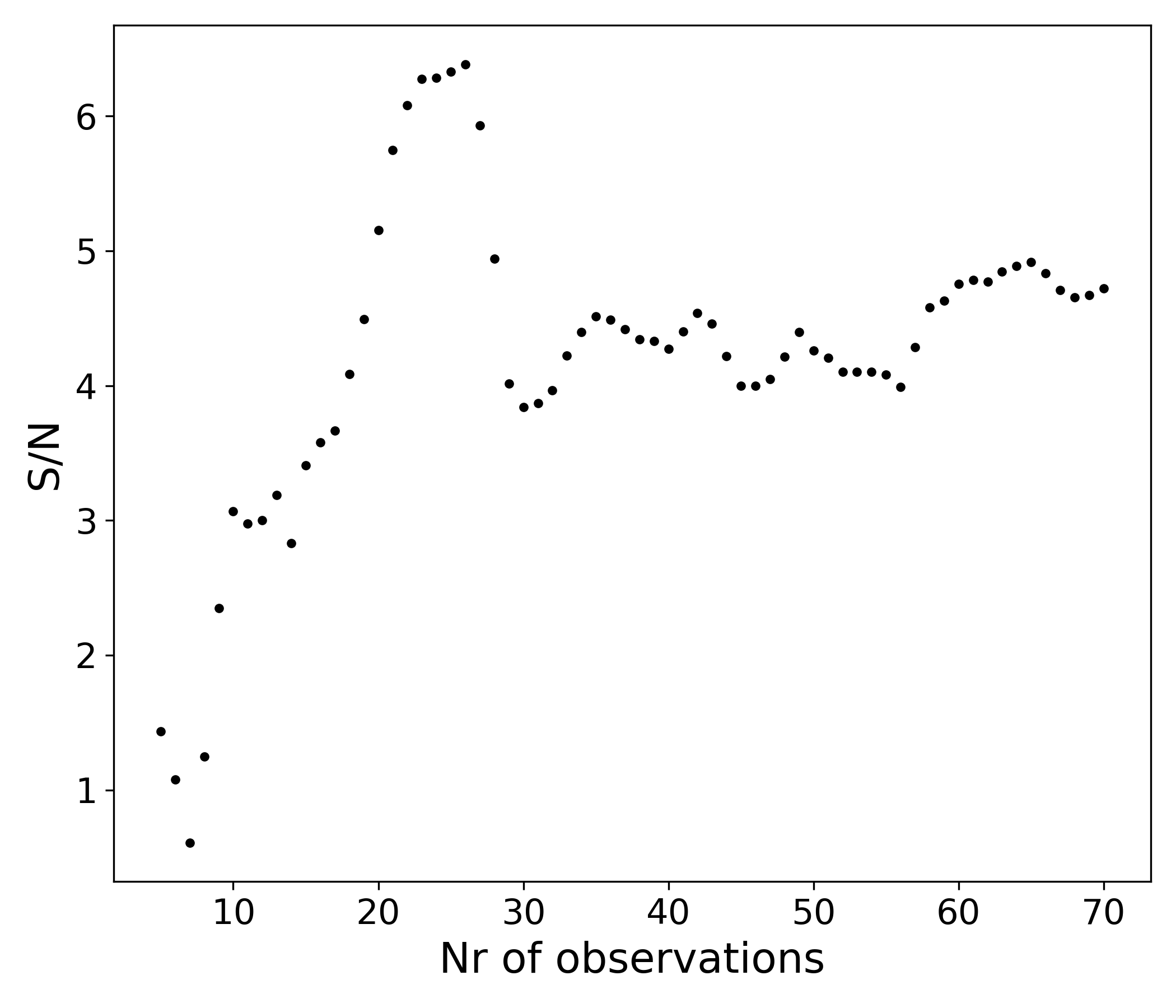}
    \caption{Top row:  From left to right, we create stacked periodograms for 1) the full set of ESPRESSO RV points, 2) the ESPRESSO RV points with the 47 day signal removed, 3) the RV points with both the 47 day and 10 day signals removed. The color scale corresponds to logarithm of the probability of a period. Bottom row: SNR of the highest peak of periodicity for the same three sets of data. The SNR increases with the number of observations for both planet signals (left and center), which is consistent with a stable planetary signal.}
    \label{fig:stack_per}
\end{figure*}

As an additional check, we created a stacked Bayesian generalized Lomb-Scargle periodogram \citep[BGLS;][]{2015A&A...573A.101M} with the RV data to determine if the strength of the periodic signal from each planet increased with increasing data points, as should be the case with a true stable planetary signal as described and implemented in \citet{Mortier_2017}. The results of this are shown in Figure \ref{fig:stack_per}. The top row shows the stacked periodograms first for the full set of ESPRESSO RV points, then with the 47-day planet signal subtracted, and finally with both the 47-day and 10-day planet signals subtracted. For the full RV set, the signal from the 47 day planet is initially blended with the suspected stellar variability signal at $\sim 25$ days. The two signals become distinct starting at $\sim 45$ observations, with the 47 day signal becoming the strongest signal overall. After subtracting out the 47-day planet's signal, the 10-day planet's signal is the strongest, with its strength increasing with the number of observations as expected. After subtracting out the signals from both planets, the $\sim 25$ days signal is the strongest, although the strength of that signal fluctuates significantly as a function of the number of observations which is consistent with the assumption of this signal resulting from stellar variability rather than a planetary candidate.

\subsection{Global Fit: Combined Photometric and Spectroscopic Analysis \label{subsec:fit}}

For a more detailed analysis, we then performed a model fitting using the \texttt{allesfitter} \citep{Gunther&Daylan_2021} package to jointly model the TESS photometry and the ESPRESSO radial velocity measurements. We do not include MINERVA-Australis, LCOGT-NRES, CHIRON, and CORALIE RVs due to their small sample sizes. We do not include PFS RVs because while they increase the observing baseline, they do not improve the significance of the mass measurement of either planet and result in a bimodal posterior for the period of planet b. We assume the inner planet is non-transiting and has a circular orbit. We treat the two ESPRESSO RV datasets separately because of the different masks used for each set of observations. We also treat the inner planet as non-transiting due to the lack of any additional transit detections using a TLS search (Section \ref{subsec:moreplanets}). We used a nested sampling algorithm consisting of 500 walkers to explore the parameter space and determine the best-fit values for the following parameters:

\begin{itemize}
    \itemsep0pt 
    \item radius ratio, $R_p/R_{\star}$, for planet c, where p denotes the planet,
    \item sum of radii divided by the orbital semi-major axis, $(R_{\star} + R_p) / a$, for planet c,
    \item cosine of the orbital inclination, $ \cos{i}$, for planet c,
    \item orbital period, $P$, for planets b and c,
    \item time of inferior conjunction, $T_{0}$, for planets b and c,
    \item radial velocity semi-amplitude, $K$, for planets b and c,
    \item eccentricity parameters $\sqrt{e}\, cos\, \omega$ and $\sqrt{e}\, sin\, \omega$, where $e$ is the orbital eccentricity and $\omega$ the argument of periastron, for planet c,
    \item RV zero point, $\gamma$,
    \item RV jitter term, $\sigma$,
    \item white noise scaling term for the TESS data, $\sigma_{TESS}$.
\end{itemize}

The values of the limb darkening parameters for TESS, $q_1$ and $q_2$, were obtained by matching the spectroscopic parameters of the primary star to the closest values of the coefficients $u_1$ and $u_2$ of the quadratic limb darkening law listed in \citet{Claret_2017}, and transforming them to the corresponding values of $q_1$ and $q_2$. These values are listed in Table \ref{tab:fitted_parameters} and held fixed in the fit.

\begin{figure}
    \centering
    \includegraphics[width=\linewidth]{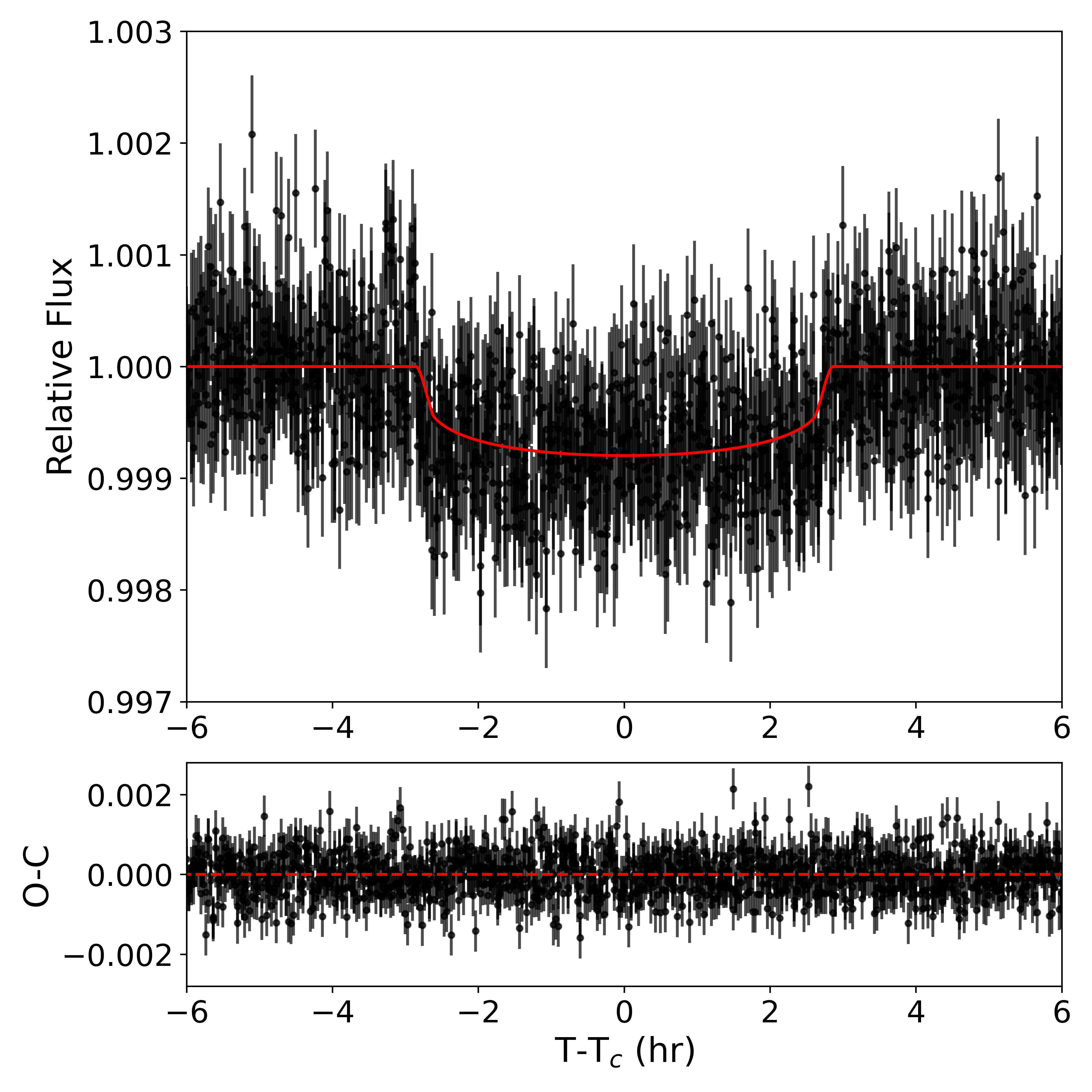}
    \caption{Phase folded TESS transit data with the modeled best-fit \texttt{allesfitter} light curve, and residuals plot on the bottom.}
    \label{fig:bestfit_tess}
\end{figure}

\begin{figure*}
    \centering
    \includegraphics[width=\linewidth]{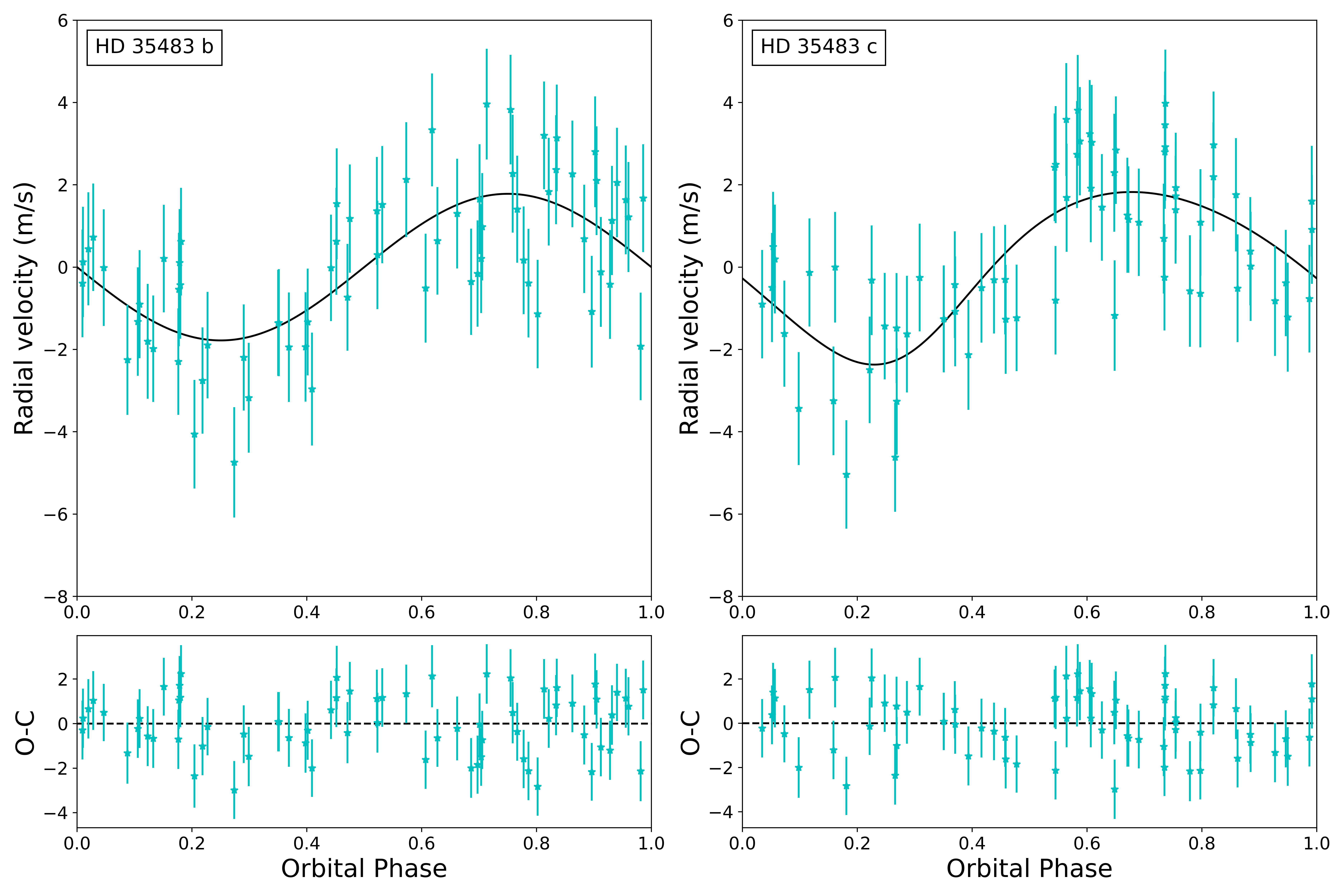}
    \caption{Phase folded and offset-subtracted ESPRESSO RVs with the modeled best-fit \texttt{allesfitter} RV curve for HD 35843 b (left) and HD 35843 c (right), with residuals plotted on the bottom. Stellar jitter has been added in quadrature with measurement uncertainties.}
    \label{fig:bestfit_rvs}
\end{figure*}

The values and uncertainties of the fitted and derived parameters listed in Tables \ref{tab:fitted_parameters} and \ref{tab:derived_parameters} are defined as the median values and 68\% confidence intervals of the posterior distributions, respectively. The best-fit light curve and RV models are shown in Figures \ref{fig:bestfit_tess} and \ref{fig:bestfit_rvs}, respectively. The corner plots for the modeled and derived parameters are shown in Figures \ref{fig:corner_fitted} and \ref{fig:corner_derived} of the Appendix, respectively. Additionally, we include the corner plot for the modeled parameters of an alternative joint fit that includes the PFS RVs in Section \ref{sec:pfsfit} of the Appendix.

\def\arraystretch{1.15}
\begin{deluxetable*}{lcc}
\tablewidth{0pc}
\tabletypesize{\scriptsize}
\tablecaption{
    Fitted system and planet parameters using ESPRESSO RVs and TESS photometry.
    \label{tab:fitted_parameters}
}
\tablehead{
    \multicolumn{1}{c}{Parameter} &
    \multicolumn{1}{c}{Prior} &
    \multicolumn{1}{c}{Value} 
}
\startdata
\multicolumn{3}{l}{\textit{Fixed parameters}} \\
$\sqrt{e_b}\, \cos{\omega_b}$ & - & 0 \\ 
$\sqrt{e_b}\, \sin{\omega_b}$ & - & 0 \\ 
 &  &  \\
\multicolumn{3}{l}{\textit{Fitted parameters}} \\
$q_{1, \textit{TESS}}$ & $N(0.32, 0.1)$ & 0.25$\pm$0.10 \\ 
$q_{2, \textit{TESS}}$ & $N(0.25, 0.1)$ & 0.22$\pm$0.10\\
$R_c / R_\star$ & $U(0, 1)$ & $0.0260 \pm 0.0006$ \\ 
$(R_c + R_\star) / a$ & $U(0, 1)$ & $0.0171 \pm 0.0005$\\
$\cos{i_c}$ & $U(0, 1)$ & $0.0074^{+0.0016}_{-0.0030}$ \\ 
$T_{0, b}$ (BJD - 2,457,000) & $U(2877,2889)$ & $2889.6910^{+0.3094}_{-0.2856} $ \\ 
$T_{0, c}$ (BJD - 2,457,000) & $U(2221.26, 2222.26)$ & $2221.7562^{+0.0020}_{-0.0016} $ \\ 
$P_b$ $\mathrm{(d)}$ & $U(8, 12)$ & $9.8991 ^{+0.0557}_{-0.0574}$ \\
$P_c$ $\mathrm{(d)}$ & $U(46, 48)$ & $46.9622 \pm 0.0002$ \\
$\sqrt{e_c}\, \cos{\omega_c}$ & $U(0, 1)$ & $-0.285^{+0.121}_{-0.097}$ \\ 
$\sqrt{e_c}\, \sin{\omega_c}$ & $U(0, 1)$ & $-0.188^{+0.221}_{-0.193} $ \\ 
$K_b$  $\mathrm{(m\, s^{-1})}$ & $U(0, 1000)$ & $1.81 \pm 0.25$ \\
$K_c$  $\mathrm{(m\, s^{-1})}$ & $U(0, 1000)$ & $2.13 \pm 0.24 $ \\
ln $\sigma_{TESS}$ & $U(-23, 0)$ & $-7.535 \pm 0.015$ \\
ln $\sigma_{E}$ $\mathrm{[km\, s^{-1}]}$ & $U(-23, 0)$ & $-6.60^{+0.10}_{-0.09}$ \\
$\gamma_{E}$ $\mathrm{(km\, s^{-1})}$ & $U(13, 14)$ & 13.76816$\pm$0.00017 \\
\enddata
\label{fitted_parameters}
\end{deluxetable*}

\def\arraystretch{1.15}
\begin{deluxetable}{lc}
\tablewidth{0pc}
\tabletypesize{\scriptsize}
\tablecaption{
    Derived planet parameters from best joint fit.
    \label{tab:derived_parameters}
}
\tablehead{
    \multicolumn{1}{c}{~~~~~~Parameter~~~~~~} &
    \multicolumn{1}{c}{~~~~~~Value~~~~~~} 
}
\startdata
$i_c$ ($^{\circ}$) & $89.58^{+0.17}_{-0.09}$ \\
$a_b$ (AU) & $0.088 \pm 0.002$ \\
$a_c$ (AU) & $0.25 \pm 0.01$ \\
$b_c^a$ & $0.47^{+0.15}_{-0.21}$ \\
$T_\mathrm{tot, c}^b$ (h) & $5.80^{+0.10}_{-0.08}$ \\
$T_\mathrm{full, c}^c$  (h) & $5.41^{+0.10}_{-0.09}$ \\
$R_c$ ($R_{\oplus}$) & $2.54^{+0.08}_{-0.07}$ \\
$M_b\, sin\, i$ ($M_{\oplus}$) & $5.84 \pm 0.84$ \\
$M_c$ ($M_{\oplus}$) & $11.32^{+1.58}_{-1.48}$ \\
$\rho_c$ $\mathrm{(g\, cm^{-3})}$ & $3.80^{+0.68}_{-0.60}$ \\
$e_b$ & 0 (fixed) \\
$e_c$ & $0.153^{+0.070}_{-0.064}$ \\
$\omega_c$ ($^{\mathrm{o}}$) & $211^{+37}_{-32}$ \\
$T_\mathrm{eq, c}$ (K) & $479 \pm 12$ 
\enddata
\tablenotetext{a}{Impact parameter.}
\tablenotetext{b}{From 1st to last (4th) contacts.}
\tablenotetext{c}{From 2nd to 3rd contacts.}

\label{derived_parameters}
\end{deluxetable}

\subsection{Potential Transit Timing Variations} \label{subsec:ttvs}

In addition to the joint fit, we also ran a fit to determine if HD 35843 c exhibited any transit timing variations that might indicate the presence of additional planets in the system. We use the best-fit parameters from the joint fit as priors for the TTV fit, holding the epoch and period fixed. While the \textit{TESS} transits are consistent with the linear ephemeris, there is evidence for TTVs in the ground-based LCOGT and NGTS observations, as shown in Figure \ref{fig:ttvs}. It is important to note, however, that none of the ground-based observations covered a full transit. Combined with the shallow transit depth of less than 1 ppt, it is possible that these apparent TTVs are not real. Future transit observations will help determine if there are indeed TTVs present, and if so, could help characterize the perturbing planet(s). Another transit of HD 35483 c should be observable by \textit{TESS} when it re-observes the star in Sector 87 and Sector 95. 

\section{Discussion} \label{sec:discussion}

We report the discovery and confirmation of two planets orbiting HD 35843: HD 35843 b is a non-transiting planet with an orbital period of 9.90$\pm$0.06 days and a minimum mass of 5.47$\pm$0.82 $M_{\oplus}$ while HD 35843 c is a transiting sub-Neptune with an orbital period of 46.9622$\pm$0.0002 days and a radius of 2.54$\pm$0.008 $R_{\oplus}$ and mass of 11.32$\pm$1.60 $M_{\oplus}$.

\begin{figure}
    \centering
    \includegraphics[width=\linewidth]{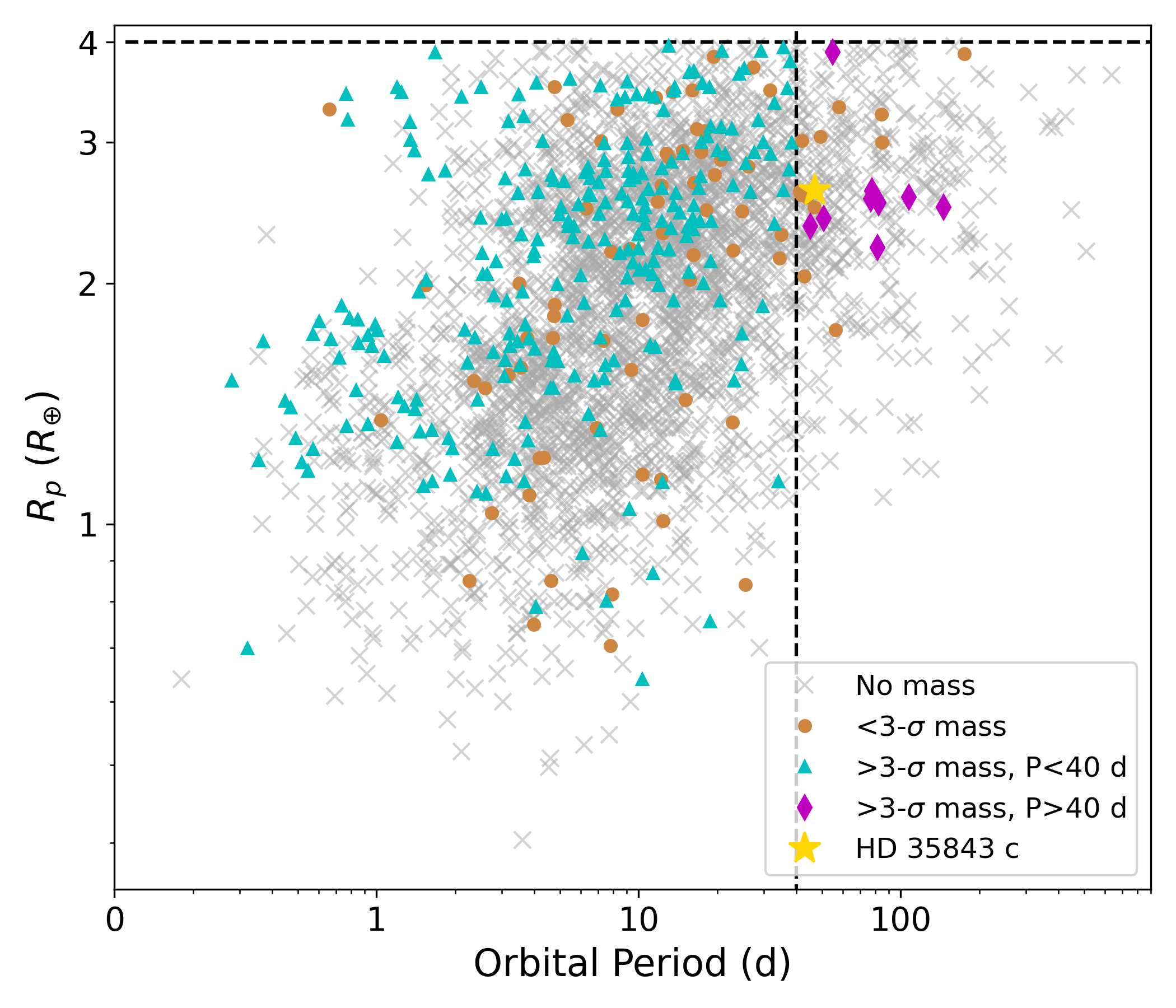}
    \caption{Period-radius diagram of confirmed and validated sub-Neptunes as of this writing. Magenta diamonds denote long period ($P> 40$ days) planets with upper limits on their masses, masses from transit timing variations, or radial velocities masses that are loosely constrained ($<3\sigma$). Cyan triangles mark well-constrained masses ($>3\sigma$). The gold star marks the subject of this paper, HD 35843 c.}
    \label{fig:per_rad_mass}
\end{figure}

The confirmation of this system adds to the small but growing list of long-period sub-Neptunes. Of the more than 4000 transiting exoplanets confirmed so far, only $\sim$12\% have a period greater than 40 days. Relatively little is known about this population of long-period, temperate planets. Sub-Neptune size planets (defined as planets with R$_p$ between 2 and 4 R$_\oplus$) are relatively plentiful, making up almost $\sim$40\% of current confirmed transiting exoplanets \citep{Sozzetti_2021}. However, $\sim$10\% of confirmed long-period (P $>$ 40 d) transiting sub-Neptunes have any type of mass measurement -- only 39 of 393. Of these 39 sub-Neptunes with mass measurements, 30 are upper limits, TTV masses, or poorly constrained mass measurements\footnote{https://exoplanetarchive.ipac.caltech.edu/}. The 9 planets with 3-$\sigma$ RV masses are marked alongside HD\,35843\ c in Figure \ref{fig:per_rad_mass}. HD\,35843\ is the third brightest host system (V = 9.4 mag) in the current sample of sub-Neptunes with a period than 40 days and a precisely measured mass. These targets are important for atmospheric follow-up, providing insight into the distinction between gas and terrestrial planet in both composition and formation.

\begin{figure}
    \centering
    \includegraphics[width=\linewidth]{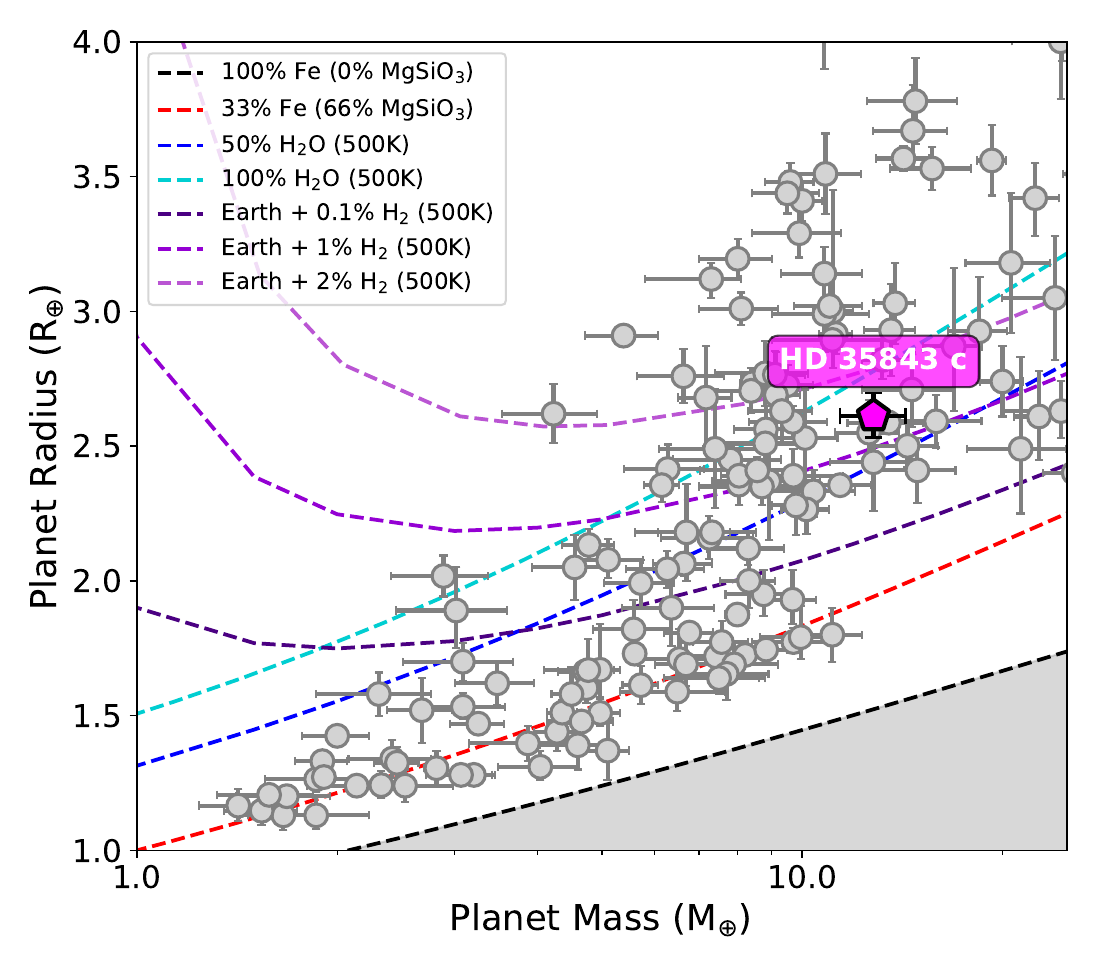}
    \caption{Mass-radius diagram made using \texttt{mr-plotter} \citep{Castro-Gonz_2023} of sub-Neptunes with measured masses as well as theoretical models from \citet{Zeng_2019}. Only planets with 5-$\sigma$ mass and radius measurements are shown for clarity. HD 35843 c lies in the overlapping region between rocky planets with a H$_2$ atmosphere and ``water worlds".}
    \label{fig:mass_rad}
\end{figure}

\begin{figure*}
    \centering
    \includegraphics[width=0.49\textwidth]{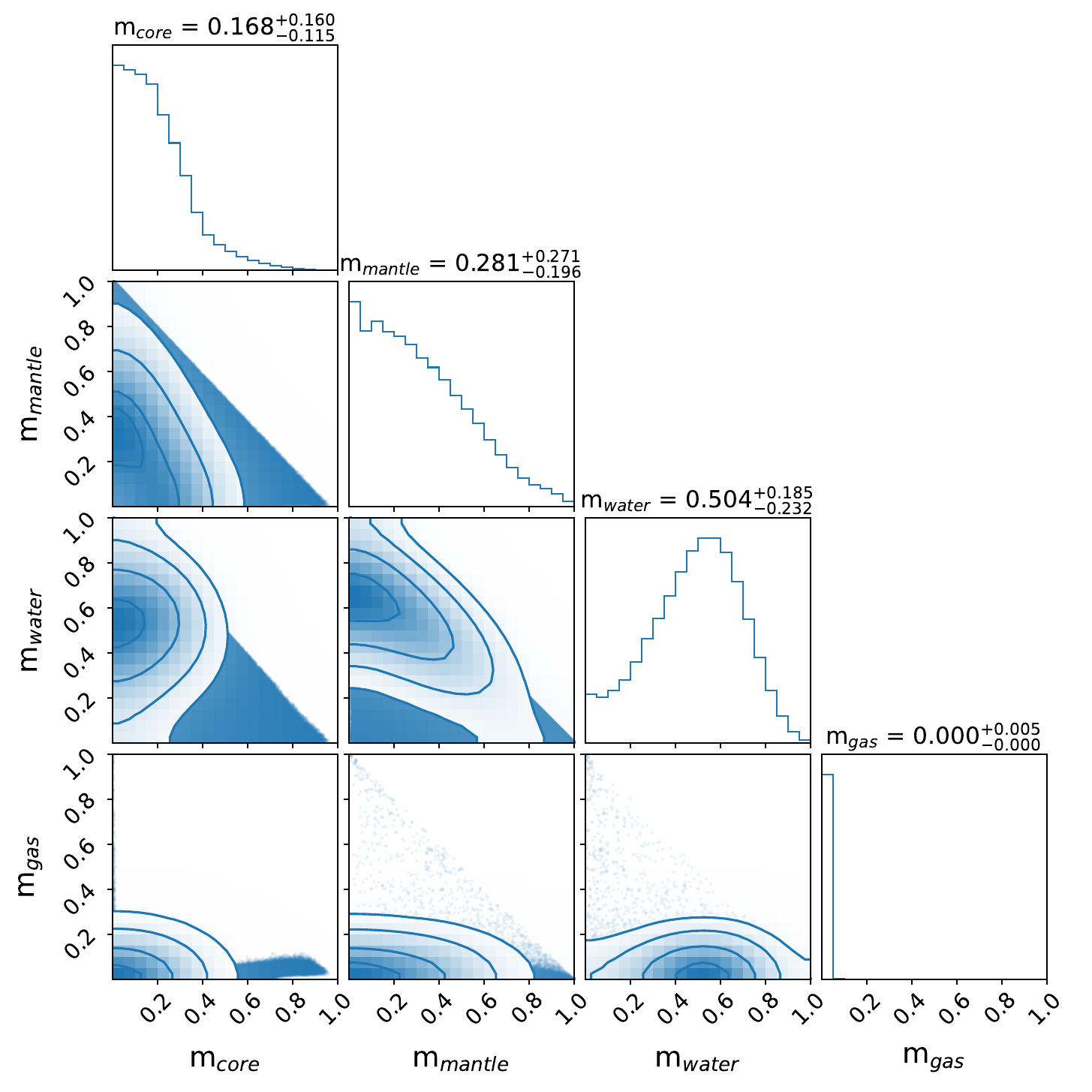}
    \includegraphics[width=0.49\textwidth]{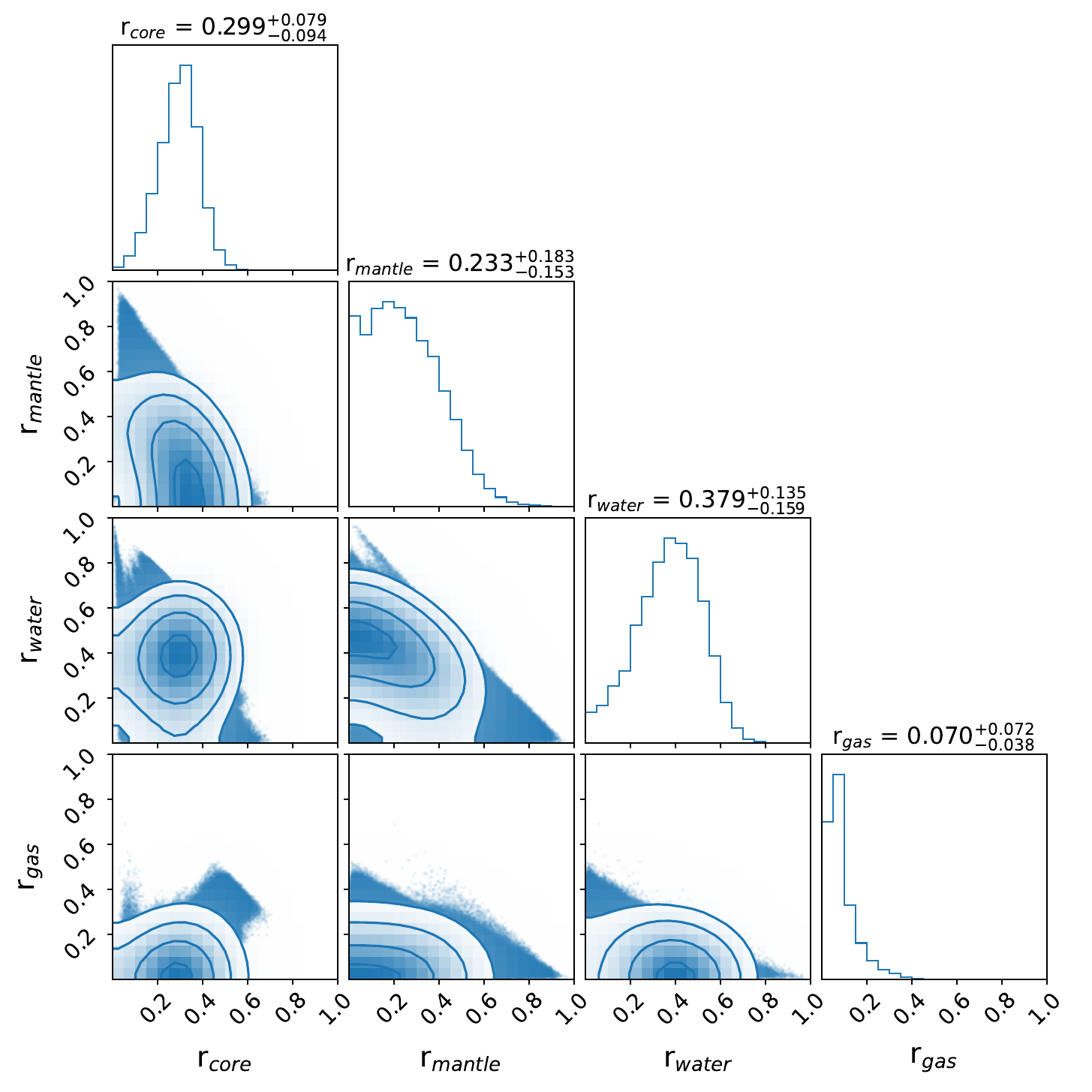}
    \caption{\textit{Left:} Mass fractions for the core, mantle, water, and gas layers calculated using ExoMDN. \textit{Right:} Radius fractions for the same layers.}
    \label{fig:mf_exomdn}
\end{figure*}

With a density of $3.80 \pm 0.70\rm\, g\, cm^{-3}$ ,  HD\,35843\ c falls within a particularly interesting portion of the radius-mass distribution of confirmed exoplanets, shown in Figure \ref{fig:mass_rad}. Based on a sample of transiting planets found by Kepler, \citet{Rogers_2015} found that planets larger in radius than $\sim$2 $R_{\oplus}$ are generally too low density to be composed of only iron and silicates. HD\,35843\ c bulk density could fit both a rocky planet with a thin hydrogen atmosphere or a water world. We analyze possible compositions by modeling the internal structure of planet c using the machine learning tool \texttt{ExoMDN} \citep{Baumeister_2023}. We run 1000 predictions using radius, mass, and temperature values drawn from normal distributions centered on the best-fit values of the joint fit, and draw 5000 samples for each prediction. ExoMDN predicts that the planet contains a significant water mass and radius fraction, as well as a non-negligible gas radius fraction (see Figure \ref{fig:mf_exomdn}). However, there exists a degeneracy between sub-Neptune planet models with H$_2$O or H-He atmospheres that cannot be broken with mass and radius alone and requires atmospheric observations \citep{Rogers_2023}. A notable example of a planet in this regime of degenerate compositions is K2-18 b, which was recently found to be a possible ``Hycean" world, with a water-rich interior and an H$_2$-rich atmosphere, along with possible bio-signature \citep{Madhusudhan_2023}. However, \citet{Wogan_2024} found that the JWST atmospheric results were also consistent with a a gas-rich mini-Neptune with 100 times solar metallicity.

With a Transmission Spectroscopy Metric (TSM; \citealt{Kempton18}) value of $\sim 25$, HD 35843 c is a promising long-period target for follow-up characterization with JWST and upcoming ELTs to determine its exact composition. While its TSM is lower than that of K2-18 b (TSM $\sim 42$), it would only require a handful of JWST observations to study its atmosphere. Its relatively cool equilibrium temperature of 479 K also makes it a promising target. Planets with equilibrium temperatures below 500 K are theorized to have clearer atmospheres with fewer clouds and hazes present, making them more amenable candidates for atmospheric observations \citep{Brande_2024}. 

In addition to its composition, the  HD\,35843\ system is rather unusual in that the outer planet  HD\,35843\ c transits but the inner planet  HD\,35843\ b may not. HD 35843 is the 6th system with a non-transiting planet interior to a transit sub-Neptune. We do not see evidence for any transits of HD\,35843\ b in the existing TESS photometry. However, based on our transit injection and recovery retrievals (Figure \ref{fig:tr_inj_rec}), there is a possibility that the transits were missed, if the inner planet's radius is below $\sim$1.4 $R_{\oplus}$. We calculate an average 1-hour Combined Differential Photometric Precision (CDPP) of $\sim$150 ppm across the existing 5 sectors of TESS data. Using the period and ephemeris from the RVs, along with the actual observation times from the TESS photometry, we determine that 14 transits would have been observed during this time span. Following Equation 7 from \citet{Kunimoto_2022}, we find that HD 35843 b would need to be at least $\sim$1.4 $R_{\oplus}$ in order to achieve an SNR of 7 for 14 transits, assuming a circular orbit and an impact parameter of 0.5. With a minimum mass of $\sim$5.5 $M_{\oplus}$, this planet could potentially be rocky in composition based on theoretical models from \citet{Zeng_2019} shown in Figure \ref{fig:per_rad_mass}). If this planet is rocky with an iron-rich core and no significant gas envelope, then its radius could be smaller than $\sim$ 1.4 $R_{\oplus}$. A planet at this radius would not have been detectable with the currently available TESS photometry. Future high-precision photometric observations could help determine if this planet is transiting or not, including upcoming TESS observations in Sectors 87 and 95. 

If HD\,35843\ b does in fact transit, then HD\,35843\ would have planets on both sides of the radius valley. An example of one such system is the HD\,22946\ (TOI-411) system which consists of 3 planets: the innermost HD\,22946\ b with 1.362 $R_{\oplus}$ orbiting at 4.04 days while HD\,22946\ c and HD\,22946\ d have radii of 2.328 $R_{\oplus}$ and 2.607 $R_{\oplus}$ respectively and orbit at 9.57 days and 47.42 days \citep{Garai_2023}. HD\,22946\ d is very similar in radius and orbit to HD\,35843\ c, though significantly denser with a mass of 26.57 $M_{\oplus}$. Depending on the size of HD\,35843\ b, the HD\,35843\ system could add another system with transiting planets on both sides of the radius valley.

If HD\,35843\ b turns out to be truly non-transiting, then this relatively rare type of system architecture may provide insight into the formation history that could also explain the eccentricity of the outer planet. Assuming an impact parameter of 1 for HD 35843 b, we can place a limit on the inclination of the planet of 87.3$^\mathrm{\circ}$. As such the mutual inclination between the two planets may be as little as $\sim 3^\mathrm{\circ}$. However, even if HD 35843 b is transiting, there could still be a significant mutual inclination. If there is a significant mutual inclination, it could be that there is a larger, longer-period planet in the system that perturbed the orbit of HD\,35843\ c \citep{Boue_Fabrycky_2014}. In this scenario, a misaligned orbit could be measured using the Rossiter-McLaughlin effect. While the expected RM effect amplitude of HD\,35843\ c is small at only $\sim 1$ m/s, such amplitudes can be detected with current instruments. In fact, such measurements have been made for planets in the HD 3167 system, which is another system with a non-transiting planet interior to a transiting sub-Neptune \citep{Christiansen_Vanderburg_2017}. In that system, the outer transiting planet is in a polar orbit, while the innermost planet, which also transits, is in a well-aligned orbit \citep{Dalal_2019, Bourrier_2021}.

In addition to the previously mentioned HD 3167 system, another example is the TOI-431 system. Like the HD\,22946\  system, the TOI-431 consists of planets on both sides of the radius valley: TOI-431 b is a super-Earth with a radius of $R_p\sim 1.28 R_{\oplus}$ and an orbital period of 0.49 days while TOI-431 d is a sub-Neptune with a radius of $R_p\sim 3.29 R_{\oplus}$ and a period of 12.46 days \citep{Osborn_2021}. Additionally, there is a third, non-transiting planet in the system between the two transiting planets: TOI-431 c with a period of 4.85 days.

\begin{figure*}
    \centering
    \includegraphics[width=0.49\textwidth]{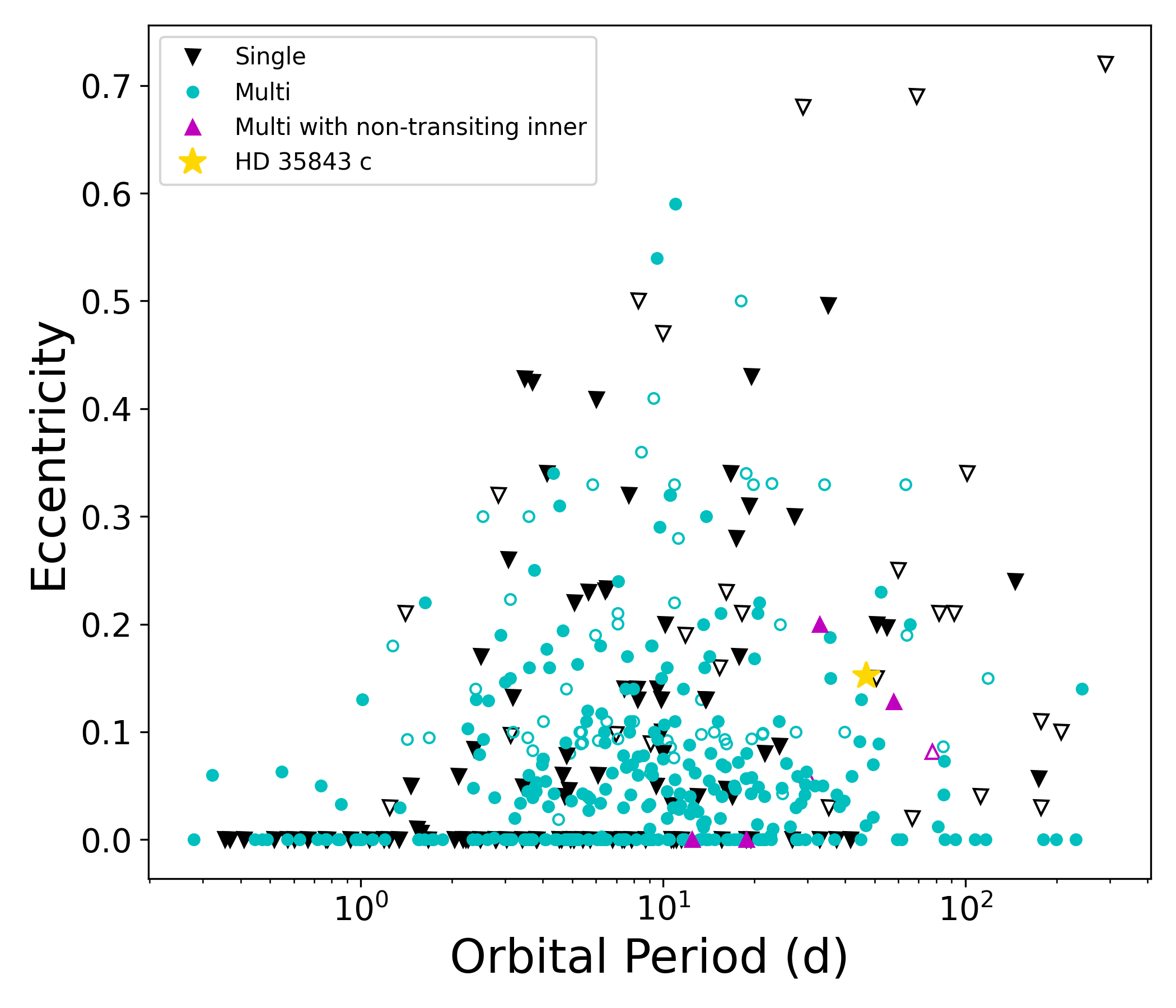}
    \includegraphics[width=0.49\textwidth]{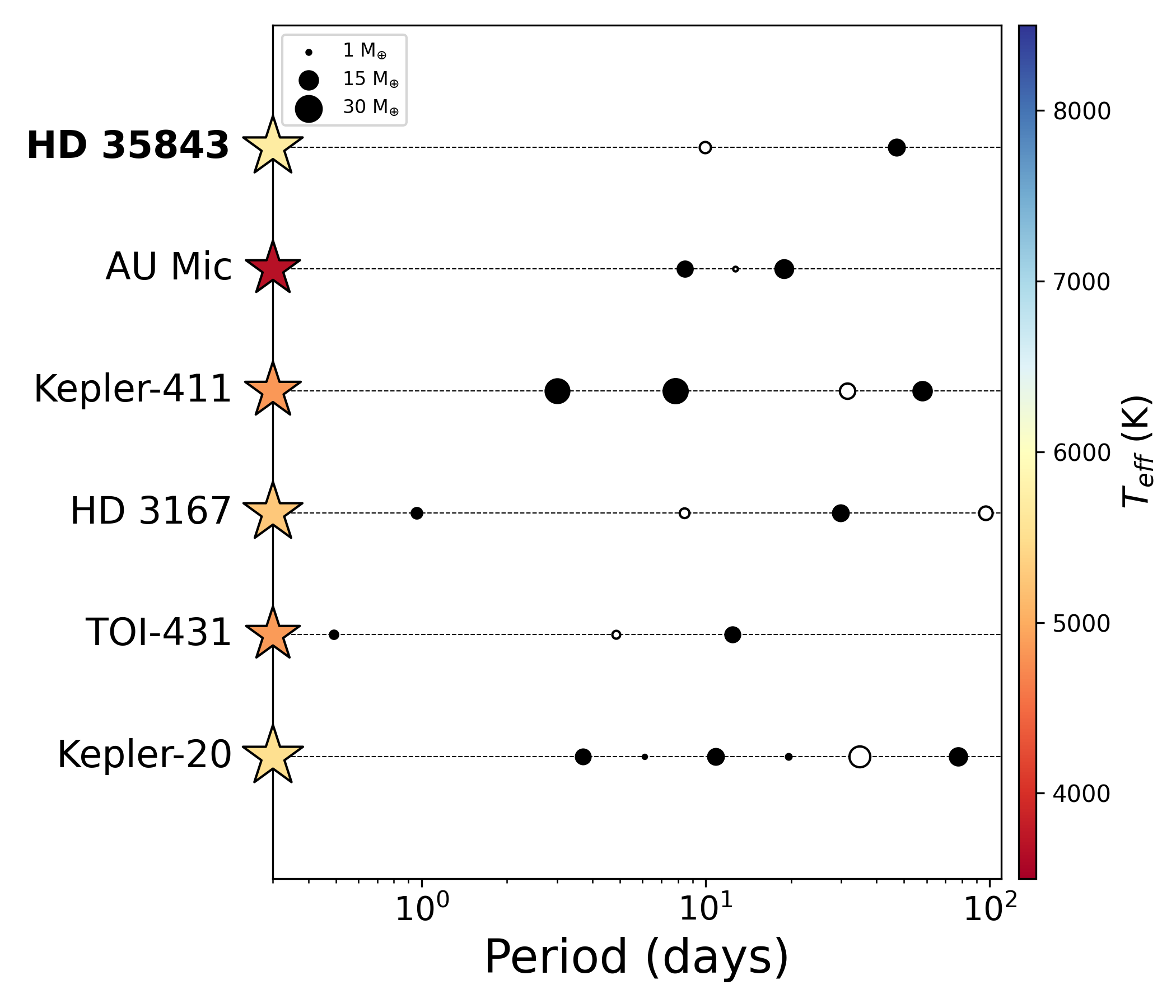}
    \caption{\textit{Left}:Eccentricity measurements of transiting sub-Neptune systems vs. orbital periods. Empty markers denote upper limits on eccentricity. HD 35843 c has the highest eccentricity of the 6 sub-Neptunes. \textit{Right:} HD 35843 and the 5 other sub-Neptune systems with a non-transiting inner planet. HD 35843 has the hottest host star and fewest number of planets of the 6 systems.}
    \label{fig:eccsubnep_multis}
\end{figure*}

This sample of 6 multi-planet systems with a transiting sub-Neptune and an inner planet with no observed transits is shown in Figure \ref{fig:eccsubnep_multis}) in magenta. Four of the six outer transiting sub-Neptunes have non-zero eccentricities. For the less compact systems like HD 35843, these eccentric orbits combined with the larger separation between planets could indicate that the systems are the remnants of resonant chains that became unstable \citep{2017MNRAS.470.1750I}. However, whether these eccentricities are indicative or not of a particular formation mechanism for these types of systems remains to be seen given the small sample size. Spin-orbit measurements would also shed light on possible formation and evolution scenarios, as a misaligned orbit could be indicative of one of the following 4 scenarios:  1) Tidal dissipation, 2) Von Zeipel–Kozai–Lidov cycles, 3) Secular resonance crossing, or 4) magnetic warping of a young protoplanetary disk \citep{Albrecht_2021}. However, a larger sample size of multiple planets systems including additional systems with truly non-transiting inner planets and well-measured orbital parameters, including eccentricity and obliquity, is needed before drawing any definitive conclusions. 

\section{Summary} \label{sec:summary}

We announce the discovery and confirmation of two planets orbiting the metal-poor G dwarf HD 35843. HD 35843 b is a planet on a 9.90$\pm$0.06 d orbit with a minimum mass of 5.84$\pm$0.84 $M_{\oplus}$ that is currently not known to be transiting although transits cannot be ruled out. HD 35843 c is a transiting sub-Neptune on a mildly eccentric (e = 0.15$\pm$0.07) 46.9622$\pm$0.0002 d orbit with a radius of 2.54$\pm$0.08 $R_{\oplus}$ mass of 11.32$\pm$1.60 $M_{\oplus}$ ($\rho_c$ = 3.80$\pm$0.70 g cm$^{-3}$). HD 35843 c was first identified by Planet Hunters TESS, after which we ruled out false positive scenarios using follow-up photometry, spectroscopy, and imaging. We used ESPRESSO RVs to characterize the mass and orbit of HD 35843 c, which also revealed the RV signal of HD 35843 b. The bulk density of HD 35843 c indicate its composition lies in the overlapping region between so-called ``water worlds" and rocky planets with thick H$_2$ atmospheres. Along with the brightness of the host star, this makes HD 35843 c a promising system for atmospheric characterization with JWST. If HD 35843 b is indeed non-transiting, then HD 35843 would be one of only 6 planet systems with a non-transiting planet interior to a transiting one.


\section*{Acknowledgments}
This work has been carried out within the framework of the NCCR PlanetS supported by the Swiss National Science Foundation under grants 51NF40$\_$182901 and 51NF40\_205606. ML acknowledges support of the Swiss National Science Foundation under grant number PCEFP2$\_$194576.

DD acknowledges support from the TESS Guest Investigator Program grant 80NSSC22K1353, and from the NASA Exoplanet Research Program grant 18-2XRP18\_2-0136.

Co-funded by the European Union (ERC, FIERCE, 101052347). Views and opinions expressed are however those of the author(s) only and do not necessarily reflect those of the European Union or the European Research Council. Neither the European Union nor the granting authority can be held responsible for them. This work was supported by FCT - Funda\c{c}\~{a}o para a Ci\^{e}ncia e a Tecnologia through national funds and by FEDER through COMPETE2020 - Programa Operacional Competitividade e Internacionaliza\c{c}\~{a}o by these grants: UIDB/04434/2020; UIDP/04434/2020.

The MEarth Team gratefully acknowledges funding from the David and Lucile Packard Fellowship for Science and Engineering (awarded to D.C.). This material is based upon work supported by the National Science Foundation under grants AST-0807690, AST-1109468, AST-1004488 (Alan T. Waterman Award), and AST-1616624, and upon work supported by the National Aeronautics and Space Administration under Grant No. 80NSSC18K0476 issued through the XRP Program. This work is made possible by a grant from the John Templeton Foundation. The opinions expressed in this publication are those of the authors and do not necessarily reflect the views of the John Templeton Foundation.

This work includes data collected under the NGTS project at the ESO La Silla Paranal Observatory. The NGTS facility is operated by a consortium institutes with support from the UK Science and Technology Facilities Council (STFC) under projects ST/M001962/1, ST/S002642/1 and ST/W003163/1.

Some of the observations in this paper made use of the High-Resolution Imaging instrument Zorro and were obtained under Gemini LLP Proposal Number: GN/S-2021A-LP-105. Zorro was funded by the NASA Exoplanet Exploration Program and built at the NASA Ames Research Center by Steve B. Howell, Nic Scott, Elliott P. Horch, and Emmett Quigley. Zorro was mounted on the Gemini South telescope of the international Gemini Observatory, a program of NSF’s OIR Lab, which is managed by the Association of Universities for Research in Astronomy (AURA) under a cooperative agreement with the National Science Foundation. on behalf of the Gemini partnership: the National Science Foundation (United States), National Research Council (Canada), Agencia Nacional de Investigaci\'{o}n y Desarrollo (Chile), Ministerio de Ciencia, Tecnolog\'{i}a e Innovaci\'{o}n (Argentina), Minist\'{e}rio da Ci\^{e}ncia, Tecnologia, Inova\c{c}\~{o}es e Comunica\c{c}\~{o}es (Brazil), and Korea Astronomy and Space Science Institute (Republic of Korea).

\textsc{Minerva}-Australis is supported by Australian Research Council LIEF Grant LE160100001, Discovery Grants DP180100972 and DP220100365, Mount Cuba Astronomical Foundation, and institutional partners University of Southern Queensland, UNSW Sydney, MIT, Nanjing University, George Mason University, University of Louisville, University of California Riverside, University of Florida, and The University of Texas at Austin.

We respectfully acknowledge the traditional custodians of all lands throughout Australia, and recognise their continued cultural and spiritual connection to the land, waterways, cosmos, and community. We pay our deepest respects to all Elders, ancestors and descendants of the Giabal, Jarowair, and Kambuwal nations, upon whose lands the \textsc{Minerva}-Australis facility at Mt Kent is situated.


This work makes use of observations from the LCOGT network. Part of the LCOGT telescope time was granted by NOIRLab through the Mid-Scale Innovations Program (MSIP). MSIP is funded by NSF.


This research has made use of the Exoplanet Follow-up Observation Program \citep{exofop_doi}
website, which is operated by the California Institute of Technology, under contract with the National Aeronautics and Space Administration under the Exoplanet Exploration Program.

Some of the data presented in this paper were obtained from the Mikulski Archive for Space Telescopes (MAST) at the Space Telescope Science Institute. The specific data sets used in this work come from the TESS all sector light curves \citep{mast_lc_doi}, and the TESS all sectors target pixel files \citep{mast_tpf_doi}.


Funding for the TESS mission is provided by NASA’s Science Mission Directorate. We acknowledge the use of public TESS data from pipelines at the TESS Science Office and at the TESS Science Processing Operations Center. Resources supporting this work were provided by the NASA High-End Computing (HEC) Program through the NASA Advanced Supercomputing (NAS) Division at Ames Research Center for the production of the SPOC data products. TESS data presented in this paper were obtained from the Mikulski Archive for Space Telescopes (MAST) at the Space Telescope Science Institute. KAC and CNW acknowledge support from the TESS mission via subaward s3449 from MIT.


%

\vspace{5mm}
\facilities{\textit{TESS}, LCOGT, Exoplanet Archive}


\software{astropy \citep{2013A&A...558A..33A,2018AJ....156..123A,2022ApJ...935..167A}, AstroImageJ \citep{Collins:2017}, TAPIR \citep{Jensen:2013}}



\bibliography{main}
\bibliographystyle{aasjournal}



\appendix

\section{RV Measurements}
\restartappendixnumbering
In this appendix, we present the values of the radial velocities from PFS, CHIRON, CORALIE, LCO-NRES, and MINERVA-Australis.

\begin{deluxetable}{llll}
\tablewidth{0pc}
\tabletypesize{\scriptsize}
\tablecaption{
    Radial Velocities
    \label{tab:recrvs}
}
\tablehead{
    \multicolumn{1}{l}{BJD} &
    \multicolumn{1}{l}{RV [km/s]} &
    \multicolumn{1}{l}{$\sigma_\mathrm{RV}$ [km/s]} &
    \multicolumn{1}{l}{Instrument}
    }
\startdata
2459470.85282  & -0.00544 & 0.00108 & PFS     \\
2459477.87206  & 0        & 0.00112 & PFS     \\
2459478.89641  & -0.00036 & 0.00100 & PFS     \\
2459501.80006  & -0.00787 & 0.00078 & PFS     \\
2459507.77986  & -0.00841 & 0.00088 & PFS     \\
2459531.85017  & -0.00025 & 0.00132 & PFS     \\
2459534.76259  & 0.0021   & 0.00129 & PFS     \\
2459593.603455 & -0.00036 & 0.00101 & PFS     \\
2459597.59076  & 0.00237  & 0.00115 & PFS     \\
2459603.636923 & 0.00049  & 0.00074 & PFS     \\
2459652.54215  & -0.00271 & 0.00123 & PFS     \\
2459653.53487  & -0.00409 & 0.00101 & PFS     \\
2459657.54543  & 0.00065  & 0.00111 & PFS     \\
2459827.89712  & 0.00246  & 0.00083 & PFS     \\
2459829.87974  & 0.00534  & 0.00092 & PFS     \\
2459832.85072  & 0.00195  & 0.00095 & PFS     \\
2459889.74822  & -0.00298 & 0.00118 & PFS     \\
2459897.84653  & 0.00204  & 0.00131 & PFS     \\
2459285.57483  & 12.348   & 0.015   & CHIRON  \\
2459340.46436  & 12.341   & 0.023   & CHIRON  \\
2459821.84363  & 13.75044 & 0.00537 & CORALIE \\
2459854.81514  & 13.75020 & 0.00505 & CORALIE \\
2459861.87825  & 13.74556 & 0.00582 & CORALIE \\
2459867.80124  & 13.74611 & 0.00555 & CORALIE \\
2459466.52146  & 14.194   & 0.084   & LCO-NRES
\enddata
\label{recrvs}
\end{deluxetable}

\section{ESPRESSO RVs and Activity Indices}
\restartappendixnumbering
In this appendix, we present the values of the ESPRESSO radial velocities and activity indices with uncertainties.

\begin{deluxetable}{llllllllll}
\rotate
\tablewidth{0pc}
\tabletypesize{\scriptsize}
\tablecaption{
    RVs and activity indices obtained from the ESPRESSO spectra under Program ID 110.243Y.001.
    \label{tab:espresso_vals_kh}
}
\tablehead{
    \multicolumn{1}{l}{BJD-2400000} &
    \multicolumn{1}{l}{RV [m/s]} &
    \multicolumn{1}{l}{FWHM} & 
    \multicolumn{1}{l}{Bisector} &
    \multicolumn{1}{l}{Contrast} &
    \multicolumn{1}{l}{S$_{MW}$} &
    \multicolumn{1}{l}{H$-{\alpha}$} &
    \multicolumn{1}{l}{Na} &
    \multicolumn{1}{l}{Ca} &
    \multicolumn{1}{l}{log(R'$_{HK}$)}
    }
\startdata
59853.7451 & 13766.546 $\pm$ 0.402 & 6659.60 $\pm$ 0.80 & -67.76 $\pm$ 0.80 & 53.372 $\pm$ 0.006 &  0.16846 $\pm$ 0.00008  &  0.20399 $\pm$ 0.00003  &  0.31514 $\pm$ 0.00003  &  0.13166 $\pm$ 0.00008  &  -5.0128 $\pm$ 0.0004   \\
59868.8646 & 13768.662 $\pm$ 0.293 & 6659.06 $\pm$ 0.59 & -68.21 $\pm$ 0.59 & 53.376 $\pm$ 0.005 &  0.17087 $\pm$ 0.00004  &  0.20212 $\pm$ 0.00002  &  0.31767 $\pm$ 0.00002  &  0.13398 $\pm$ 0.00004  &  -5.0003 $\pm$ 0.0002   \\
59876.8567 & 13768.649 $\pm$ 0.413 & 6663.68 $\pm$ 0.83 & -64.68 $\pm$ 0.83 & 53.377 $\pm$ 0.007 &  0.17000 $\pm$ 0.00008     &  0.20282 $\pm$ 0.00004  &  0.31091 $\pm$ 0.00003  &  0.13314 $\pm$ 0.00008  &  -5.0048 $\pm$ 0.0004   \\
59878.845 & 13770.082 $\pm$ 0.404 & 6663.85 $\pm$ 0.81 & -66.14 $\pm$ 0.81 & 53.351 $\pm$ 0.006 &  0.17064 $\pm$ 0.00008  &  0.20325 $\pm$ 0.00004  &  0.31208 $\pm$ 0.00003  &  0.13376 $\pm$ 0.00008  &  -5.0015 $\pm$ 0.0004   \\
59880.8392 & 13766.161 $\pm$ 0.309 & 6663.07 $\pm$ 0.62 & -65.22 $\pm$ 0.62 & 53.315 $\pm$ 0.005 &  0.17404 $\pm$ 0.00005  &  0.20326 $\pm$ 0.00003  &  0.31681 $\pm$ 0.00002  &  0.13701 $\pm$ 0.00005  &  -4.9844 $\pm$ 0.0002   \\
59883.8385 & 13763.764 $\pm$ 0.313 & 6660.77 $\pm$ 0.63 & -64.37 $\pm$ 0.63 & 53.361 $\pm$ 0.005 &  0.17137 $\pm$ 0.00005  &  0.20249 $\pm$ 0.00003  &  0.31708 $\pm$ 0.00002  &  0.13445 $\pm$ 0.00005  &  -4.9978 $\pm$ 0.0003   \\
59891.8522 & 13763.154 $\pm$ 0.297 & 6656.76 $\pm$ 0.59 & -68.69 $\pm$ 0.59 & 53.378 $\pm$ 0.005 &  0.16961 $\pm$ 0.00005  &  0.20210 $\pm$ 0.00002   &  0.31345 $\pm$ 0.00002  &  0.13277 $\pm$ 0.00004  &  -5.0068 $\pm$ 0.0002   \\
59904.8492 & 13771.005 $\pm$ 0.285 & 6662.56 $\pm$ 0.57 & -66.80 $\pm$ 0.57 & 53.316 $\pm$ 0.005 &  0.17300 $\pm$ 0.00004    &  0.20234 $\pm$ 0.00002  &  0.31144 $\pm$ 0.00002  &  0.13601 $\pm$ 0.00004  &  -4.9896 $\pm$ 0.0002   \\
59905.7077 & 13772.945 $\pm$ 0.325 & 6660.93 $\pm$ 0.65 & -65.89 $\pm$ 0.65 & 53.346 $\pm$ 0.005 &  0.17080 $\pm$ 0.00006   &  0.20153 $\pm$ 0.00003  &  0.30888 $\pm$ 0.00002  &  0.13391 $\pm$ 0.00005  &  -5.0007 $\pm$ 0.0003   \\
59906.6481 & 13773.702 $\pm$ 0.314 & 6662.03 $\pm$ 0.63 & -65.71 $\pm$ 0.63 & 53.334 $\pm$ 0.005 &  0.17160 $\pm$ 0.00005   &  0.20241 $\pm$ 0.00002  &  0.30770 $\pm$ 0.00002   &  0.13468 $\pm$ 0.00005  &  -4.9966 $\pm$ 0.0003   \\
59907.7187 & 13771.675 $\pm$ 0.447 & 6658.64 $\pm$ 0.89 & -65.49 $\pm$ 0.89 & 53.387 $\pm$ 0.007 &  0.16473 $\pm$ 0.00010   &  0.20097 $\pm$ 0.00004  &  0.31046 $\pm$ 0.00003  &  0.12809 $\pm$ 0.00009  &  -5.0328 $\pm$ 0.0005   \\
59909.7664 & 13770.688 $\pm$ 0.437 & 6658.31 $\pm$ 0.87 & -65.18 $\pm$ 0.87 & 53.395 $\pm$ 0.007 &  0.16425 $\pm$ 0.00009  &  0.20111 $\pm$ 0.00004  &  0.30245 $\pm$ 0.00003  &  0.12763 $\pm$ 0.00009  &  -5.0354 $\pm$ 0.0005   \\
59910.7034 & 13768.184 $\pm$ 0.344 & 6657.73 $\pm$ 0.69 & -65.84 $\pm$ 0.69 & 53.393 $\pm$ 0.006 &  0.16743 $\pm$ 0.00006  &  0.20020 $\pm$ 0.00003   &  0.30375 $\pm$ 0.00002  &  0.13068 $\pm$ 0.00006  &  -5.0182 $\pm$ 0.0003   \\
59917.7452 & 13771.888 $\pm$ 0.299 & 6661.02 $\pm$ 0.60 & -68.76 $\pm$ 0.60 & 53.359 $\pm$ 0.005 &  0.17029 $\pm$ 0.00005  &  0.20299 $\pm$ 0.00002  &  0.30848 $\pm$ 0.00002  &  0.13342 $\pm$ 0.00004  &  -5.0033 $\pm$ 0.0002   \\
59919.5808 & 13769.698 $\pm$ 0.401 & 6657.01 $\pm$ 0.80 & -67.36 $\pm$ 0.80 & 53.367 $\pm$ 0.006 &  0.16718 $\pm$ 0.00008  &  0.20166 $\pm$ 0.00003  &  0.30343 $\pm$ 0.00003  &  0.13044 $\pm$ 0.00008  &  -5.0195 $\pm$ 0.0004   \\
59928.6052 & 13768.414 $\pm$ 0.551 & 6657.59 $\pm$ 1.10 & -67.11 $\pm$ 1.10 & 53.436 $\pm$ 0.009 &  0.16170 $\pm$ 0.00014   &  0.19905 $\pm$ 0.00005  &  0.30241 $\pm$ 0.00004  &  0.12519 $\pm$ 0.00014  &  -5.0497 $\pm$ 0.0008   \\
59928.6933 & 13769.312 $\pm$ 0.383 & 6660.35 $\pm$ 0.77 & -66.79 $\pm$ 0.77 & 53.374 $\pm$ 0.006 &  0.16781 $\pm$ 0.00008  &  0.20087 $\pm$ 0.00003  &  0.30384 $\pm$ 0.00002  &  0.13104 $\pm$ 0.00007  &  -5.0162 $\pm$ 0.0004   \\
59937.8231 & 13768.078 $\pm$ 0.374 & 6658.87 $\pm$ 0.75 & -66.19 $\pm$ 0.75 & 53.396 $\pm$ 0.006 &  0.16681 $\pm$ 0.00007  &  0.20206 $\pm$ 0.00003  &  0.30909 $\pm$ 0.00002  &  0.13008 $\pm$ 0.00007  &  -5.0215 $\pm$ 0.0004   \\
59938.7900 & 13767.122 $\pm$ 0.374 & 6659.65 $\pm$ 0.75 & -66.26 $\pm$ 0.75 & 53.410 $\pm$ 0.006 &  0.16738 $\pm$ 0.00007  &  0.20108 $\pm$ 0.00003  &  0.31301 $\pm$ 0.00002  &  0.13064 $\pm$ 0.00007  &  -5.0184 $\pm$ 0.0004   \\
59940.6806 & 13766.463 $\pm$ 0.321 & 6658.60 $\pm$ 0.64 & -67.70 $\pm$ 0.64 & 53.385 $\pm$ 0.005 &  0.16960 $\pm$ 0.00005   &  0.20180 $\pm$ 0.00003   &  0.30732 $\pm$ 0.00002  &  0.13276 $\pm$ 0.00005  &  -5.0068 $\pm$ 0.0003   \\
59942.6622 & 13765.466 $\pm$ 0.467 & 6656.62 $\pm$ 0.93 & -65.93 $\pm$ 0.93 & 53.446 $\pm$ 0.007 &  0.16468 $\pm$ 0.00010   &  0.19999 $\pm$ 0.00004  &  0.30601 $\pm$ 0.00003  &  0.12805 $\pm$ 0.00010   &  -5.0330 $\pm$ 0.0006    \\
59943.5614 & 13767.103 $\pm$ 0.363 & 6659.74 $\pm$ 0.73 & -67.50 $\pm$ 0.73 & 53.399 $\pm$ 0.006 &  0.16744 $\pm$ 0.00007  &  0.19956 $\pm$ 0.00003  &  0.30710 $\pm$ 0.00002   &  0.13069 $\pm$ 0.00007  &  -5.0182 $\pm$ 0.0004   \\
59946.7701 & 13769.616 $\pm$ 0.313 & 6658.07 $\pm$ 0.63 & -68.09 $\pm$ 0.63 & 53.388 $\pm$ 0.005 &  0.16858 $\pm$ 0.00005  &  0.20127 $\pm$ 0.00002  &  0.30903 $\pm$ 0.00002  &  0.13178 $\pm$ 0.00005  &  -5.0121 $\pm$ 0.0003   \\
59953.5527 & 13770.355 $\pm$ 0.346 & 6659.76 $\pm$ 0.69 & -67.83 $\pm$ 0.69 & 53.364 $\pm$ 0.006 &  0.16753 $\pm$ 0.00006  &  0.19929 $\pm$ 0.00003  &  0.31961 $\pm$ 0.00002  &  0.13078 $\pm$ 0.00006  &  -5.0176 $\pm$ 0.0003   \\
59954.7578 & 13771.983 $\pm$ 0.300 & 6662.11 $\pm$ 0.60 & -67.40 $\pm$ 0.60 & 53.326 $\pm$ 0.005 &  0.17196 $\pm$ 0.00005  &  0.20242 $\pm$ 0.00002  &  0.30937 $\pm$ 0.00002  &  0.13501 $\pm$ 0.00005  &  -4.9948 $\pm$ 0.0002   \\
59960.7366 & 13769.360 $\pm$ 0.340 & 6663.19 $\pm$ 0.68 & -63.73 $\pm$ 0.68 & 53.326 $\pm$ 0.005 &  0.17018 $\pm$ 0.00006  &  0.20201 $\pm$ 0.00003  &  0.30670 $\pm$ 0.00002   &  0.13332 $\pm$ 0.00006  &  -5.0038 $\pm$ 0.0003   \\
59960.7489 & 13770.005 $\pm$ 0.357 & 6660.48 $\pm$ 0.71 & -62.29 $\pm$ 0.71 & 53.340 $\pm$ 0.006 &  0.16924 $\pm$ 0.00007  &  0.20201 $\pm$ 0.00003  &  0.30652 $\pm$ 0.00002  &  0.13241 $\pm$ 0.00007  &  -5.0087 $\pm$ 0.0004   \\
59960.7615 & 13769.473 $\pm$ 0.373 & 6660.75 $\pm$ 0.75 & -64.00 $\pm$ 0.75 & 53.339 $\pm$ 0.006 &  0.16906 $\pm$ 0.00008  &  0.20173 $\pm$ 0.00003  &  0.30618 $\pm$ 0.00002  &  0.13224 $\pm$ 0.00007  &  -5.0096 $\pm$ 0.0004   \\
59960.7739 & 13770.519 $\pm$ 0.373 & 6660.22 $\pm$ 0.75 & -64.21 $\pm$ 0.75 & 53.345 $\pm$ 0.006 &  0.16871 $\pm$ 0.00008  &  0.20130 $\pm$ 0.00003   &  0.30567 $\pm$ 0.00002  &  0.13190 $\pm$ 0.00007   &  -5.0115 $\pm$ 0.0004   \\
59963.6427 & 13768.919 $\pm$ 0.461 & 6656.66 $\pm$ 0.92 & -66.41 $\pm$ 0.92 & 53.435 $\pm$ 0.007 &  0.16477 $\pm$ 0.00011  &  0.20022 $\pm$ 0.00004  &  0.30499 $\pm$ 0.00003  &  0.12814 $\pm$ 0.00011  &  -5.0325 $\pm$ 0.0006   \\
59966.6801 & 13769.400 $\pm$ 0.392 & 6655.25 $\pm$ 0.78 & -67.69 $\pm$ 0.78 & 53.417 $\pm$ 0.006 &  0.16610 $\pm$ 0.00008   &  0.20001 $\pm$ 0.00003  &  0.30751 $\pm$ 0.00003  &  0.12941 $\pm$ 0.00008  &  -5.0253 $\pm$ 0.0004   \\
59967.7251 & 13769.738 $\pm$ 0.507 & 6654.86 $\pm$ 1.01 & -67.53 $\pm$ 1.01 & 53.461 $\pm$ 0.008 &  0.15871 $\pm$ 0.00014  &  0.19792 $\pm$ 0.00004  &  0.30539 $\pm$ 0.00003  &  0.12233 $\pm$ 0.00013  &  -5.0671 $\pm$ 0.0008   \\
59969.7486 & 13766.414 $\pm$ 0.361 & 6665.57 $\pm$ 0.72 & -69.12 $\pm$ 0.72 & 53.233 $\pm$ 0.006 &  -  &  -  &  -  &  -  &  -  \\
59970.6278 & 13766.176 $\pm$ 0.320 & 6658.17 $\pm$ 0.64 & -67.37 $\pm$ 0.64 & 53.365 $\pm$ 0.005 &  0.16783 $\pm$ 0.00005  &  0.19724 $\pm$ 0.00002  &  0.30567 $\pm$ 0.00002  &  0.13106 $\pm$ 0.00005  &  -5.0161 $\pm$ 0.0003   \\
59978.6391 & 13768.187 $\pm$ 0.422 & 6660.96 $\pm$ 0.84 & -66.81 $\pm$ 0.84 & 53.349 $\pm$ 0.007 &  0.16814 $\pm$ 0.00009  &  0.19768 $\pm$ 0.00004  &  0.30492 $\pm$ 0.00003  &  0.13136 $\pm$ 0.00009  &  -5.0145 $\pm$ 0.0005   \\
59991.6372 & 13764.334 $\pm$ 0.320 & 6657.87 $\pm$ 0.64 & -68.19 $\pm$ 0.64 & 53.381 $\pm$ 0.005 &  0.16919 $\pm$ 0.00006  &  0.20056 $\pm$ 0.00003  &  0.30653 $\pm$ 0.00002  &  0.13237 $\pm$ 0.00005  &  -5.0089 $\pm$ 0.0003   \\
59994.6880 & 13768.002 $\pm$ 0.562 & 6653.52 $\pm$ 1.12 & -66.95 $\pm$ 1.12 & 53.438 $\pm$ 0.009 &  0.15330 $\pm$ 0.00017   &  0.19620 $\pm$ 0.00005   &  0.30834 $\pm$ 0.00004  &  0.11715 $\pm$ 0.00016  &  -5.1005 $\pm$ 0.0011   \\
59995.5821 & 13768.608 $\pm$ 0.371 & 6658.89 $\pm$ 0.74 & -69.41 $\pm$ 0.74 & 53.358 $\pm$ 0.006 &  0.16522 $\pm$ 0.00007  &  0.19737 $\pm$ 0.00003  &  0.31321 $\pm$ 0.00002  &  0.12856 $\pm$ 0.00007  &  -5.0301 $\pm$ 0.0004   \\
60007.6066 & 13769.793 $\pm$ 0.304 & 6658.97 $\pm$ 0.61 & -65.98 $\pm$ 0.61 & 53.321 $\pm$ 0.005 &  0.17065 $\pm$ 0.00005  &  0.20000 $\pm$ 0.00002      &  0.31010 $\pm$ 0.00002   &  0.13376 $\pm$ 0.00005  &  -5.0015 $\pm$ 0.0003   \\
60008.5808 & 13769.967 $\pm$ 0.360 & 6661.80 $\pm$ 0.72 & -66.89 $\pm$ 0.72 & 53.332 $\pm$ 0.006 &  0.16887 $\pm$ 0.00007  &  0.19853 $\pm$ 0.00003  &  0.30753 $\pm$ 0.00002  &  0.13206 $\pm$ 0.00007  &  -5.0106 $\pm$ 0.0004   \\
60027.5622 & 13765.686 $\pm$ 0.479 & 6660.00 $\pm$ 0.96 & -65.48 $\pm$ 0.96 & 53.392 $\pm$ 0.008 &  0.16438 $\pm$ 0.00012  &  0.19680 $\pm$ 0.00004  &  0.30222 $\pm$ 0.00003  &  0.12776 $\pm$ 0.00012  &  -5.0347 $\pm$ 0.0007   \\
60030.5236 & 13763.900 $\pm$ 0.410 & 6660.59 $\pm$ 0.82 & -64.87 $\pm$ 0.82 & 53.375 $\pm$ 0.007 &  0.16547 $\pm$ 0.00010  &  0.19812 $\pm$ 0.00003  &  0.30617 $\pm$ 0.00003  &  0.12881 $\pm$ 0.00009  &  -5.0287 $\pm$ 0.0005  
\enddata
\label{espresso_vals_kh}
\end{deluxetable}

\begin{deluxetable}{llllllllll}
\rotate
\tablewidth{0pc}
\tabletypesize{\scriptsize}
\tablecaption{
    RVs and activity indices obtained from the ESPRESSO spectra under Program ID 110.2481. The log(R'$_{HK}$) values were derived under an assumption of a G8V star.
    \label{tab:espresso_vals_fb}
}
\tablehead{
    \multicolumn{1}{l}{BJD-2400000} &
    \multicolumn{1}{l}{RV [m/s]} &
    \multicolumn{1}{l}{FWHM} & 
    \multicolumn{1}{l}{Bisector} &
    \multicolumn{1}{l}{Contrast} &
    \multicolumn{1}{l}{S$_{MW}$} &
    \multicolumn{1}{l}{H$-{\alpha}$} &
    \multicolumn{1}{l}{Na} &
    \multicolumn{1}{l}{Ca} &
    \multicolumn{1}{l}{log(R'$_{HK}$)}
    }
\startdata
59857.8728 & 13769.095 $\pm$ 0.430 & 6656.11 $\pm$ 0.86 & -63.46 $\pm$ 0.86 & 53.339 $\pm$ 0.007 &  -  &  -  &  -  &  -  &  -   \\
59862.7025 & 13765.218 $\pm$ 0.438 & 6655.06 $\pm$ 0.88 & -68.37 $\pm$ 0.88 & 53.379 $\pm$ 0.007 &  -  &  -  &  -  &  -  &  -   \\
59866.7833 & 13769.552 $\pm$ 0.415 & 6653.12 $\pm$ 0.83 & -68.51 $\pm$ 0.83 & 53.414 $\pm$ 0.007 &  0.16565 $\pm$ 0.00008  &  0.20154 $\pm$ 0.00004  &  0.31494 $\pm$ 0.00003  &  0.12898 $\pm$ 0.00008  &  -4.6361 $\pm$ 0.0005   \\
59869.7052 & 13767.729 $\pm$ 0.550 & 6652.27 $\pm$ 1.10 & -67.25 $\pm$ 1.10 & 53.448 $\pm$ 0.009 &  0.15913 $\pm$ 0.00013  &  0.20383 $\pm$ 0.00005  &  0.31583 $\pm$ 0.00004  &  0.12273 $\pm$ 0.00013  &  -4.6780 $\pm$ 0.0009    \\
59873.8282 & 13767.109 $\pm$ 0.381 & 6660.57 $\pm$ 0.76 & -66.65 $\pm$ 0.76 & 53.326 $\pm$ 0.006 &  0.17056 $\pm$ 0.00007  &  0.20338 $\pm$ 0.00003  &  0.31657 $\pm$ 0.00003  &  0.13368 $\pm$ 0.00006  &  -4.6070 $\pm$ 0.0004    \\
59878.8207 & 13770.787 $\pm$ 0.480 & 6659.34 $\pm$ 0.96 & -64.34 $\pm$ 0.96 & 53.365 $\pm$ 0.008 &  0.16503 $\pm$ 0.00010   &  0.20364 $\pm$ 0.00004  &  0.31109 $\pm$ 0.00003  &  0.12838 $\pm$ 0.00010   &  -4.6399 $\pm$ 0.0006   \\
59882.6616 & 13764.816 $\pm$ 0.440 & 6655.10 $\pm$ 0.88 & -63.11 $\pm$ 0.88 & 53.353 $\pm$ 0.007 &  0.16659 $\pm$ 0.00009  &  0.20208 $\pm$ 0.00004  &  0.30992 $\pm$ 0.00003  &  0.12987 $\pm$ 0.00009  &  -4.6304 $\pm$ 0.0006   \\
59887.7272 & 13764.806 $\pm$ 0.412 & 6649.31 $\pm$ 0.82 & -67.58 $\pm$ 0.82 & 53.391 $\pm$ 0.007 &  0.16399 $\pm$ 0.00008  &  0.20061 $\pm$ 0.00004  &  0.31071 $\pm$ 0.00003  &  0.12739 $\pm$ 0.00008  &  -4.6464 $\pm$ 0.0005   \\
59891.7139 & 13761.835 $\pm$ 0.458 & 6653.92 $\pm$ 0.92 & -69.11 $\pm$ 0.92 & 53.385 $\pm$ 0.008 &  0.16362 $\pm$ 0.00010   &  0.20217 $\pm$ 0.00004  &  0.31097 $\pm$ 0.00003  &  0.12703 $\pm$ 0.00009  &  -4.6487 $\pm$ 0.0006   \\
59896.6254 & 13768.783 $\pm$ 0.659 & 6653.00 $\pm$ 1.32 & -65.03 $\pm$ 1.32 & 53.456 $\pm$ 0.011 &  0.15302 $\pm$ 0.00018  &  0.20277 $\pm$ 0.00006  &  0.31290 $\pm$ 0.00005   &  0.11688 $\pm$ 0.00017  &  -4.7213 $\pm$ 0.0013   \\
59904.7560 & 13770.829 $\pm$ 0.363 & 6656.72 $\pm$ 0.73 & -66.21 $\pm$ 0.73 & 53.321 $\pm$ 0.006 &  0.17073 $\pm$ 0.00006  &  0.20287 $\pm$ 0.00003  &  0.30939 $\pm$ 0.00002  &  0.13384 $\pm$ 0.00006  &  -4.6060 $\pm$ 0.0004    \\
59907.6364 & 13773.043 $\pm$ 0.455 & 6655.93 $\pm$ 0.91 & -66.64 $\pm$ 0.91 & 53.354 $\pm$ 0.008 &  0.16477 $\pm$ 0.00010   &  0.20227 $\pm$ 0.00004  &  0.30601 $\pm$ 0.00003  &  0.12813 $\pm$ 0.00009  &  -4.6415 $\pm$ 0.0006   \\
59910.7990 & 13767.999 $\pm$ 0.437 & 6651.79 $\pm$ 0.87 & -65.61 $\pm$ 0.87 & 53.387 $\pm$ 0.007 &  0.16489 $\pm$ 0.00009  &  0.20098 $\pm$ 0.00004  &  0.30387 $\pm$ 0.00003  &  0.12825 $\pm$ 0.00009  &  -4.6408 $\pm$ 0.0006   \\
59914.6641 & 13770.143 $\pm$ 0.421 & 6651.20 $\pm$ 0.84 & -66.62 $\pm$ 0.84 & 53.387 $\pm$ 0.007 &  0.16152 $\pm$ 0.00008  &  0.19824 $\pm$ 0.00004  &  0.31399 $\pm$ 0.00003  &  0.12502 $\pm$ 0.00008  &  -4.6621 $\pm$ 0.0005   \\
59917.7566 & 13772.663 $\pm$ 0.369 & 6656.29 $\pm$ 0.74 & -67.11 $\pm$ 0.74 & 53.357 $\pm$ 0.006 &  0.16859 $\pm$ 0.00006  &  0.20345 $\pm$ 0.00003  &  0.30949 $\pm$ 0.00002  &  0.13179 $\pm$ 0.00006  &  -4.6184 $\pm$ 0.0004   \\
59925.5992 & 13768.672 $\pm$ 0.667 & 6656.78 $\pm$ 1.33 & -67.74 $\pm$ 1.33 & 53.448 $\pm$ 0.011 &  0.15623 $\pm$ 0.00018  &  0.20180 $\pm$ 0.00006   &  0.30036 $\pm$ 0.00005  &  0.11995 $\pm$ 0.00018  &  -4.6980 $\pm$ 0.0013    \\
59928.8433 & 13768.846 $\pm$ 0.336 & 6657.50 $\pm$ 0.67 & -67.13 $\pm$ 0.67 & 53.313 $\pm$ 0.006 &  0.16981 $\pm$ 0.00006  &  0.20056 $\pm$ 0.00003  &  0.30668 $\pm$ 0.00002  &  0.13296 $\pm$ 0.00006  &  -4.6113 $\pm$ 0.0003   \\
59933.7624 & 13767.626 $\pm$ 0.476 & 6658.13 $\pm$ 0.95 & -64.36 $\pm$ 0.95 & 53.325 $\pm$ 0.008 &  0.16451 $\pm$ 0.00011  &  0.20213 $\pm$ 0.00004  &  0.30629 $\pm$ 0.00003  &  0.12789 $\pm$ 0.00010   &  -4.6431 $\pm$ 0.0007   \\
59936.7580 & 13769.622 $\pm$ 0.387 & 6657.99 $\pm$ 0.77 & -64.93 $\pm$ 0.77 & 53.351 $\pm$ 0.007 &  0.16918 $\pm$ 0.00007  &  0.20285 $\pm$ 0.00003  &  0.30667 $\pm$ 0.00003  &  0.13235 $\pm$ 0.00007  &  -4.6150 $\pm$ 0.0004    \\
59939.6448 & 13766.016 $\pm$ 0.609 & 6653.33 $\pm$ 1.22 & -65.25 $\pm$ 1.22 & 53.456 $\pm$ 0.010 &  0.15870 $\pm$ 0.00016   &  0.20040 $\pm$ 0.00006   &  0.30627 $\pm$ 0.00005  &  0.12232 $\pm$ 0.00015  &  -4.6809 $\pm$ 0.0011   \\
59942.6498 & 13765.459 $\pm$ 0.694 & 6652.83 $\pm$ 1.39 & -66.63 $\pm$ 1.39 & 53.484 $\pm$ 0.012 &  0.15435 $\pm$ 0.00019  &  0.20033 $\pm$ 0.00007  &  0.30613 $\pm$ 0.00005  &  0.11815 $\pm$ 0.00018  &  -4.7115 $\pm$ 0.0014   \\
59945.7360 & 13769.167 $\pm$ 0.615 & 6650.24 $\pm$ 1.23 & -67.17 $\pm$ 1.23 & 53.466 $\pm$ 0.011 &  0.15852 $\pm$ 0.00016  &  0.20180 $\pm$ 0.00006   &  0.30707 $\pm$ 0.00005  &  0.12215 $\pm$ 0.00015  &  -4.6821 $\pm$ 0.0011   \\
59953.7859 & 13770.941 $\pm$ 0.449 & 6654.97 $\pm$ 0.90 & -64.81 $\pm$ 0.90 & 53.351 $\pm$ 0.008 &  0.16617 $\pm$ 0.00010   &  0.20289 $\pm$ 0.00004  &  0.31134 $\pm$ 0.00003  &  0.12947 $\pm$ 0.00009  &  -4.6329 $\pm$ 0.0006   \\
59956.5932 & 13772.233 $\pm$ 0.565 & 6654.83 $\pm$ 1.13 & -63.22 $\pm$ 1.13 & 53.392 $\pm$ 0.010 &  0.16033 $\pm$ 0.00015  &  0.20025 $\pm$ 0.00005  &  0.30758 $\pm$ 0.00004  &  0.12388 $\pm$ 0.00014  &  -4.6700 $\pm$ 0.001      \\
59999.6580 & 13768.717 $\pm$ 0.395 & 6653.50 $\pm$ 0.79 & -66.89 $\pm$ 0.79 & 53.339 $\pm$ 0.007 &  0.16792 $\pm$ 0.00008  &  0.20013 $\pm$ 0.00003  &  0.30928 $\pm$ 0.00003  &  0.13115 $\pm$ 0.00008  &  -4.6224 $\pm$ 0.0005   \\
60002.5551 & 13768.577 $\pm$ 0.429 & 6653.66 $\pm$ 0.86 & -66.77 $\pm$ 0.86 & 53.342 $\pm$ 0.007 &  -  &  -  &  -  &  -  &  --   \\
60005.5627 & 13770.959 $\pm$ 0.458 & 6656.11 $\pm$ 0.92 & -65.76 $\pm$ 0.92 & 53.355 $\pm$ 0.008 &  0.16567 $\pm$ 0.00010   &  0.19971 $\pm$ 0.00004  &  0.3063 $\pm$ 0.00003   &  0.12899 $\pm$ 0.00010   &  -4.6360 $\pm$ 0.0006    \\
60008.5693 & 13769.449 $\pm$ 0.419 & 6655.66 $\pm$ 0.84 & -64.69 $\pm$ 0.84 & 53.329 $\pm$ 0.007 &  0.16790 $\pm$ 0.00009   &  0.19819 $\pm$ 0.00003  &  0.30694 $\pm$ 0.00003  &  0.13113 $\pm$ 0.00008  &  -4.6225 $\pm$ 0.0005  
\enddata

\label{espresso_vals_fb}
\end{deluxetable}

\section{Allesfitter Corner Plots}
\restartappendixnumbering

In this appendix, we show the corner plots from our \texttt{allesfitter} fit.

\begin{figure*}[!h]
    \centering
    \includegraphics[width=\textwidth]{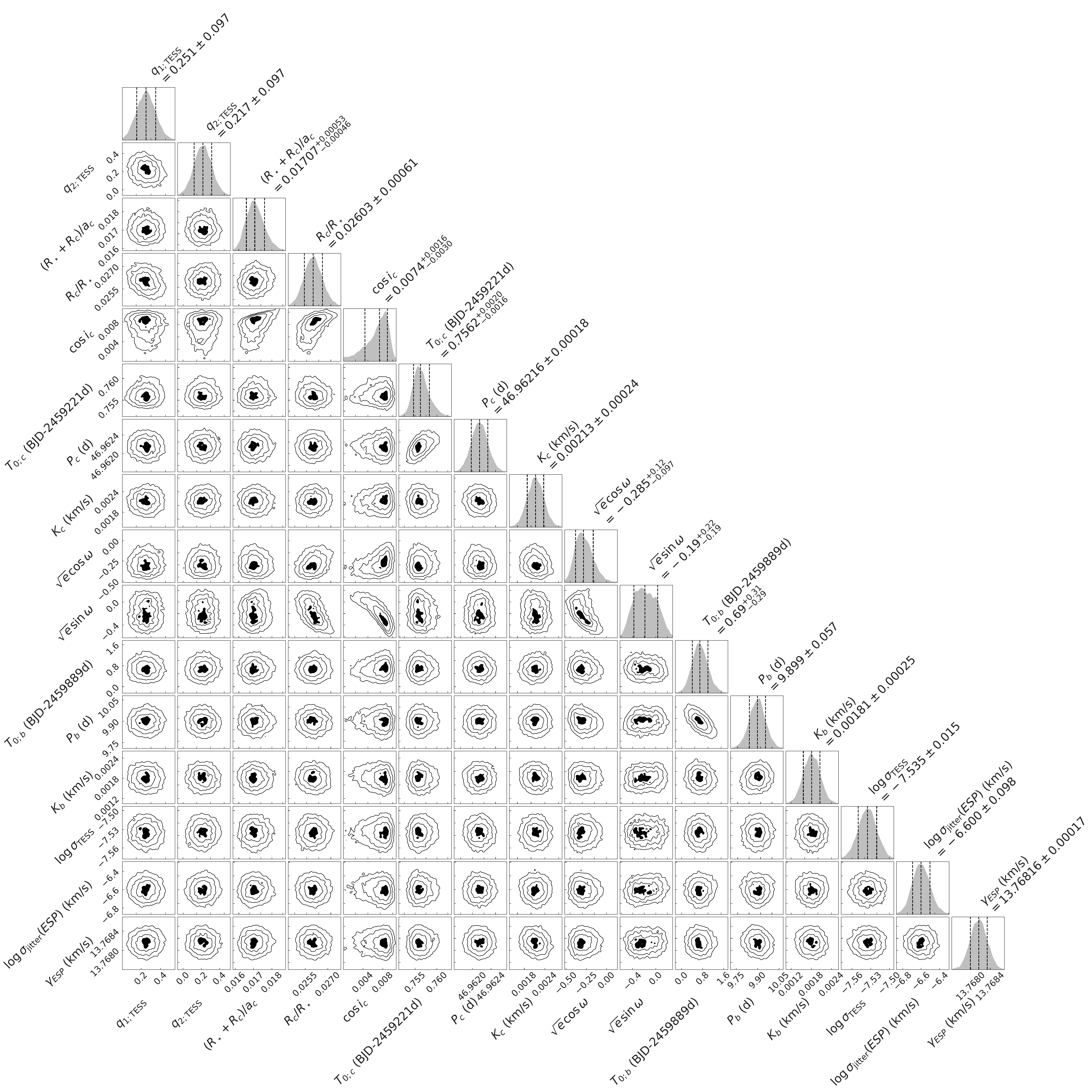}
    \caption{Corner plots of modeled parameters obtained from \texttt{allesfitter}.}
    \label{fig:corner_fitted}
\end{figure*}

\begin{figure*}[!h]
    \centering
    \includegraphics[width=\textwidth]{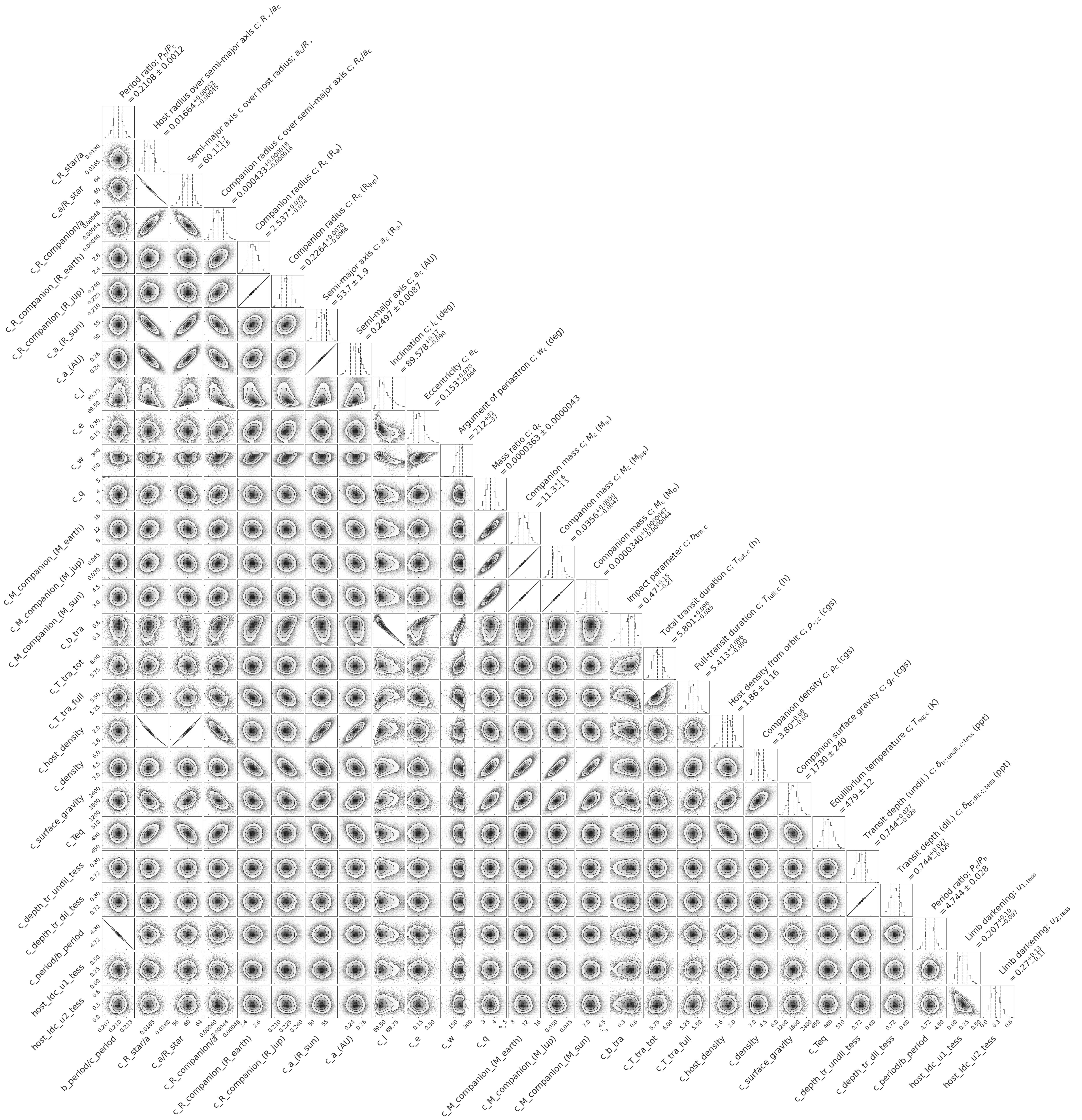}
    \caption{Corner plots of derived parameters obtained from \texttt{allesfitter}.}
    \label{fig:corner_derived}
\end{figure*}

\section{Joint Fit with PFS} \label{sec:pfsfit}
\restartappendixnumbering

In this appendix, we show the corner plot for the fitted parameters of an alternative joint fit that includes the PFS RVs. We include all of the parameters from the joint fit described in Section \ref{subsec:fit} and also include jitter term and offset term for the PFS dataset. While the parameters are largely the same between this fit and the final fit, the posterior distribution for $P_b$ is bimodal here.

\begin{figure*}[!h]
    \centering
    \includegraphics[width=\textwidth]{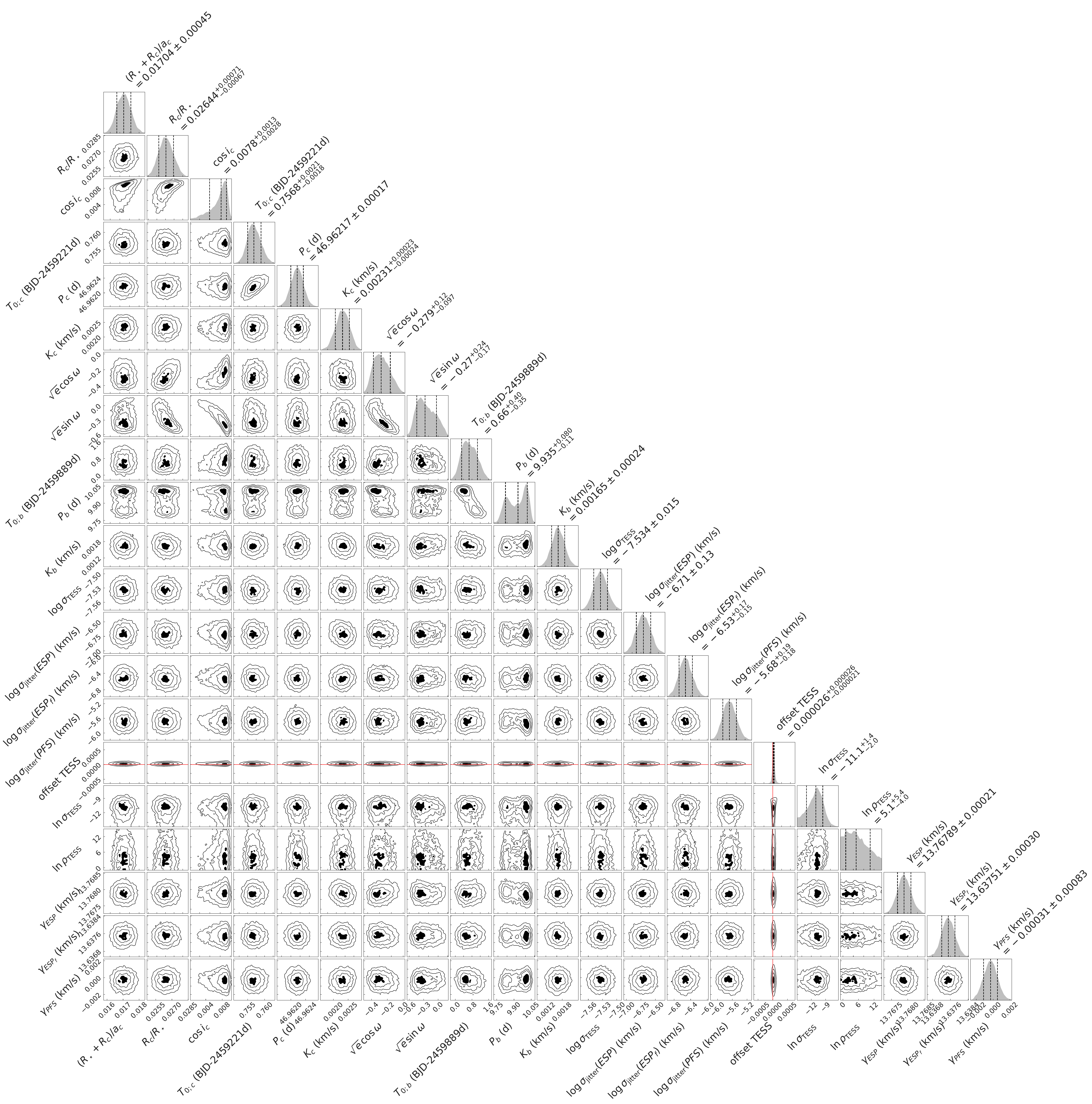}
    \caption{Corner plots of modeled parameters obtained from an \texttt{allesfitter} joint fit including the PFS RVs.}
    \label{fig:corner_fitted_pfs}
\end{figure*}

\section{Transit Timing Variations}
\restartappendixnumbering

In this appendix, we show the potential TTVs of HD 35843 c.

\begin{figure*}[!h]
    \centering
    \includegraphics[width=\textwidth]{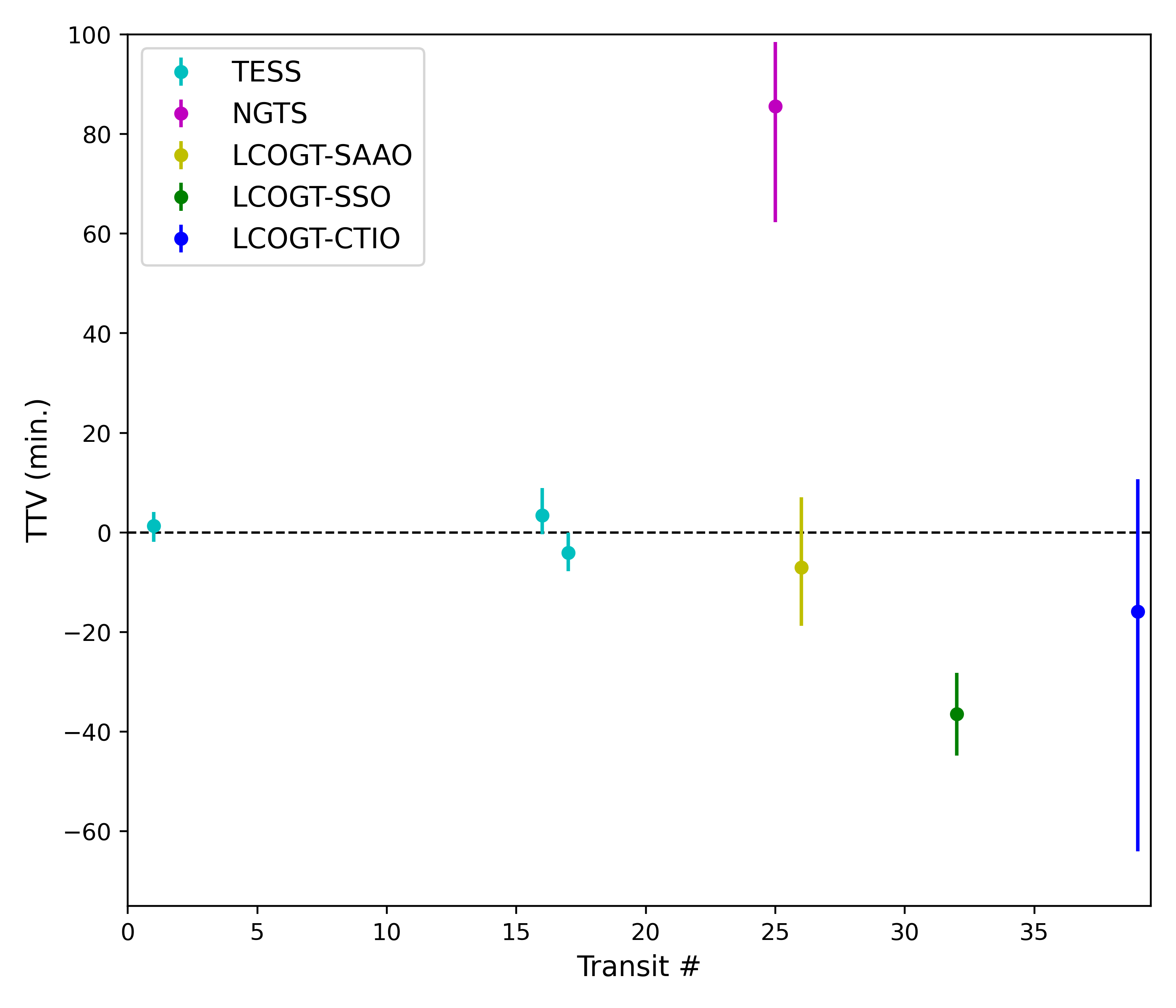}
    \caption{HD 35843 c TTVs for \textit{TESS}, LCOGT, and NGTS observations.}
    \label{fig:ttvs}
\end{figure*}

\end{document}